\documentclass{jfm}

\usepackage{graphicx}
\graphicspath{{Figures/}}
\usepackage{newtxtext}
\usepackage{amsmath}
\usepackage{caption}
\usepackage{newtxmath}
\usepackage{subfigure}

\usepackage{booktabs}
\usepackage{array}

\usepackage{xcolor}
\usepackage[normalem]{ulem} 
\newcommand\hl{\bgroup\markoverwith
	{\textcolor{yellow}{\rule[-.5ex]{2pt}{2.5ex}}}\ULon}

\DeclareMathAlphabet{\mathpzc}{OT1}{pzc}{m}{it}
\usepackage{bm}      
\usepackage{bbm}     
\usepackage{dsfont}  
\usepackage{yfonts}  

\usepackage{natbib}
\usepackage{hyperref}
\usepackage{lineno}
\hypersetup{
    colorlinks = true,
    urlcolor   = blue,
    citecolor  = blue,
    linkcolor=blue
}

\newcommand{\RomanNumeralCaps}[1]

\title{Role of acoustic metasurface in the nonlinear mode--mode interaction and breakdown of hypersonic boundary layer}

\author{Yifeng Chen\aff{1},
  Peixu Guo\aff{1}\thanks{Email address of the corresponding author: peixu.guo@polyu.edu.hk},   \and Chihyung Wen\aff{1}}

\affiliation{\aff{1} Department of Aeronautical and Aviation Engineering,The Hong Kong Polytechnic University, Kowloon, Hong Kong SAR, China}

\begin{document}
	\maketitle

\begin{abstract} 
Boundary-layer instability and transition control have drawn extensive attention from the hypersonic community. The acoustic metasurface has become a promising passive control method, owing to its straightforward implementation and lack of requirement for external energy input. Currently, the effects of the acoustic metasurface on the early and late transitional stages remain evidently less understood than the linear instability stage. In this study,  the transitional stage of a flat-plate boundary layer at Mach 6 is investigated, with a particular emphasis on the nonlinear mode--mode interaction. The acoustic metasurface is modelled by the well-validated time domain impedance boundary condition (TDIBC). First, the resolvent analysis is performed to obtain the optimal disturbances, which reports two peaks corresponding to the oblique first mode and the planar Mack second mode. These two most amplified responses are regarded as the dominant primary instabilities that trigger the transition. Subsequently, both optimal forcings are introduced upstream in the direct numerical simulation, which leads to pronounced detuned modes before breakdown. The takeaway is that the location of the acoustic metasurface is significant in minimizing skin friction and delaying transition onset simultaneously. The placement of the metasurface in the linearly unstable region of the second mode delays the transition, which is due to the suppressed streak in the oblique breakdown scenario. However, in the late stage of the transition, the acoustic metasurface induces an undesirable increment of skin friction overshoot due to the augmented shear-induced dissipation work, which mainly arises from reinforced detuned modes related to the combination resonance. Meanwhile, by restricting the location of the metasurface upstream of the overshoot region, this undesirable augment of skin friction can be eliminated. As a result, the reasonable placement of metasurface is crucial to damping the early instability while causing less negative impacts on the late transitional stage.
\end{abstract}


\section{Introduction}\label{sec:intro}

One of the key obstacles to the development of hypersonic vehicles has been the prediction and control of the laminar-turbulent boundary layer transition. The performance of hypersonic vehicles may be compromised by the transition, as it would escalate skin friction and heat flux (Fedorov \citeyear{Fedorov2011}). Therefore, accurate transition prediction and effective control (or delay) technologies are crucial for the design of hypersonic vehicles, particularly regarding the thermal protection system. Transition induced by freestream disturbances with low amplitudes usually follows four stages: receptivity (excitation of initial disturbance inside the boundary layer), linear instability (including modal and nonmodal growth), nonlinear instability and parametric resonance, and breakdown to turbulence (Morkovin \etal \space \citeyear{morkovin1994transition}). Hypersonic boundary layer transition over two-dimensional or axisymmetric models at a zero angle of attack is typically triggered by the growth of unstable modes, known as the first mode, associated with the vorticity disturbance (Smith \citeyear{smith1989first}), and Mack second mode of acoustic nature (Mack \citeyear{Mack1975}; Fedorov\citeyear{Fedorov2003a}; Ma \& Zhong \citeyear{Ma2003}; Chen \etal \space \citeyear{Chen2023b}). In realistic flight or wind tunnel conditions, these two types of instabilities tend to coexist and compete with one another. 

Regarding the breakdown or parametric resonance, several scenarios have been confirmed by numerous direct numerical simulations (DNS) and experimental studies. These include the first-mode dominated subharmonic resonance (Kosinov \etal \space \citeyear{kosinov1994experiments}), the first-mode dominated oblique breakdown (Mayer \etal \space \citeyear{Mayer2011}), the second-mode oblique breakdown (Hartman \etal \space \citeyear{Hartman2021}), the second-mode dominated fundamental resonance (Kennedy \etal \space \citeyear{Kennedy2022}; Hader  \& Fasal \citeyear{Hader2019}; Unnikrishnan \& Gaitonde \citeyear{Kennedy2022}), and the second-mode subharmonic resonance (Bountin  \etal\space\citeyear{Bountin2008}; Franko \& Lele \citeyear{Franko2013}). These conventional transition scenarios are characterised by tuned modes with frequencies that are integer times $f_{1/2}$. Here, $f_{1/2}$ refers to half of the frequency of the primary instability wave. The dominating transition scenario depends on the specific numerical forcing with given flow and geometric conditions. In a realistic disturbance environment, several primary instabilities can be excited simultaneously and coexist. In this case, complex mode--mode interactions emerge, and detuned modes may be pronounced (Fezer \& Kloker 2000 \citeyear{fezer2000spatial}). As reported by Chen \etal \space(\citeyear{Chen2017}) for the flared-cone model at Peking University, multiple primary instabilities give rise to complex nonlinear interactions and significant energy transfer between low- and high-frequency components. A possible combination resonance is also indicated, which differs from that induced by a single primary instability (Hader  \& Fasal \citeyear{Hader2019}; Franko \& Lele \citeyear{Franko2013}). Nonetheless, direct evidence in the transitional stage was not provided. Recently, Guo \etal\space  (\citeyear{Guo2023a}) have demonstrated the significance of the non-conventional combination resonance associated with modal detuning in the high-speed boundary layer transition. The nonlinear interaction between low-frequency first mode and high-frequency second mode can produce a new breakdown scenario. This multiple-primary-instabilities scenario is believed to be noteworthy, since multiple instabilities are seeded in realistic boundary layers. As an extension, the effect of flow control on the breakdown scenario is of particular interest.

Boundary-layer transition control strategies are useful in delaying the occurrence of transition. These strategies can be categorised into active methods, such as wall cooling and heating (Zhao \etal \space \citeyear{Zhao2018}; Zhou \etal \space \citeyear{zhou2022control}), and passive methods, such as the acoustic metasurface implemented by the porous coating (Fedorov \etal \space \citeyear{Fedorov2001}; Zhao \etal \space \citeyear{Zhao2022}). The acoustic metasurface, including the absorptive acoustic metasurface, the impedance-near-zero acoustic metasurface, and the reflection-controlled acoustic metasurface, was proposed mainly to suppress the second mode of acoustic nature (Zhao \etal \space \citeyear{Zhao2022}). The passive approach is flexible for implementation and requires no additional energy input, which has attracted growing interest from both academia and industry. Experimental observations have shown that the porous coating can successfully postpone the transition onset (Rasheed \etal \space \citeyear{Rasheed2002}; Fedorov \etal \space \citeyear{Fedorov2003b}; Wagner \etal \space \citeyear{Wagner2013}; Wagner \etal \space \citeyear{Wagner2014}). However, porous coating leads to an undesirable increment of wall friction and flux during the late stage of transitional flow. This phenomenon has been observed experimentally (Wartemann \etal \space \citeyear{Wartemann2015}) but remains poorly understood. 

In the linear stability stage, the effect of the porous coating can be theoretically described and satisfactorily predicted using linear stability theory (LST) with the deduced acoustic impedance boundary condition (Fedorov \etal \space \citeyear{Fedorov2001}). The physical mechanism by which the porous coating influences the linear stage has been widely studied (Fedorov \etal \space \citeyear{Fedorov2003b}; Tian \& Wen \citeyear{tian2021growth}; Chen \etal \space \citeyear{Chen2023a}; Ji \etal \space \citeyear{ji2023impact}). It is known that viscous effects within the pores of the porous coating can dissipate the second-mode instability. Acoustic wave reflection is also linked to regimes with phase cancellation or reinforcement (Bres \etal \space \citeyear{Bres2013}). The porous coating effect in the nonlinear regime has also been the subject of early research (Hader \etal \space \citeyear{Hader2013}; Hader \etal \space \citeyear{Hader2014}). It was discovered that the second-mode fundamental breakdown caused a delayed transition onset. However, these investigations are restricted to single-primary-mode perturbations and the resulting typical breakdown situations, which may not accurately represent the transitional flow with broadband disturbances in wind tunnels or flight tests. Recently, Sousa \etal \space (\citeyear{Sousa2024}) investigated the hypersonic transition delay over a broadband wall impedance using dynamic large-eddy simulation. Good agreement was observed with experimental data in the literature. The corresponding wind tunnel experiment features a high-enthalpy flow with a highly cooled wall condition, where the first mode is significantly suppressed. Consequently, the nonlinear interaction between the low-frequency first mode and the high-frequency second mode was not considered in their simulation. The resulting breakdown is more likely to be a fundamental breakdown induced by the second mode, which represents the only dominant instability under such cold-wall conditions.

The current work is to examine the impact of porous coatings on the nonlinear mode--mode interaction and the resulting breakdown process. Two research objectives are included: 1) to reveal the transition delay mechanism by the acoustic metasurface subject to multiple primary instabilities; 2) to explain why the skin friction is augmented in the late transitional stage. To this end, the effect of the metasurface is evaluated using a modelled impedance boundary condition (IBC) to avoid meshing the cavity, which can save computational costs. The time-domain impedance boundary condition (TDIBC) (Fung \& Ju \citeyear{Fung2004}; Douasbin \etal \space \citeyear{Douasbin2018}; Fi{\'e}vet \etal \space \citeyear{fievet2021numerical}; Guo \etal \space \citeyear{Guo2023b}) can be efficiently incorporated into a Navier-Stokes solver and handle broadband disturbances. The accuracy and efficiency of TDIBC have been shown in the simulation of late transitional and turbulent boundary layers (Scalo \etal \space \citeyear{Scalo2015}; Chen \& Scalo \citeyear{Chen2021b}; Sousa \etal \space \citeyear{Sousa2024}). It is anticipated that the mode--mode interaction and the resulting breakdown scenario with the acoustic metasurface will be revealed to understand why the transition onset is delayed. Moreover, regarding the enhanced skin friction in the late transition stage,  which has also been observed in the experiments by Wagner \etal \space (\citeyear{Wagner2013}; \citeyear{Wagner2014}), the energy budget analysis and bi-Fourier analysis are performed to explore the underlying physics.

The remainder of this paper is organised as follows. \S\space \ref{Methodology} provides an overview of the investigated model and flow parameters, the formulation of TDIBC, stability analysis tools, and numerical methods for direct numerical simulation (DNS). \S\space \ref{Validation} presents the fitting results of TDIBC using a theoretical model, the results of the stability analysis, and the effect of TDIBC on the linear stage. Comparative transition simulations induced by multiple disturbances with and without the acoustic metasurface are shown in \S\space \ref{Results}. The instantaneous and time-averaged flow fields, Bi-Fourier analysis and energy budget analysis are included. The summary of this work is concluded in \S\space \ref{Conclusion}.

\section{Flow condition and methodology}\label{Methodology}

The freestream flow condition investigated is based on a wind tunnel experiment by Bountin \etal \space(\citeyear{Bountin2013}) at Mach 6 and unit Reynolds number $1.05 \times {10^{7}}{\text{m}^{ - 1}}$. The freestream temperature $T_\infty$ is 43.18 K and the isothermal wall temperature (room temperature) $T_w$ = 293 K are employed. In this study, the subscripts "$\infty $" and "$w$"  refer to the values at freestream and on the wall, respectively. The freestream variables are utilised for nondimensionalisation, except that ${\rho _\infty }u_\infty ^2$ is utilised for pressure. Unless otherwise stated, the reference length ${L_\textit{ref}}$ is taken as 1 mm. Without unit markings, the following physical quantities are presented in their dimensionless form.

\subsection{Direct numerical simulation}
The governing equation for the computation of flow  fields is the compressible Navier--Stokes equation in the conservation form:

\begin{equation}\label{eq2.1}
\frac{{\partial {\boldsymbol{U}}}}{{\partial t}} + \frac{{\partial {\boldsymbol{F}}}}{{\partial x}} + \frac{{\partial {\boldsymbol{G}}}}{{\partial y}} + \frac{{\partial {\boldsymbol{H}}}}{{\partial z}} =  {\mathsfbi{B}\boldsymbol{f'}},
\end{equation}
where ${\boldsymbol{U}}{\rm{ = (}}\rho {\rm{, \ }}\rho u{\rm{, \ }}\rho v{\rm{, \ }}\rho w{\rm{, \ }}\rho e{{\rm{)}}^{\rm T}}$ is the vector of conservative variables, and ${\boldsymbol{F}}$, ${\boldsymbol{G}}$, and ${\boldsymbol{H}}$ are the vectors of inviscid and viscous fluxes. Detailed expressions can be found in Anderson (\citeyear{anderson1995computational}). Here, $\rho $ is density, and $u$, $v$, and $w$ are velocities in the Cartesian $x$, $y$, and $z$ directions, respectively. Total energy per unit mass is denoted by $e$. The matrix ${\mathsfbi{B}}$ constrains the forcing vector (${\boldsymbol{f'}}$) to be added at a certain region, which is set to be $x$ = 0.04 m in this study. In the DNS, the optimal forcings obtained by resolvent analysis are employed to initiate the unsteady flow. 

The perfect gas model is employed with a constant specific heat ratio of 1.4. Furthermore, the dynamic viscosity is calculated using Sutherland's law, and the thermal conductivity coefficient is calculated based on a constant Prandtl number 0.72. 
The simulation is constituted by two steps. In the first step, the base flow is calculated, and the right-hand-side term of equation (\ref{eq2.1}) is set to zero. Subsequently, the resolvent analysis or DNS is performed on the converged base flow. It is assumed that the vector of conservation variable can be decomposed into its base-flow and perturbation parts:
\begin{equation}\label{eq2.2}
{\boldsymbol{U}}(x,y,z,t) = {\boldsymbol{\bar U}}(x,y) + {\boldsymbol{U'}}(x,y,z,t).
\end{equation}
Hereafter, the overbar and prime represent mean variables and perturbation variables, respectively.

The DNS is performed using a multi-block parallel finite difference solver called OpenCFD, which has been successfully employed in high-speed transitional and turbulent simulations (Li \etal \space \citeyear{xinliang2002direct}; Li \etal \space \citeyear{xin2009acoustic}). The inviscid flux splitting is implemented by the Steger-Warming scheme, and a seven-order weight essential non-oscillation (WENO) scheme is employed for the reconstruction of the split flux. The six-order central difference scheme is utilised for viscous flux discretisation, and a third-order Runge-Kutta method is employed for temporal integration. With regard to the boundary condition, the left boundary of the computational domain is freestream, and the wall is set to be isothermal, non-slip and no-penetration or with TDIBC in a certain region. The upper and outer boundaries are addressed by extrapolation. The spanwise boundary is set to be periodic.  

\subsection{Time domain impedance boundary condition}
DNS with realistic metasurface microstructures can be conducted to investigate the transition control mechanism. However, it is computationally expensive to mesh numerous micro-cavities. To efficiently and simultaneously simulate the interactions between acoustic metasurfaces and multi-modal waves, the application of an impedance boundary condition is recommended. The early acoustic impedance models of metasurface were all built in the frequency domain and applied with a single disturbance frequency in the hypersonic boundary-layer transition studies (Fedorov \etal \space \citeyear{Fedorov2001}; M{\"o}ser \& Michael \citeyear{moser2009engineering}), which are not suitable for transition considering broadband disturbances. In this study, a multi-pole broadband impedance model with a piecewise linear recursive convolution technique proposed by Fung \& Ju (\citeyear{Fung2004}) will be revised and incorporated into the current DNS code afterwards.  This embedded boundary condition enables efficient simulations of broadband-disturbance propagation by transforming the acoustic impedance boundary condition from the frequency domain into the time domain.

The reflective coefficient $\tilde R$ in the frequency domain is defined as
\begin{equation}\label{eq2.33}
\tilde R(\omega ) = {{{A^{\textit{out}}}(\omega )} \mathord{\left/
		{\vphantom {{{A^{\textit{out}}}(\omega )} {{A^{\textit{in}}}}}} \right.
		\kern-\nulldelimiterspace} {{A^{\textit{in}}}}}(\omega ),
\end{equation}
where ${{A^{\textit{in}}}}$ and ${{A^{\textit{out}}}}$ are the amplitudes of ingoing and outgoing waves, respectively. The wall softness $\tilde S$ is given by
\begin{equation}\label{eq2.11}
\tilde S(\omega ) = 1 + \tilde R(\omega ).
\end{equation}
 Physically $\tilde S$ = 2 and $\tilde S$ = 1 when the wall is totally reflective (rigid wall) and absorptive (soft wall), respectively. Combining equations (\ref{eq2.33}) and (\ref{eq2.11}) gives rise to 
\begin{equation}\label{eq2.12}
{A^{\textit{out}}}(\omega ) =  - {A^{\textit{in}}}(\omega ) + \tilde S(\omega ){A^{\textit{in}}}(\omega ).
\end{equation}
The softness can be approximated by a multi-oscillator model proposed by Fung \& Ju (\citeyear{Fung2004})
\begin{equation}\label{eq2.13}
\tilde S(s) = \mathop \sum \limits_{k = 1}^{{n_0}} \left[ {\frac{{{\mu _k}}}{{s - {p_k}}} + \frac{{\mu _k^\dag }}{{s - p_k^\dag }}} \right] + {C_0},\: s = {\rm{i}}\omega,
\end{equation}
where ${p_k}$  and $p_k^\dag $ are poles of pole base function, and ${\mu _k} = {\rm{i}} \cdot {\rm{Residue}}\left[ {\tilde S(s),{p_k}} \right]$. The superscript $\dag $  refers to complex conjugate, and $n_0$ is the total number of pole pairs. In addition, ${C_0}$ is a constant real number. The utilisation of this constant number to improve the accuracy of the model was also reported by Fi{\'e}vet \etal \space (\citeyear{fievet2021numerical}). Due to the requirement of the reality of the signal (Rienstra \citeyear{Rienstra2006}), the softness in the time domain must be real and ${\tilde S^\dag }(s) = \tilde S( - s)$. The values of poles and residue are obtained from a nonlinear fitting process (Douasbin \etal \space \citeyear{Douasbin2018}) to approximate the softness calculated from a well-validated impedance model, which considers the effect of high-order diffraction (Zhao \etal \space \citeyear{Zhao2018a}). The authors have fully validated the impedance model by comparing it with the DNS result over meshed metasurface (Zhao  \etal \space \citeyear{Zhao2019}; Guo \etal \space \citeyear{Guo2023b}).
Then, the softness in the time domain can be obtained using the inverse Fourier transform:
\begin{equation}\label{eq2.14}
\tilde S(t) = \frac{1}{{2\pi }}\int_{ - \infty }^\infty  {\tilde S(\omega ){e^{{\rm{i}}\omega t}}} d\omega .
\end{equation}
Using the Residue theorem, one may obtain
\begin{equation}\label{eq2.15}
\tilde S(t) = \left( {\mathop \sum \limits_{k = 1}^{{n_0}} {\mu _k}{e^{{p_k}t}} + \mathop \sum \limits_{k = 1}^{{n_0}} \mu _k^\dag {e^{p_k^\dag t}}} \right)H(t) + {C_0}\delta (t).
\end{equation}
Here, the Heaviside function $H$($t$) indicates that the causality condition is satisfied, and $\delta $ is the Dirac delta function. Taking the inverse Fourier transform of equation (\ref{eq2.12}), one may obtain
\begin{equation}\label{eq2.16}
{A^{\textit{out}}}(t) = ({C_0} - 1){A^{in}}(t) + \int_0^\infty  {\tilde S(\tau ){A^{in}}(t - \tau )} d\tau .
\end{equation}
Substituting equation (\ref{eq2.15}) into equation (\ref{eq2.16}) it yields
\begin{equation}\label{eq2.17}
{A^{\textit{out}}}(t) = ({C_0} - 1){A^{\textit{in}}}(t) + \int_0^\infty  {\mathop \sum \limits_{k = 1}^{{n_0}} ({\mu _k}{e^{{p_k}\tau }} + \mu _k^\dag {e^{p_k^\dag \tau }}){A^{\textit{in}}}(t - \tau )} d\tau .
\end{equation}
For the next time at step $t + \Delta t$, equation (\ref{eq2.17}) can be written as
\begin{equation}\label{eq2.18}
{A^{\textit{out}}}(t + \Delta t) = ({C_0} - 1){A^{\textit{in}}}(t + \Delta t) + \int_0^\infty  {\mathop \sum \limits_{k = 1}^{{n_0}} ({\mu _k}{e^{{p_k}\tau }} + \mu _k^\dag {e^{p_k^\dag \tau }}){A^{\textit{in}}}(t + \Delta t - \tau )} d\tau .
\end{equation}
Here, the convolution in equation (\ref{eq2.18}) can be written as $G_k^{\textit{in}}$, that is
\begin{equation}\label{eq2.19}
G_k^{\textit{in}}(t + \Delta t) = \int_0^\infty  {{\mu _k}{e^{{p_k}\tau }}{A^{\textit{in}}}(t + \Delta t - \tau )} d\tau .
\end{equation}
According to Fung \& Ju (\citeyear{Fung2004}), equation (\ref{eq2.19}) can be calculated by a recursive scheme
\begin{equation}\label{eq2.520}
G_k^{\textit{in}}(t + \Delta t) = {z_k}G_k^{\textit{in}}(t) + {\mu _k}\Delta t[{w_{k0}}{A^{\textit{in}}}(t + \Delta t) + {w_{k1}}{A^{\textit{in}}}(t)].
\end{equation}
Here ${z_k} = {e^{{p_k}\Delta t}}$ , and 

\begin{equation}\label{eq2.21}
\left\{
\begin{aligned}
w_{k0} &= \frac{z_k - 1}{p_k^2 \Delta t^2} - \frac{1}{p_k \Delta t}, \\
w_{k1} &= -\frac{z_k - 1}{p_k^2 \Delta t^2} + \frac{z_k}{p_k \Delta t}.
\end{aligned}
\right.
\end{equation}
With a first-order approximation (Scalo \etal \space \citeyear{Scalo2015}), the ingoing wave at $t + \Delta t$ can be expressed by
\begin{equation}\label{eq2.22}
{A^{\textit{in}}}(t + \Delta t,y) \simeq {A^{\textit{in}}}(t,y) + \sqrt {{T_w}} /M{a_\infty }\Delta t\frac{\partial }{{\partial y}}\left( { - v'(t) + M{a_\infty }\sqrt {{T_w}} p'(t)} \right).
\end{equation}
Then, combining equation (\ref{eq2.18}) and equation (\ref{eq2.22}) gives rise to the fluctuating wall-normal velocity at $t + \Delta t$:
\begin{equation}\label{eq2.23}
v'(t + \Delta t) = \left( {{A^{\textit{in}}}(t + \Delta t){\rm{ + }}{A^{out}}(t + \Delta t)} \right)/2 = {\rm{Real}}\left( {\mathop \sum \limits_{k = 1}^{{n_0}} G_k^{\textit{in}}(t + \Delta t)} \right) + {C_0}{A^{\textit{in}}}(t + \Delta t).
\end{equation}
Finally, the fluctuation pressure at $t + \Delta t$ is
\begin{equation}\label{eq2.244}
p'(t + \Delta t) = {{\left[ {{A^{\textit{in}}}(t + \Delta t) + v'(t + \Delta t)} \right]} \mathord{\left/
		{\vphantom {{\left[ {{A^{in}}(t + \Delta t) + v'(t + \Delta t)} \right]} {\left( {M{a_\infty }\sqrt {{T_{\rm{w}}}} } \right)}}} \right.
		\kern-\nulldelimiterspace} {\left( {M{a_\infty }\sqrt {{T_w}} } \right)}}.
\end{equation}
In the simulation where the metasurface exists locally, the pressure and normal velocity on the wall are updated based on equations (\ref{eq2.23}) and (\ref{eq2.244}) at each time step. The density is calculated by the updated pressure from the equation of state for perfect gases.

\subsection{Resolvent analysis}
The resolvent analysis provides the most energetic response of the flow field due to per unit energy of external forcing. Substituting equation (\ref{eq2.2}) into equation (\ref{eq2.1}),  and subtracting the base-flow equation yield the following form: 

\begin{equation}\label{eq2.3}
\frac{{\partial {\boldsymbol{U'}}}}{{\partial t}} + \frac{{\partial {\boldsymbol{F'}}}}{{\partial x}} + \frac{{\partial {\boldsymbol{G'}}}}{{\partial y}} + \frac{{\partial {\boldsymbol{H'}}}}{{\partial z}} =  {\mathsfbi{B}\boldsymbol{f'}}.
\end{equation}
Considering a small amplitude forcing term to study the instability, one may obtain
\begin{equation}\label{eq2.4}
\frac{{\partial {\boldsymbol{U'}}}}{{\partial t}} = {\boldsymbol{{\mathsfbi A}U'}} + {\mathsfbi{B}\boldsymbol{f'}},
\end{equation}
where matrix ${\mathsfbi{A}}$ is the Jacobian matrix constituted by the base-flow variables. The harmonic assumption is utilised for the small-amplitude perturbation vector  
\begin{equation}\label{eq2.5}
{\boldsymbol{U'}}(x,y,z,t) = {\boldsymbol{\hat U}}(x,y)\exp ({\rm{i}}\beta z - {\rm{i}}\omega t) + {\rm{c}}{\rm{.c}}{\rm{.}},
\end{equation}
where ${\boldsymbol{\hat U}}$ is the complex eigenfunction, $\beta $ is the spanwise wavenumber, $\omega$ is the angular frequency, and ${\rm{c}}{\rm{.c}}{\rm{.}}$ denotes complex conjugate. Similarly, the harmonic forcing can be written as
\begin{equation}\label{eq2.6}
{\boldsymbol{f'}}(x,y,z,t) = {\boldsymbol{\hat f}}(x,y)\exp ({\rm{i}}\beta z - {\rm{i}}\omega t) + {\rm{c}}{\rm{.c}}{\rm{.}}
\end{equation}
Substituting equations (\ref{eq2.5}) and (\ref{eq2.6}) into equation (\ref{eq2.4}) gives
\begin{equation}\label{eq2.7}
{\boldsymbol{\hat U}} = {\mathsfbi{RB}\boldsymbol{\hat f}},{\begin{array}{*{20}{c}}
	{}&{{\mathsfbi{R}} = ( - {\rm{i}}\omega {\mathsfbi{I}}{\rm{ - }}{\mathsfbi{A}})}
	\end{array}^{ - 1}},
\end{equation}
where ${\mathsfbi{R}}$ represents the response matrix, indicating the relationship between the external forcing and the linear response of the system. Here, the identity operator is represented by ${\mathsfbi{I}}$.
In resolvent analysis, the maximal amplification of the energy, i.e. the optimal gain ${\sigma ^2}$, is sought in the parameter space of ($\omega$, $\beta $). The optimal gain is defined by the energy ratio of the output (response) to the input (forcing) of the system

\begin{equation}\label{eq2.8}
{\sigma ^2}(\beta ,\omega ) = \mathop {\max }\limits_{{\boldsymbol{\hat f}}} \frac{{{{\left\| {{\boldsymbol{\hat U}}} \right\|}_E}}}{{{{\left\| {{\mathsfbi{B}\boldsymbol{\hat f}}} \right\|}_E}}}.
\end{equation}
In this study, Chu's energy (Chu \& Kovasznay \citeyear{Chu1958}) is employed to calculate the energy norm

\begin{equation}\label{eq2.9}
{\left\| {{\boldsymbol{\hat U}}} \right\|_E} = \iint\limits_{\Omega} {{{\boldsymbol{U}}^\bot }{\mathsfbi{M}\boldsymbol{\hat U}}}dxdy,
\end{equation}
where $\Omega $ represents the computational domain for resolvent analysis, the superscript $ \bot $ refers to the conjugate transpose, and ${\mathsfbi{M}}$ is the weight operator given by Bugeat  \etal \space (\citeyear{bugeat20193d}). A local energy integral in a dimensionless form is defined by

\begin{equation}\label{eq2.10}
{E_{Chu}}(x) = \frac{1}{2}\int_0^\infty  {\left[ {\bar \rho ({{u'}^2} + {{v'}^2} + {{w'}^2}) + \frac{{\bar T}}{{\gamma Ma_\infty ^2\bar \rho }}{{\rho '}^2} + \frac{{\bar \rho }}{{\gamma (\gamma  - 1)Ma_\infty ^2\bar T}}{{T'}^2}} \right]} {\rm{ }}dy.
\end{equation}

  The optimisation problem in equation equation (\ref{eq2.8}) can be transformed into an eigenvalue problem with respect to $\sigma ^2$, as demonstrated by  Sipp \& Marquet  (\citeyear{sipp2013characterization}). The resulting discrete eigenvalue problem is solved using ARPACK software for given values of $\beta $ and $\omega$  associated with the regular mode (Sorensen \etal \space \citeyear{Sorensen_ARPACK_1996}). Additional details regarding the resolvent analysis solver and the associated validation cases can be found in Hao \etal \space (\citeyear{Hao2023}) and Guo \etal \space (\citeyear{Guo2023b}).

\subsection{Linear stability theory and parabolized stability equation}
To identify transient growth and modal growth captured by resolvent analysis, linear stability theory (LST) and parabolized stability equation (PSE) are employed. PSE further considers the non-parallel effect into compared to LST. Specifically, the LST provides the initial inlet profiles (eigenfunctions) for PSE, and PSE can obtain the non-local evolution of eigenmodes, including the first and second modes of interest here. In PSE, the disturbance ${\boldsymbol{\psi '}}$ is expressed by
\begin{equation}\label{eq2.25}
{\boldsymbol{\psi '}}(x,y,z,t) = {\boldsymbol{\hat \psi }}(x,y)\exp \left( {{\rm{i}}\int_{{x_0}}^x {\alpha d\tilde x}  + {\rm{i}}\beta z - {\rm{i}}\omega t} \right)+ {\rm{c}}{\rm{.c}}{\rm{.}},
\end{equation}
where the vector ${\boldsymbol{\psi }} = {(\rho,u,v,w,T)^{\rm T}}$,  ${\boldsymbol{\hat \psi }}$ and $\alpha $ are the shape function and the complex streamwise wavenumber, respectively, and ${x_0}$ is the initialization location of PSE marching.
Substituting equation (\ref{eq2.25}) into equation (\ref{eq2.3}) and dropping the forcing terms give rise to the PSE governing equation
\begin{equation}\label{eq2.26}
({{\mathsfbi{L}}_0} + {{\mathsfbi{L}}_1}){\boldsymbol{\hat \psi }} + {{\mathsfbi{L}}_2}\frac{{\partial {\boldsymbol{\hat \psi }}}}{{\partial x}} + \frac{{\partial \alpha }}{{\partial x}}{{\mathsfbi{L}}_3}{\boldsymbol{\hat \psi }} = \boldsymbol{0}.
\end{equation}
The effects of the locally parallel flow, the non-parallel base flow, the non-local shape function, and the streamwise-varying wavenumber are absorbed in the base-flow-related operators ${{\mathsfbi{L}}_0}$, ${{\mathsfbi{L}}_1}$, ${{\mathsfbi{L}}_2}$, and ${{\mathsfbi{L}}_3}$, respectively. Detailed expressions of these operators can be found in Paredes (\citeyear{paredes2014advances}). An eigenvalue problem is solved in LST when keeping only the local operator ${{\mathsfbi{L}}_0}$ in equation (\ref{eq2.26}). The calculation is performed by an in-house code CHASES, which integrates the LST, LPSE, NPSE, and sensitivity analyses. The code has been validated by a series of cases for LST and LPSE compared with both theoretical (Guo \etal \space \citeyear{Guo2020}; Guo \etal \space \citeyear{Guo2021}; Guo \etal \space \citeyear{Guo2022a}) and DNS (Cao \etal \space \citeyear{Cao2023}; Hao \etal \space \citeyear{Hao2023}; Guo \etal \space \citeyear{Guo2023a}) results. The detailed formulation and the numerical method
 can be found in the references.

\subsection{Case initialisation for direct numerical simulation}
To explore the effect of the acoustic metasurface in hypersonic transitional flows, three cases subject to the same forcing yet different wall boundary conditions are considered here. In the present Mach 6 state, Guo \etal \space (\citeyear{Guo2023a}) reported two dominant optimal disturbances: low-frequency oblique wave and high-frequency planar wave. These two types of forcing are considered in the transition simulation. The mathematical form of the forcing in DNS is given by

\begin{equation}\label{eq2.24}
{\boldsymbol{f'}}({x_0},y,z,t) = {\boldsymbol{f'}_p}({x_0},y,z,t) + {\boldsymbol{f'}_o}({x_0},y,z,t),
\end{equation}

where ${x_0}$ = 0.04 m is the specified forcing location, and subscripts "$p$" and "$o$" represent the optimal planar wave and oblique wave obtained by resolvent analysis, respectively. According to our previous study (Guo \etal \space \citeyear{Guo2023a}), the additional background noise term produces a neglectable effect on the flowfield, which is thus not considered here. The planar wave forcing is detailed as
\begin{equation}\label{eq2.24}
{\boldsymbol{f'}_p}({x_0},y,z,t) = {\varepsilon _{_p}}{{{\boldsymbol{\hat f}}}_p}({x_0},y)\exp ({\rm{i}}{\beta _p}z - {\rm{i}}{\omega _p}t) + {\rm{c}}{\rm{.c}}{\rm{.}},
\end{equation}
where $\varepsilon$ is the amplitude coefficient. A pair of oblique waves with an opposite spanwise wavenumber is employed, which is expressed by
\begin{equation}\label{eq2.24}
{\boldsymbol{f'}_o}({x_0},y,z,t) = {\varepsilon _{_o}}\left[ {{{{\boldsymbol{\hat f}}}_o}({x_0},y)\exp ({\rm{i}}{\beta _o}z - {\rm{i}}{\omega _o}t) + {{{\boldsymbol{\hat f}}}_{ - o}}({x_0},y)\exp ( - {\rm{i}}{\beta _o}z - {\rm{i}}{\omega _o}t)} \right] + {\rm{c}}{\rm{.c}}{\rm{.}}
\end{equation}

In all transition simulation cases, the initial amplitude ${(\rho u)'_{max}}$ = 0.002 keeps the same for planar wave $({\omega _p},{\beta _p})$ and a pair of oblique waves $({\omega _o}, \pm {\beta _o})$ at $x$ = 0.045 m. This setup excites the first mode and second mode of equal importance near the forcing. In this case, strong mutual interactions can occur with the presence of two primary instabilities. As reported by Guo \etal \space (\citeyear{Guo2023a}), the resulting nonlinear mechanism (combination resonance) is independent of both the initial amplitude ratio and the absolute amplitude of the oblique and planar waves. Therefore, this study focuses on one specified setup of wave amplitudes without loss of generality.

The wall boundary conditions for DNS cases are presented in table \ref{label1}. The boundary condition is of primary interest in this study. In Case 1, only the solid wall boundary condition is employed. For Case 2 with the acoustic metasurface, TDIBC is employed starting from $x$ = 0.115 m to the end of the computational domain, i.e., $x$ = 0.6 m. It is important to note that $x = 0.115$ m corresponds to the neutral point of the optimal plane wave (10, 0), downstream of which the modal growth begins. In Case 3, during the late stage of the transitional flow ($x$ > 0.34m), the TDIBC is replaced by solid walls to mitigate undesirable higher wall friction and flux in the late transitional stage. The selection of $x$ = 0.34 m in Case 3 also considers the modal growth region of the second mode. As illustrated in figure \ref{fig2} ($b$), the modal growth of the optimal plane wave (10, 0) terminates at approximately  $x$ = 0.34 m. A detailed discussion and explanation will be provided in the subsequent sections.

 In the DNS, the computational domain ranges from $x$ = 0 to $x$ = 0.6 m. The number of the grid points is $3060 \times 260 \times 60$, denoting the grid numbers in the streamwise, wall-normal, and spanwise directions, respectively. The corresponding dimensionless grid spacings are $\Delta {x^ + \approx}$  3.13, $\Delta z^ +  \approx $ 5.9, and  $\Delta y_{min}^ +  \approx $ 0.30, which are evaluated based on the procedure described by Guo \etal \space (\citeyear{Guo2022b}, \citeyear{Guo2023a}). Grid convergence of the amplitude evolution of the optimal disturbances has been confirmed (Guo \etal \space \citeyear{Guo2023a}). A good agreement is also reached between the DNS and the stability analysis.
 
	\begin{table} 
	\centering
	\caption{Streamwise range of different wall boundary conditions for transitional DNS cases}
	\label{label1}
	\begin{tabular}{>{\centering\arraybackslash}c>{\centering\arraybackslash}p{4.5cm}>{\centering\arraybackslash}p{4.5cm}}
		\hline
		Case & Solid wall
		& TDIBC \\
		1  & 0--0.6m & -- \\
		2 & 0--0.115m & 0.115m--0.6m \\
		3 & 0--0.115m, 0.34--0.6 m & 0.115m--0.34m \\
	\end{tabular}
\end{table}

\section{Linear instability stage}\label{Validation} 
In this section, the linear stage of the dominant primary instabilities and the effect of the acoustic metasurface are of interest. The multi-pole fitting result is used in the TDIBC to model the acoustic metasurface. The correctness of the embedded TDIBC code is examined in comparison with the result obtained by directly meshing the cavities. The resolvent analysis is employed to capture the optimal response of the boundary layer. The effect of the metasurface on single-frequency disturbances in the low- and high-frequency ranges is demonstrated, respectively. The preferred starting location of TDIBC is further discussed.

\subsection{Fitting result of softness and validation of TDIBC}
\begin{figure}
	\vspace{1em}
	\centerline{
		\subfigure{\includegraphics[width=8cm]{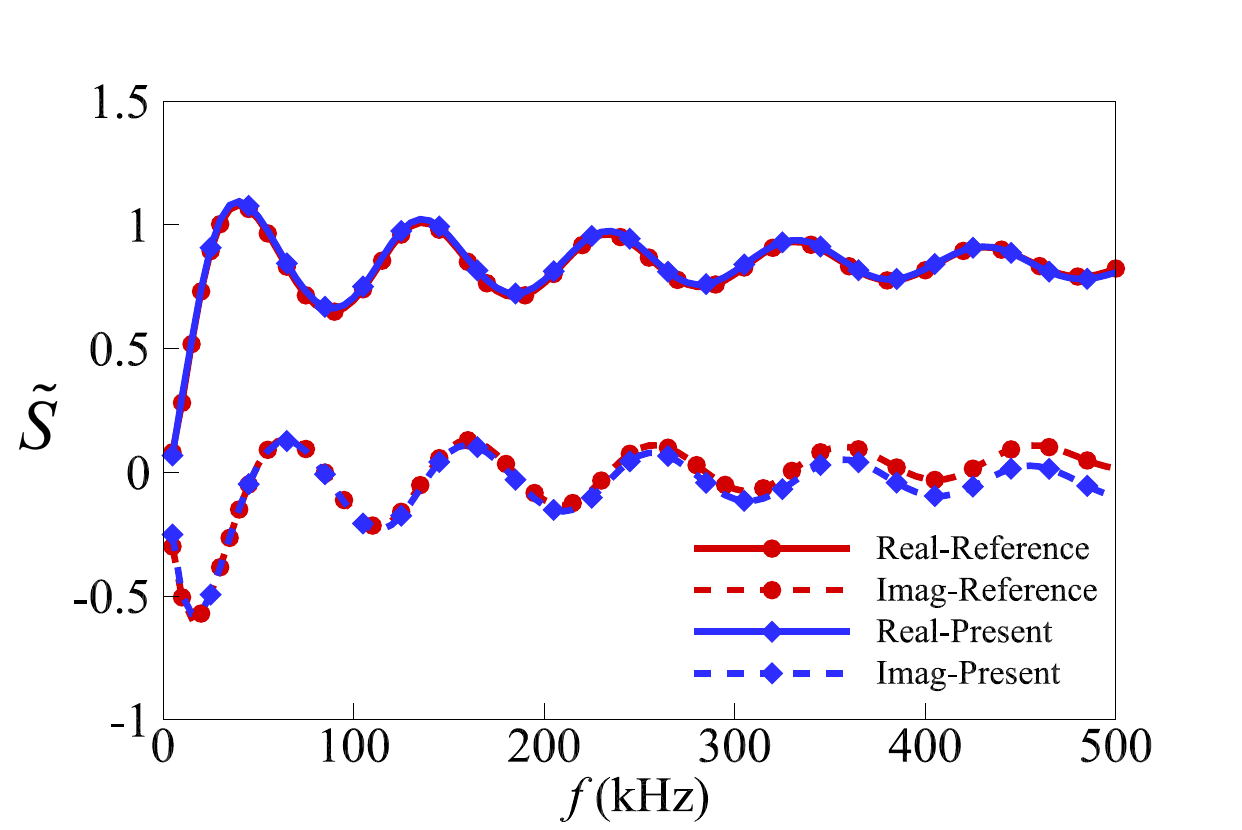}}}
	\caption{Comparison of wall softness between the reference impedance model (Zhao \etal \space \citeyear{Zhao2018a}) and the multi-pole fitting results (present).}
	\label{fig1}
\end{figure}

Regarding the metasurface employed, the dimensionless half width $b/{L_\textit{ref}} = 0.196$, the unit-cell periods are $s/{L_\textit{ref}} = 0.52$, and depth $H/{L_\textit{ref}} = 1.642$ are all consistent with those in Zhao \etal \space (\citeyear{Zhao2019}). The geometric setting corresponds to a porosity of $\varphi  = {{2b} \mathord{\left/{\vphantom {{2b} s}} \right.\kern-\nulldelimiterspace} s} = 0.75$. These parameters were designed to control the second mode and have demonstrated their evident effectiveness in both LST and two-dimensional DNS (Zhao \etal \space \citeyear{Zhao2019}). Figure \ref{fig1} compares the softness derived from a fitting procedure based on the theoretical model, which considers high-order diffraction (Zhao \etal \space \citeyear{Zhao2018a}). The fitting result was based on the method proposed by Douasbin \etal \space (\citeyear{Douasbin2018}) using equation (\ref{eq2.13}), which finally include ${C_0}$ = -0.05 and poles and residues shown in Appendix \ref{Appendix A}. It is indicated that the fitting result agrees well with the theoretical model within a broadband frequency domain. This agreement enables the accurate simulation subject to both high- and low-frequency disturbances using the acoustic metasurface. The comparison among the cases with TDIBC, with meshed cavities and with a constant wall impedance model ($v' = Ap'$, and $A$ is a constant acoustic admittance), can be found in Appendix \ref{Appendix B}. A good agreement is reached, validating the reliability of the TDIBC code in this study.

\begin{figure}
	\vspace{1em}
	\centerline{
		\subfigure{\includegraphics[width=7cm]{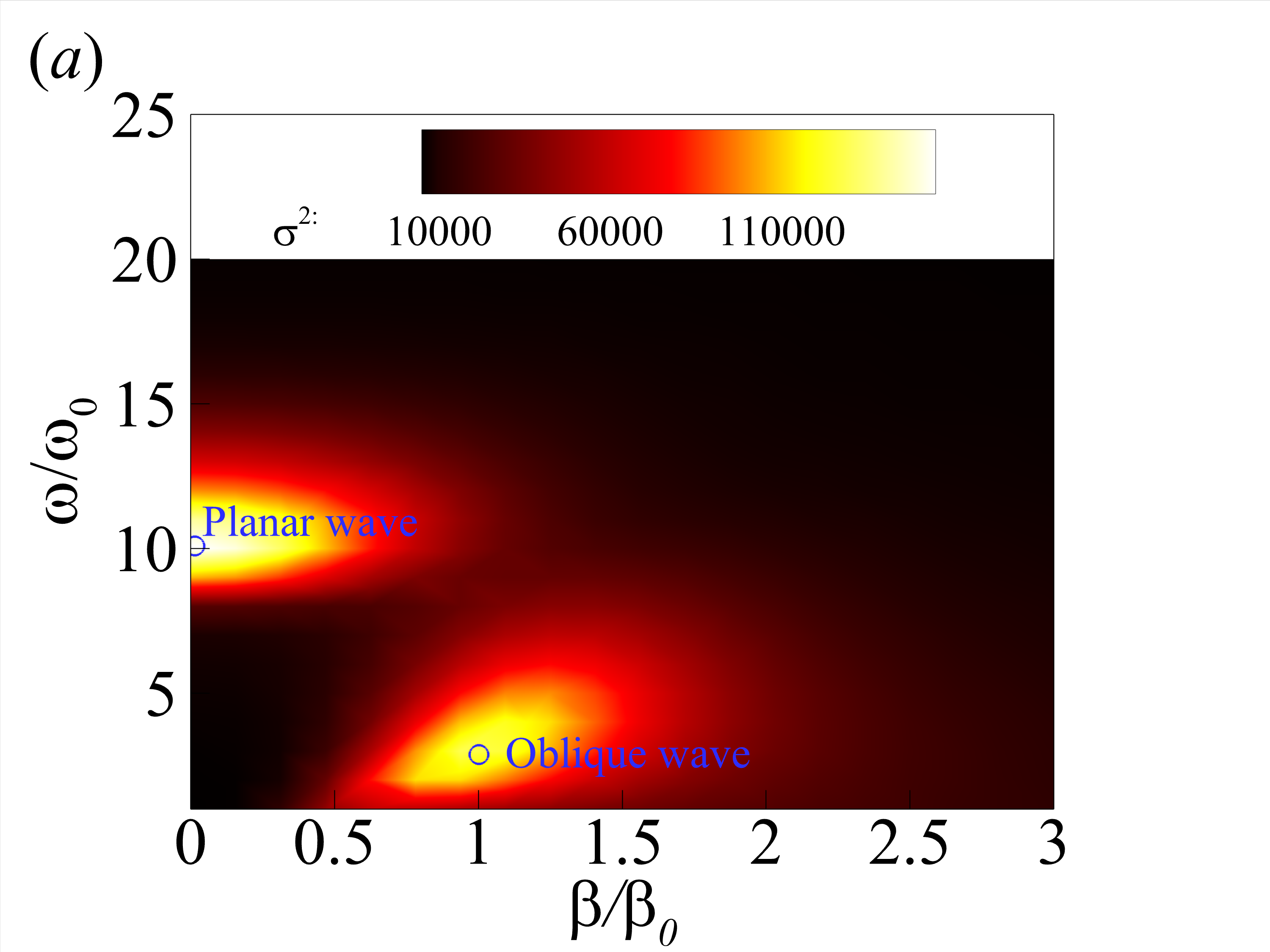}}
		\subfigure{\includegraphics[width=7cm]{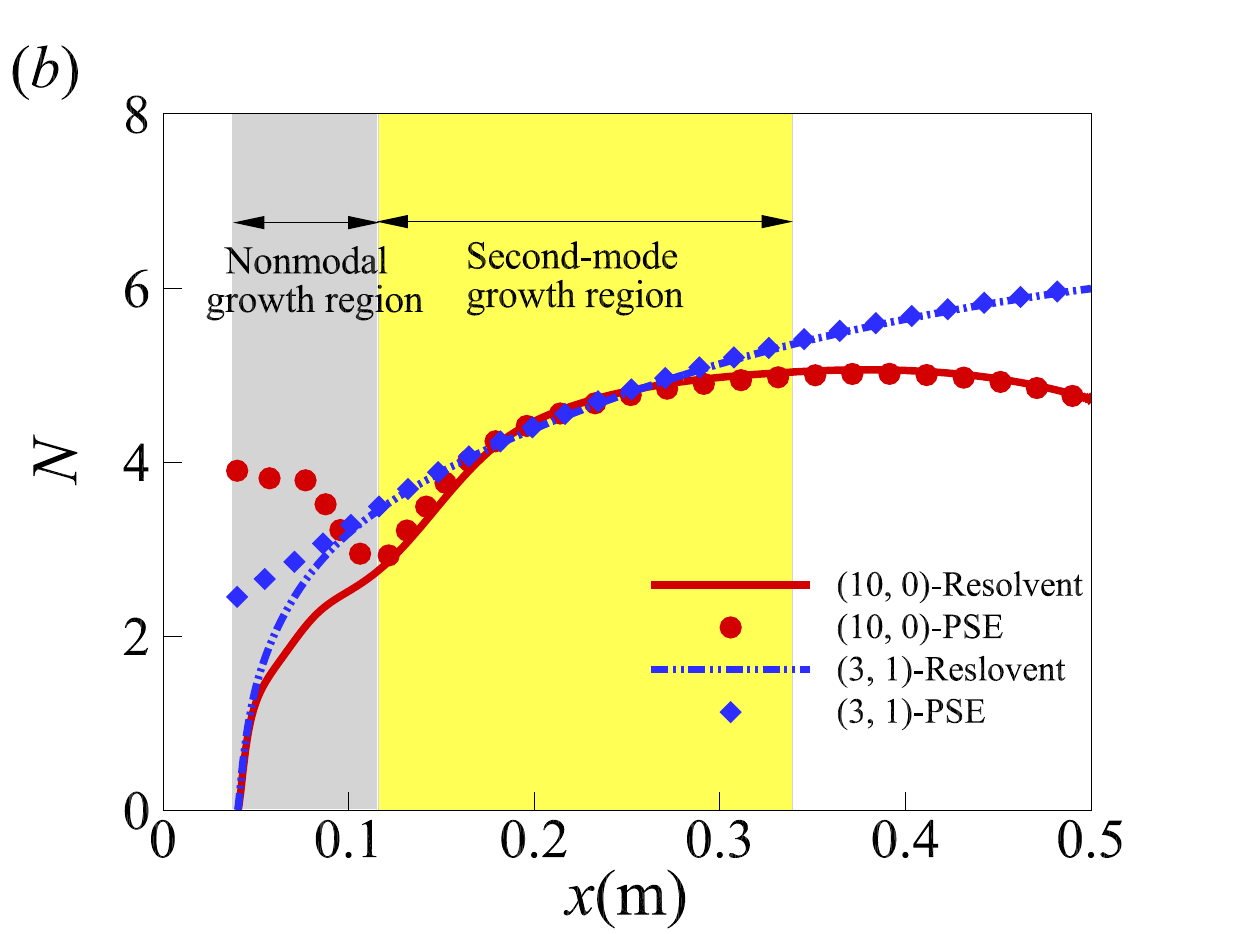}}}
	\caption{($a$) Contours of optimal gain in the parameter space of the spanwise wavenumber and the angular frequency, where ${\omega _0}{L_\textit{ref}}/{u_\infty }$ = 0.1 and ${\beta _0}{L_\textit{ref}}$ = 0.8. ($b$) Comparison of $N$-factors between PSE and resolvent analysis. $N$-factor curves of PSE are shifted to be compared with resolvent analysis.}
	\label{fig2}
\end{figure}

\subsection{Result of stability analysis}\label{stability analysis}
In resolvent analysis, the interval 0--0.5 m is utilised for the optimization problem, and the forcing is introduced at ${x_0}$ = 0.04 m. The $N$-factor is defined by $N = 0.5\log ({{{E_{Chu}}} \mathord{\left/{\vphantom {{{E_{\textit{Chu}}}} {{E_{\textit{Chu,0}}}}}} \right.\kern-\nulldelimiterspace} {{E_{Chu,0}}}})$, where ${{E_{\textit{Chu,0}}}}$ is measured at $x$ = ${x_0}$ and $N$ is 0 at ${x_0}$.

The gain contour for a wide range of spanwise wavenumbers and angular frequencies is shown in figure \ref{fig2}($a$). Clearly, the low-frequency oblique wave related to the first mode (${\omega /{\omega _0}}$, $\beta /{\beta _0}$) = (3, 1) and the high-frequency planar wave related to the second mode (${\omega /{\omega _0}}$, $\beta /{\beta _0}$) = (10, 0) are prominent in the gain contour. In this study, ${\omega /{\omega _0}}$ = 10 corresponds to a dimensional frequency of 125.8 kHz, which has also been shown to be the peak frequency in the energy spectrum under the considered experiment condition (Bountin \etal \space \citeyear{Bountin2013}).  For brevity, the optimal response for a Fourier mode with ${\omega /{\omega _0}}$ = 10 and $\beta /{\beta _0}$ = 0 referred to mode (10, 0), and the same applies to other optimal disturbances. In the following DNS, the three-dimensional transition to turbulence will be initiated using the optimal forcings determined by resolvent analysis. As shown in figure \ref{fig2}($b$), the optimal response captured by resolvent analysis undergoes a transient growth downstream of the forcing location, which can not be captured in PSE when a purely modal profile is introduced at the inlet. The transient growth region is between the forcing location and the position where the growth rate ${{dN} \mathord{\left/{\vphantom {{dN} {dx}}} \right.\kern-\nulldelimiterspace} {dx}}$ from the resolvent analysis converges with that from the PSE. For instance, the interval $x$$ \in $[0.04 m, 0.115 m] corresponds to the transient growth region of Fourier mode (10, 0). The PSE results indicate that modal growth begins at approximately $x$ = 0.115 m and ends at about $x$ = 0.34 m for (10, 0), as shown in figure \ref{fig2}($b$). As for the optimal oblique wave (3, 1), its $N$-factor exceeds that of the planar wave (10, 0) downstream. This is due to a longer growth region of (3, 1), despite a lower maximum growth rate.

In the numerical simulation, both modal and nonmodal growths are included. Given the broadband nature of freestream disturbances in wind tunnels and realistic flight conditions, both the first and second modes should be considered in the investigation of hypersonic boundary-layer instability and transition. In this study, optimal disturbances $({\omega _p},  {\beta _p})$ = (10, 0) and $({\omega _o},  {\beta _o})$ = (3, 1) were utilised to initiate the unsteady flow for cases listed in table \ref{label1}. These disturbances enable complicated nonlinear mode--mode interactions and spectral broadening rapidly (Guo \etal \space \citeyear{Guo2023a}). The two are also representative building blocks of the various primary instabilities with different frequencies and wavenumbers. The selection of only two of them enables the fundamental observation of the excited secondary instabilities downstream.

\begin{figure}
	\vspace{1em}
	\centerline{
		\subfigure{\includegraphics[width=7cm]{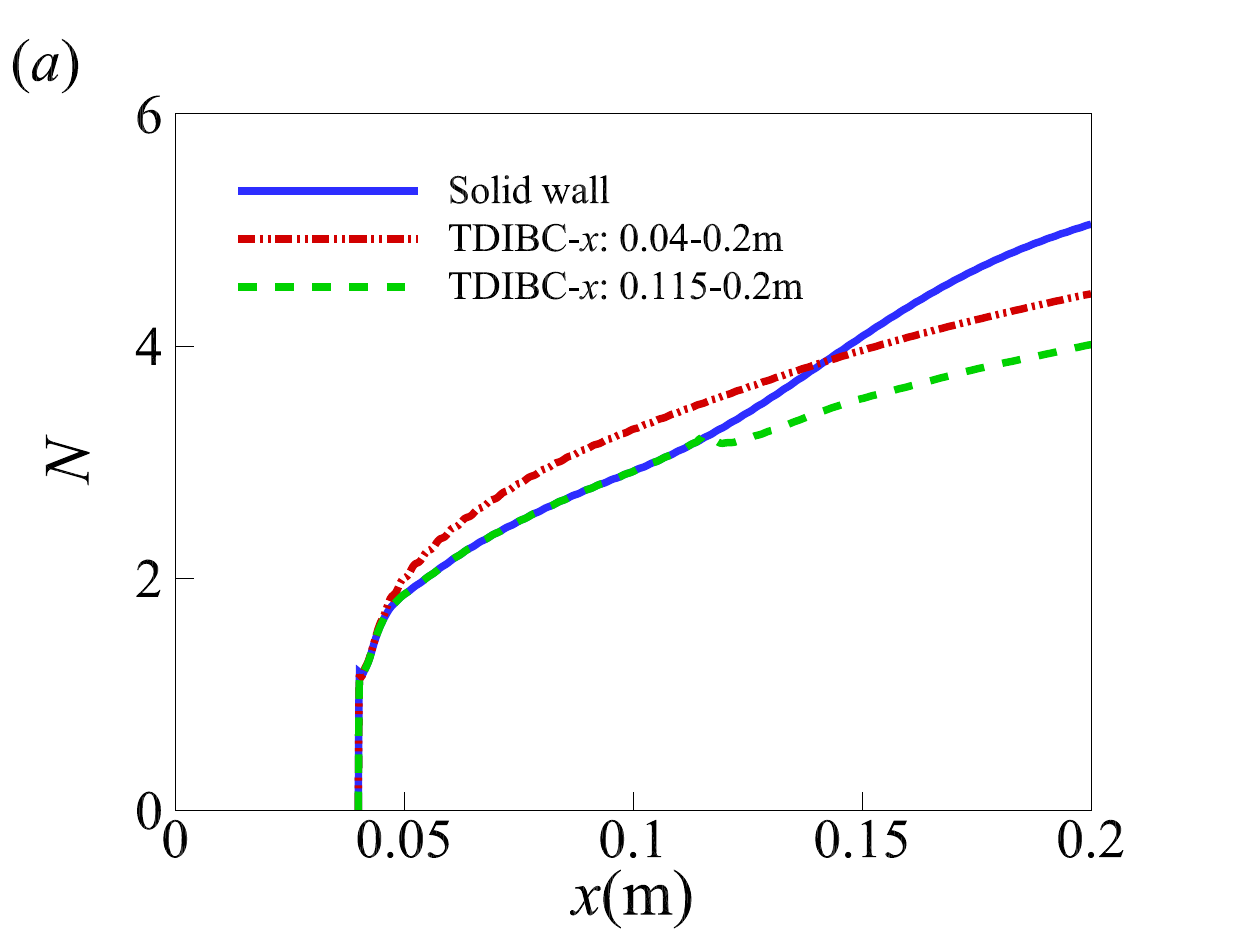}}
		\subfigure{\includegraphics[width=7cm]{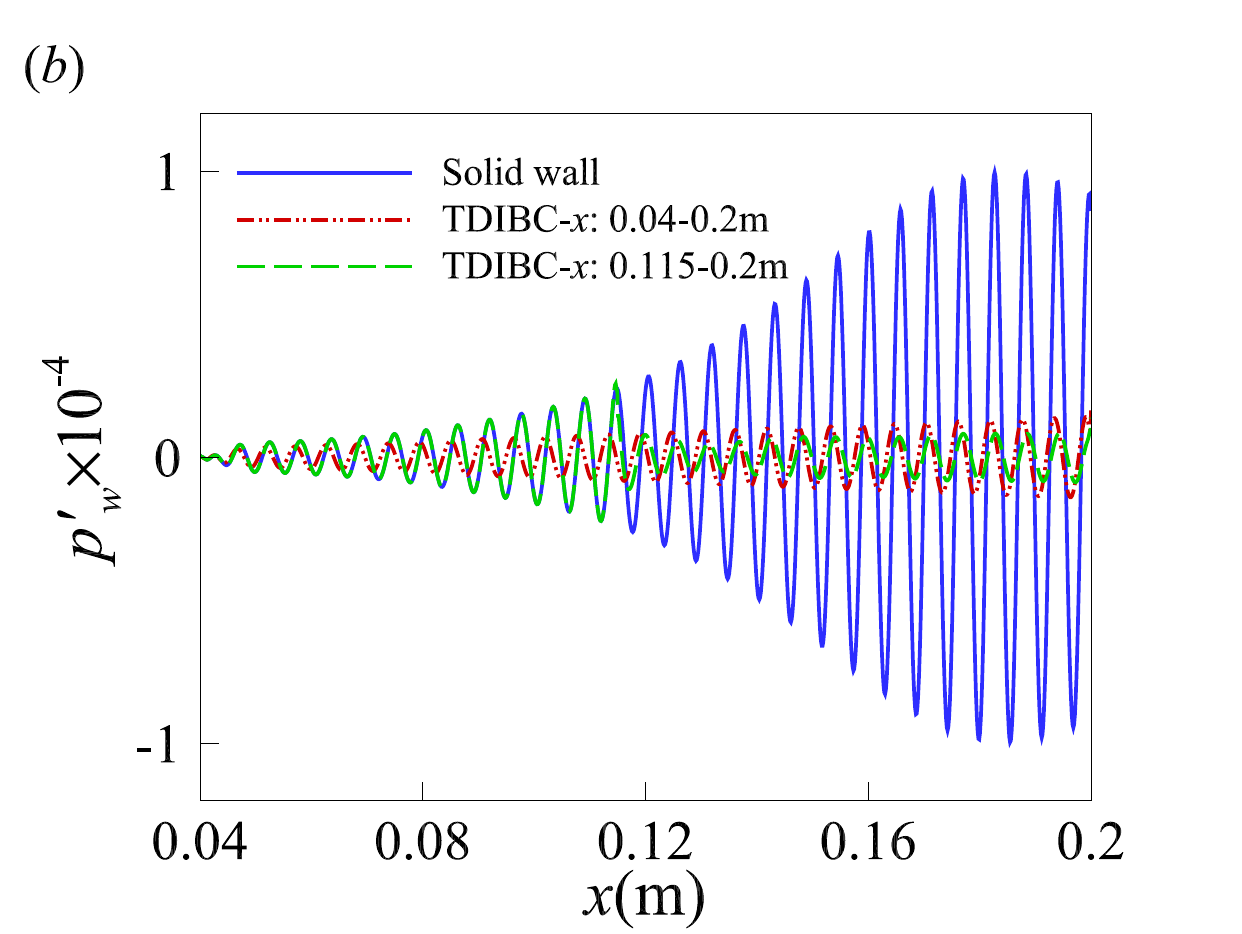}}}
	\caption{ ($a$) $N$-factor evolution and ($b$) dimensionless wall pressure fluctuation for different wall boundary conditions subject to the optimal forcing of mode (10, 0).}
	\label{fig3}
\end{figure}

\subsection{TDIBC for single-frequency disturbance}\label{single-frequency disturbance}
In this section, the singer-frequency disturbance with small amplitude is introduced in a precursor simulation to study its linear evolution individually. The verification of initialisation for the optimal forcing in the DNS is depicted in Appendix \ref{Appendix C}.

\begin{figure}
	\vspace{1em}
	\centerline{
		\subfigure{\includegraphics[width=6.5cm]{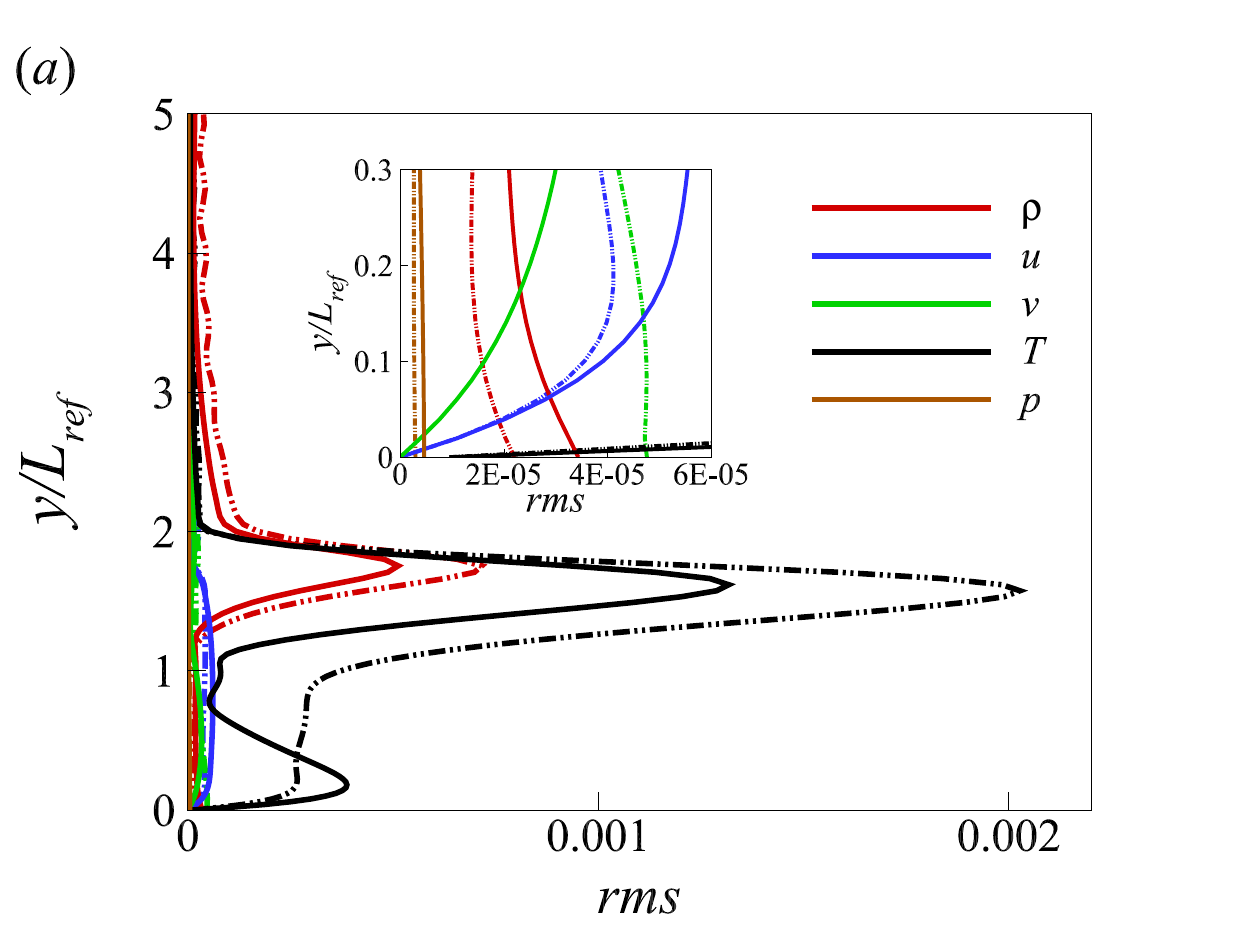}}
		\subfigure{\includegraphics[width=6.5cm]{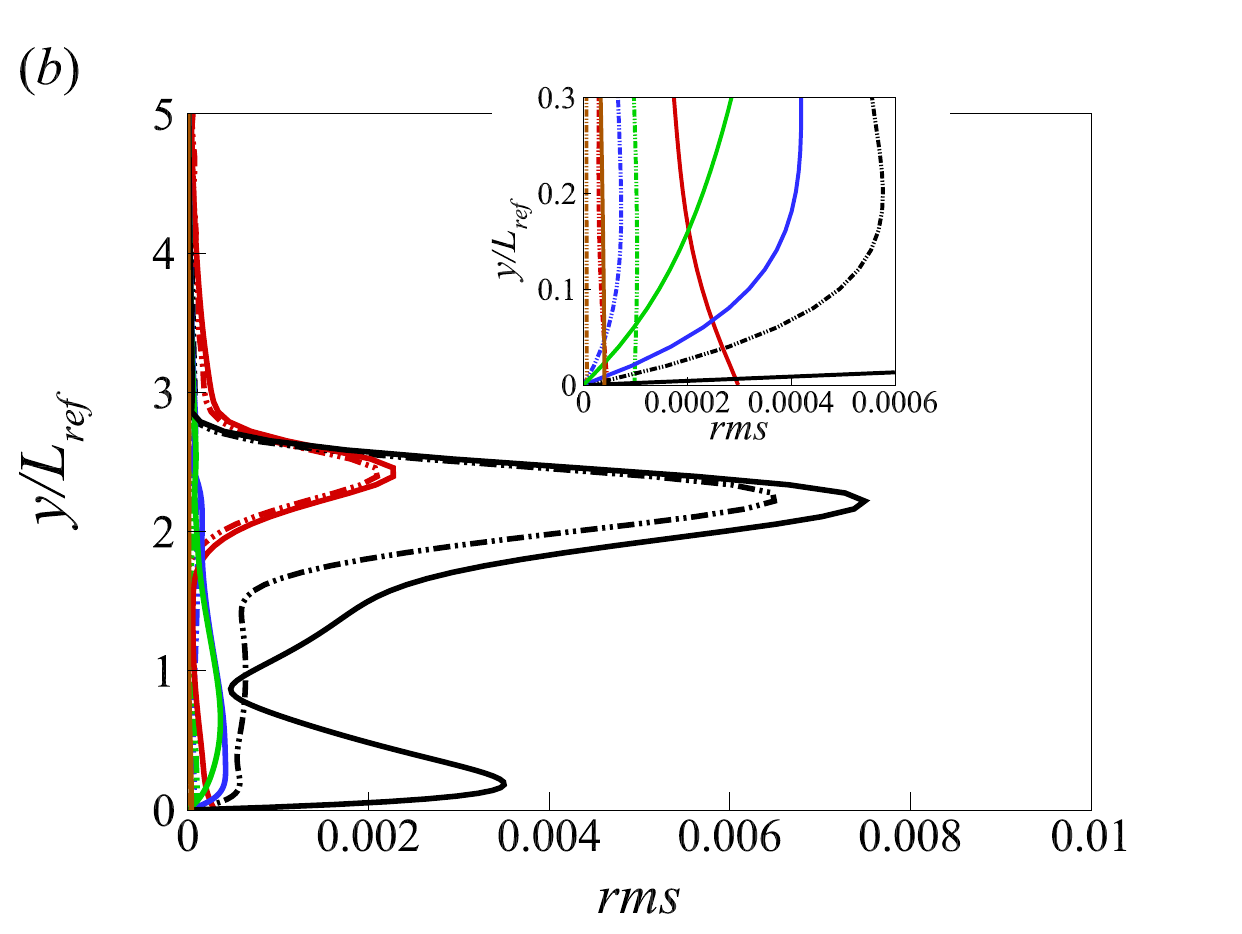}}}
	\caption{Comparison of dimensionless $rms$ magnitude between linear-stage cases using solid wall (solid line) and TDIBC (dash dot-dot line) at ($a$) $x$ = 0.08m and ($b$) $x$ = 0.16m initialised by optimal mode (10, 0). The TDIBC is applied within the range of 0.04--0.2 m.}
	\label{fig16}
\end{figure}

Figure \ref{fig3} compares the evolution of the Fourier modes (10, 0) with a solid wall and TDIBC, starting from $x$ = 0.04 m or $x$ = 0.115 m. The acoustic metasurface is observed to suppress the modal growth of the second mode while enhancing the transient growth, based on the definitions of transient-growth regions in section \ref{stability analysis}. Note that $x$ = 0.115 m is the starting point of the modal growth of Fourier mode (10, 0), according to the PSE result in figure \ref{fig2}($b$). When TDIBC is employed in the range of 0.04--0.2 m, the total Chu's energy is augmented at around $x$ = 0.1m, while the wall pressure fluctuation is suppressed there in figure \ref{fig3}($b$). This indicates that the wall pressure fluctuation may not reflect the overall impact of the metasurface. The result agrees with the numerical simulation of Wang \& Zhong (\citeyear{Wang2012}) and the wind tunnel experiment of Lukashevich \etal \space (\citeyear{lukashevich2016experimental}), which reported that the acoustic metasurface is effective only within the unstable region of the second mode. The detailed effect of the acoustic metasurface on each primitive variable downstream and upstream of the synchronisation point of the second mode is compared in figure \ref{fig16}. 

\begin{figure}
	\centerline{
		\subfigure{\includegraphics[width=7cm]{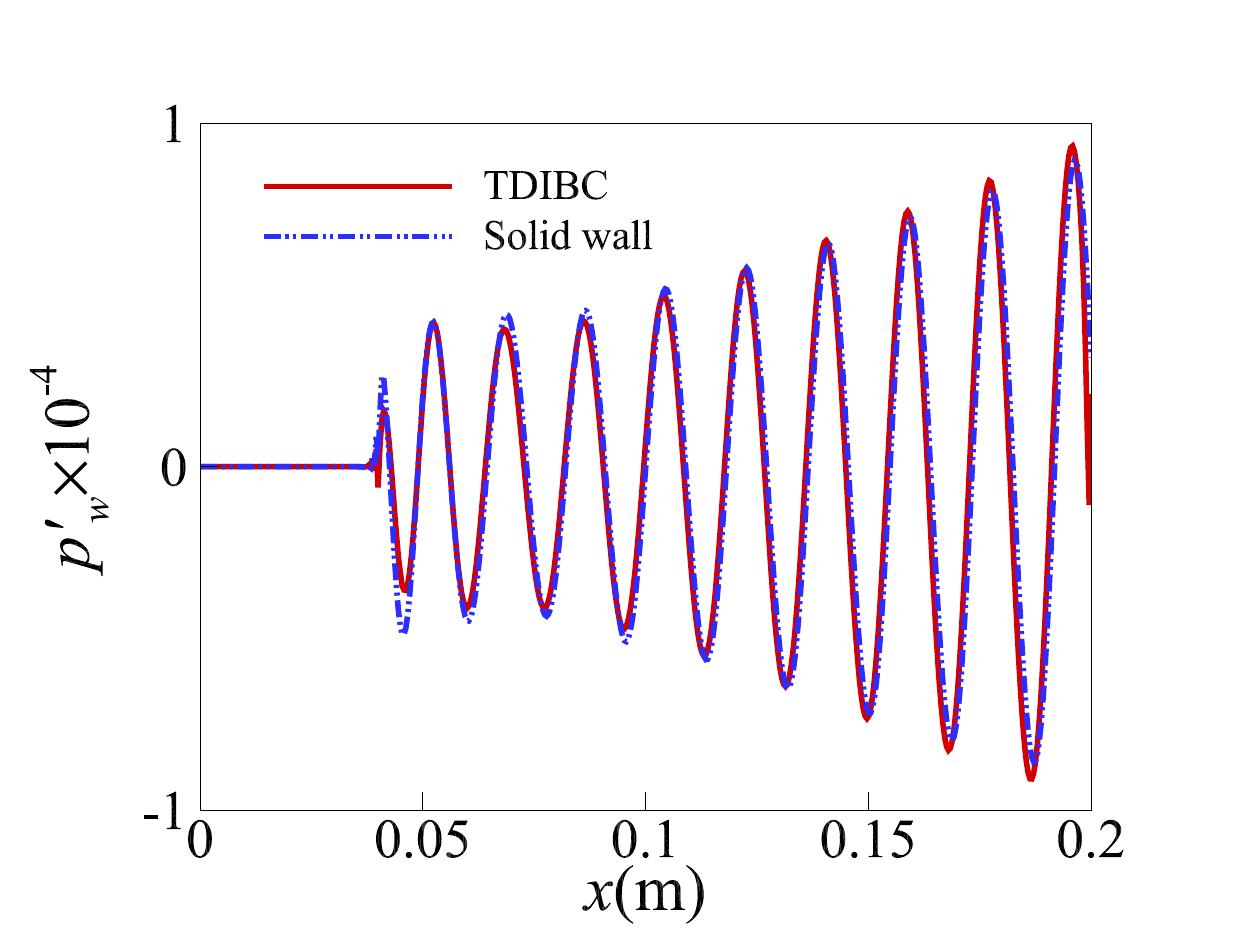}}}
	\caption{Comparison of dimensionless pressure fluctuation at the wall for the optimal wave (3, 1) between solid wall condition and TDIBC in the linear stage.}
	\label{fig4}
\end{figure}

Figure \ref{fig16} compares the root-mean-square ($rms$) values of primitive variables between the results with a solid wall and the acoustic metasurface (realised by TDIBC). It illustrates that the $rms$ of the wall pressure fluctuation is lower with TDIBC at $x$ = 0.08 m, which agrees with figure \ref{fig3}($b$). However, the density and temperature are significantly enhanced near the boundary layer edge (around $y/{L_\textit{ref}}$ = 2). This leads to a higher total energy of the planar wave (10, 0) in this region, namely a larger $N$-factor shown in figure \ref{fig3}($a$). Downstream of the neutral point of the second mode, the disturbances are completely suppressed regarding both the total energy and the wall pressure fluctuation, as shown in figure \ref{fig16}($b$).
On the contrary, the acoustic metasurface slightly destabilises the oblique optimal disturbance (3, 1), as shown in figure \ref{fig4}. 

Based on the above finding, the starting location of TDIBC for both Case 2 and Case 3 was selected as $x$ = 0.115 m to avoid unnecessary amplification of unstable waves. Besides, TDIBC ends at $x$= 0.34 m in Case 3 because the second mode is stable downstream of $x$ = 0.34 m (see figure \ref{fig2} ($b$)). It should be noted that no prior reports have examined the impact of the metasurface in the transient growth stage. More research is necessary to determine how the metasurface affects the Orr/Lift-up processes that underpin the transient growth. More importantly, it is not yet understood how the transition initiated by the first and second modes responds to the acoustic metasurface. In the next section, the effect of the acoustic metasurface in combination resonance (Guo \etal \space \citeyear{Guo2023a}) is focused on, which requires the involvement of both first and second modes.

\begin{figure}
	\vspace{1em}
	\centerline{
		\subfigure{\includegraphics[width=12.5cm]{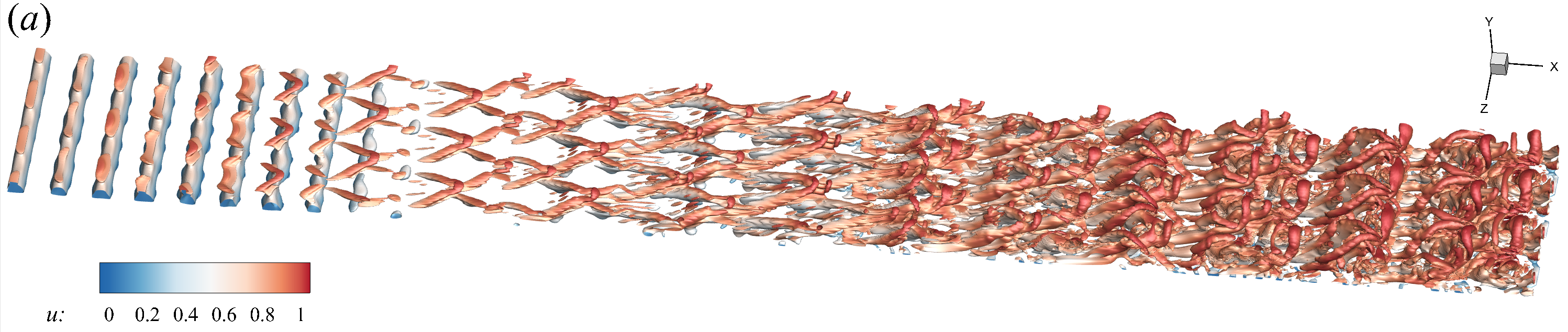}}}
	\centerline{
		\subfigure{\includegraphics[width=12.5cm]{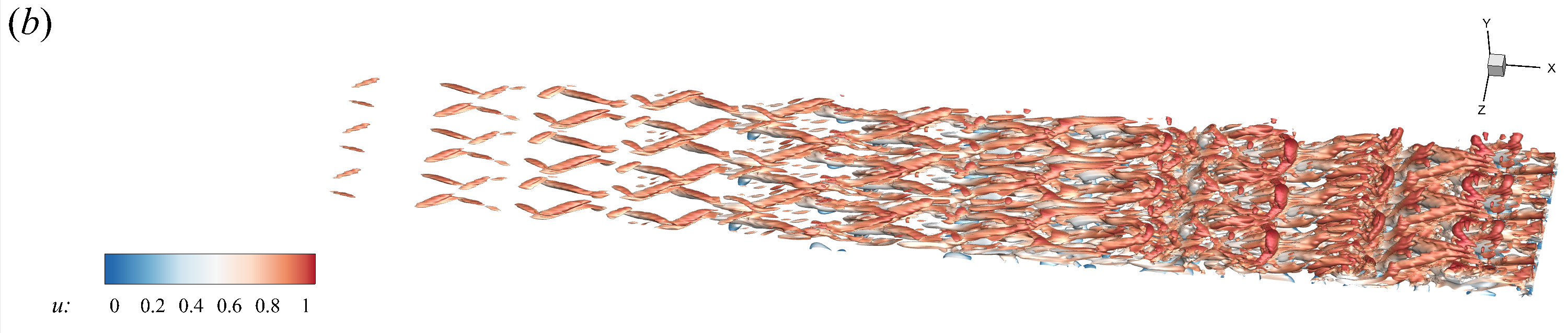}}}
	\centerline{
		\subfigure{\includegraphics[width=12.5cm]{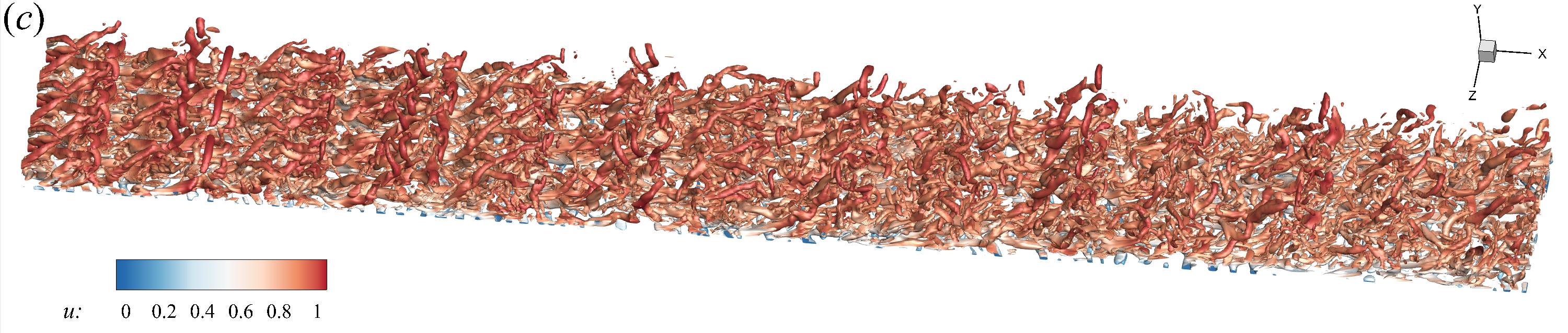}}}
	\centerline{
		\subfigure{\includegraphics[width=12.5cm]{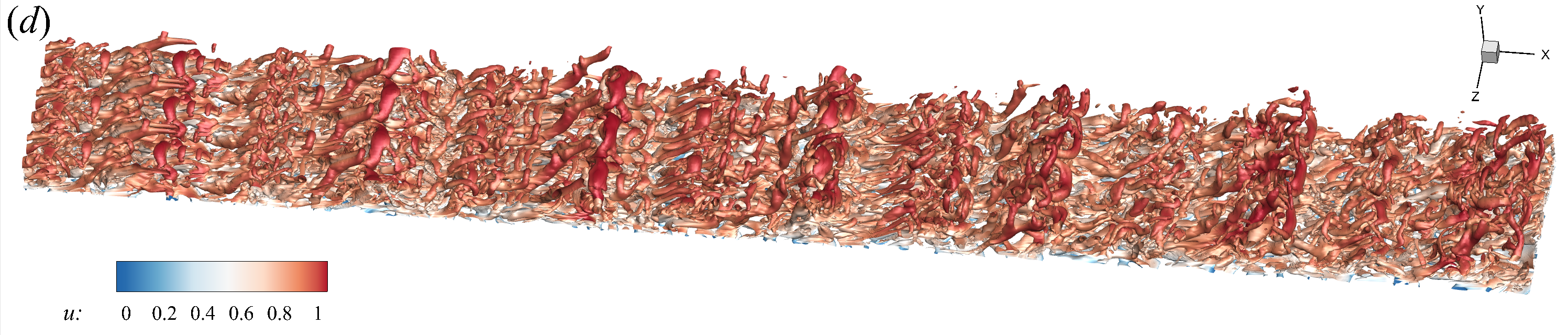}}}
	\caption{The $Q$-criterion iso-surface ${({L_\textit{ref}}/{u_\infty })^2}Q = 0.005$ coloured by the dimensionless streamwise velocity in the range of 0.2 m < $x$ < 0.4 m for ($a$) Case 1 and ($b$) Case 2, and of 0.4 m < $x$ < 0.6 m for ($c$) Case 1 and ($d$) Case 2.}
	\label{fig5}
\end{figure}

\section{Nonlinear mode--mode interaction and breakdown}\label{Results}
\subsection{Transition onset and end}

Following the verification of the initialisation of optimal responses and the embedded TDIBC code, the nonlinear interaction between the low-frequency and high-frequency components and the resulting transition process with the acoustic metasurface are of concern. In this section, optimal disturbances with two prominent frequencies and the same amplitude are applied to initiate the transition. Three cases with different wall boundary conditions listed in table \ref{label1} are simulated and compared. The effect of the acoustic metasurface on the combination resonance will be evaluated in detail for statistical average and instantaneous flow fields, especially for mode--mode interaction mechanisms using bi-Fourier analysis and energy budget analysis.

\begin{figure}
	\vspace{1em}
	\centerline{
		\subfigure{\includegraphics[width=12cm]{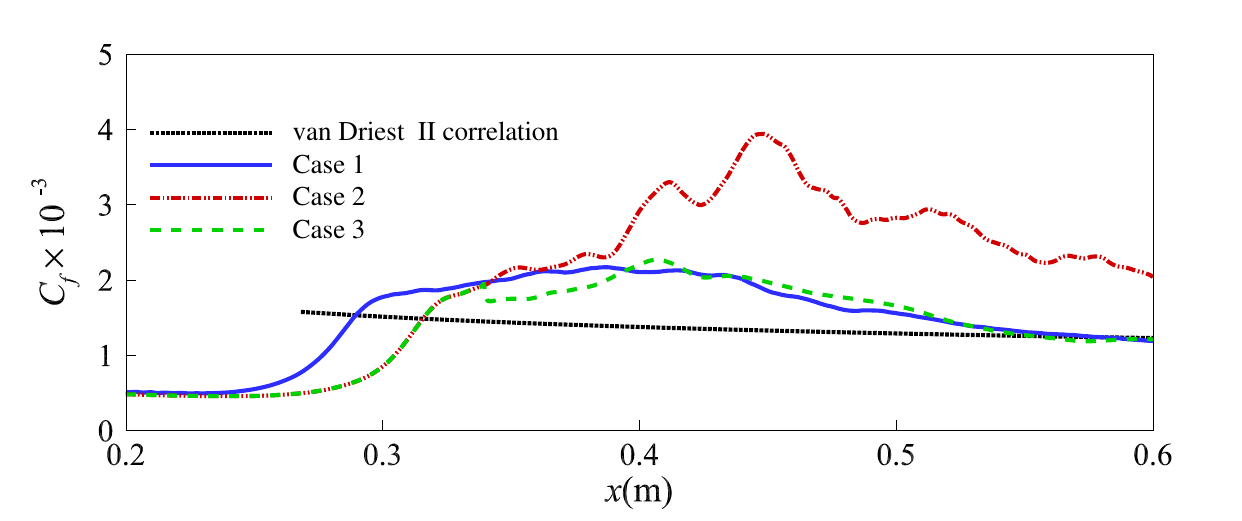}}}
	\caption{Quantitative results of spanwise- and time-averaged skin friction coefficient and the van Driest II formula for ${C_f}$. The van Driest II formula is applied following the procedure of Guo \etal \space (\citeyear{Guo2022b}).}
	\label{fig6}
\end{figure}

The comparison of vortex structures between Case 1 and Case 2 is depicted in figure \ref{fig5}. The spanwise-aligned structure observed in Case 1 (figure \ref{fig5}($a$)), representing the planar-wave mode (10, 0), vanishes in Case 2 (figure \ref{fig5}($b$)) due to the presence of the acoustic metasurface. This indicates that mode (10, 0) is highly suppressed by the acoustic metasurface. Further downstream, the staggered structure appears later in Case 2 than in Case 1, which is caused by the oblique wave $(3, \pm 1)$. Additionally, the hairpin vortex, indicating the late stage of transition, is also delayed regarding its appearance in Case 2 compared to Case 1.  Until the end of the computational domain, the vortex structure in Case 1 has broken down into small-scale structures, indicating fully developed turbulence. This is consistent with the result of the skin friction coefficient illustrated in figure \ref{fig6}, in which the curve of Case 1 collapses into the van Driest correlation near the end of the computational domain.

Figure \ref{fig6} shows the spanwise- and time-averaged skin friction coefficient for three cases, accompanied by the van Driest II correlation. The increment of the skin friction coefficient is delayed for Case 2 and Case 3 compared with Case 1. By estimating the transition onset location based on the minimum skin friction coefficient, the transition onset occurs at approximately ${x_\textit{onset}}$ = 0.22 m in Case 1, while it shifts downstream to around  ${x_\textit{onset}}$ = 0.25 m in Case 2 and Case 3. The transition ends at around $x$ = 0.57 for Case 1 and Case 3, as the skin friction coefficient collapses to van Driest II correlation. The transition delay efficiency is evaluated by $\eta$, which is defined by 
\begin{equation}\label{eq2.24}
\eta  = \frac{{{x_\textit{onset,{\rm{ }}Case 2 or 3}} - {x_\textit{onset,{\rm{ }}Case 1}}}}{{{x_\textit{onset,{\rm{ }}Case 1}}}}.
\end{equation}	
The transition delay efficiency is about $14\% $ based on the onset location for Cases 2 and 3 with the acoustic metasurface. This value is not prominent compared with the wind tunnel experiment of Wagner \etal \space (\citeyear{Wagner2013}), where the second mode is dominant under a cooling wall condition at Mach 7.5 with the first mode highly suppressed. Nevertheless, this finding confirms the effectiveness of the acoustic metasurface in delaying the transition induced jointly by the low-frequency first mode and the high-frequency second mode.  However, in Case 2, the skin friction coefficient is higher than its counterpart in Case 1 in the late transitional region, which starts at around $x$ = 0.34 m. This phenomenon was also reported in the wind tunnel experiment in DLR conducted by Wagner \etal \space (\citeyear{Wagner2013}; \citeyear{Wagner2014}; \citeyear{Wartemann2015}). The reason for higher skin friction in the late transitional region is of general interest and will be explored using bi-Fourier analysis and energy budget equation later. In Case 3, the undesirable increase in wall friction is eliminated when the wall boundary condition reverts to a solid wall downstream of $x$ = 0.34 m. The comparative study demonstrates that positioning the acoustic metasurface within the unstable region of the second mode is more effective in controlling flow transition.

\begin{figure}
	\vspace{1em}
	\centerline{
		\subfigure{\includegraphics[width=13cm]{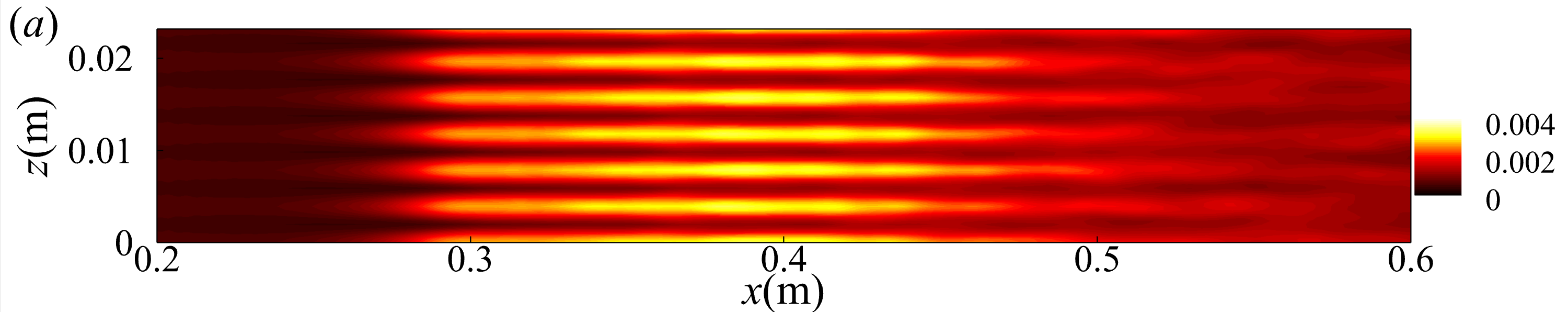}}}
	\centerline{
		\subfigure{\includegraphics[width=13cm]{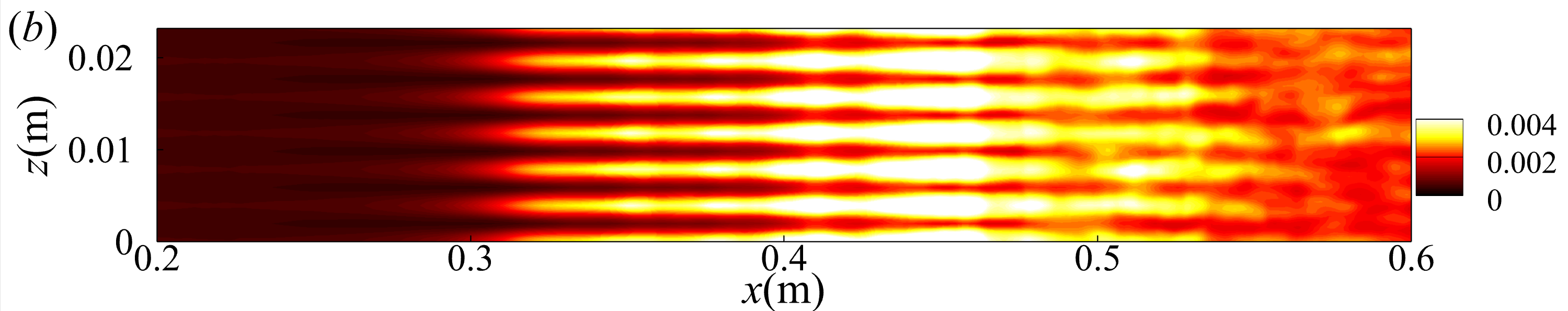}}}
	\centerline{
		\subfigure{\includegraphics[width=13cm]{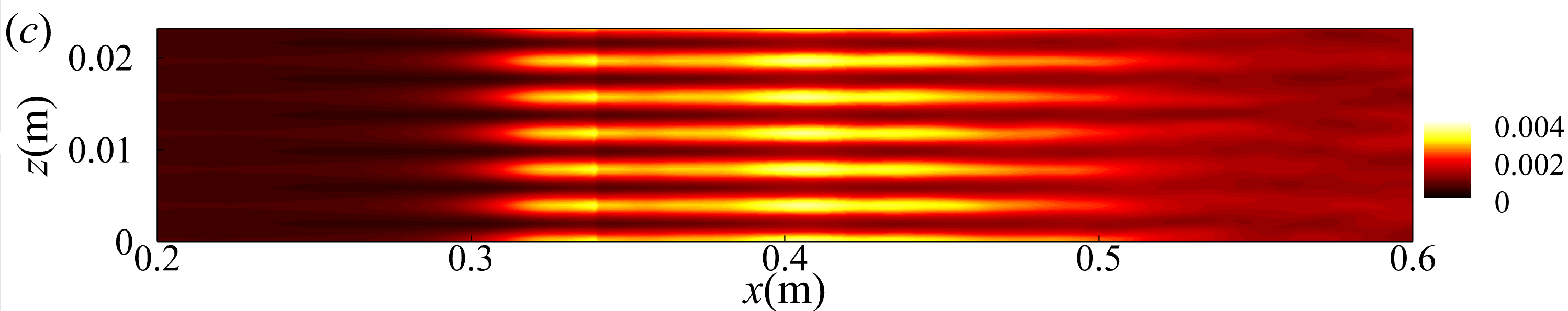}}}
	\caption{Contour of time-averaged skin friction coefficient for ($a$) Case 1, ($b$) Case 2, and ($c$) Case 3.}
	\label{fig7}
\end{figure}

Figure \ref{fig7} provides the contour of the time-averaged skin friction coefficient. The streak spacing corresponds to the spanwise wavenumber of the steady streamwise mode (0, 2), which can be nonlinearly produced by the interaction of the forced oblique waves $(3,  \pm 1)$. It is evident that, compared to the solid-wall case, the appearance of streaks is delayed in the cases with TDIBC. Meanwhile, the maximum skin friction induced by streaks is more pronounced for Case 2, which is consistent with the spanwise-averaged result shown in figure \ref{fig6}. Compared to the results of Case 1, Case 3 exhibits a purely delayed effect in the contour of time-averaged skin friction. Meanwhile, these two cases converge to fully developed turbulence nearly at the same location, as shown in figure \ref{fig6}. Interestingly, the skin friction drops slightly where the acoustic metasurface is replaced by the solid-wall condition at $x$ = 0.34 m. Further downstream, it gradually develops into a consistent result with Case 1. This indicates that the acoustic metasurface only produces a local effect. The undesirable increment of wall friction in Case 2 downstream of $x$ = 0.34 m is thus caused by the acoustic metasurface. The underlying physics may relate to the strengthened mean shear near the wall, as reported in the permeable wall over the supersonic boundary layer (Chen \& Scalo \citeyear{Chen2021a}; \citeyear{Chen2021b}). This issue will be discussed later.

\subsection{Mode--mode interaction}\label{interaction}
\begin{figure}
	\vspace{1em}
	\centerline{
		\subfigure{\includegraphics[width=13cm]{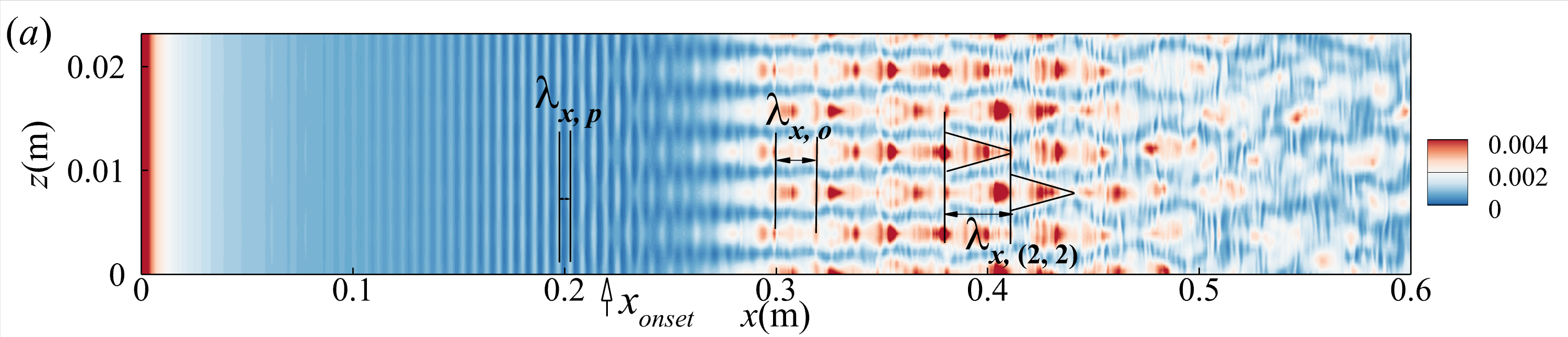}}}
	\centerline{
		\subfigure{\includegraphics[width=13cm]{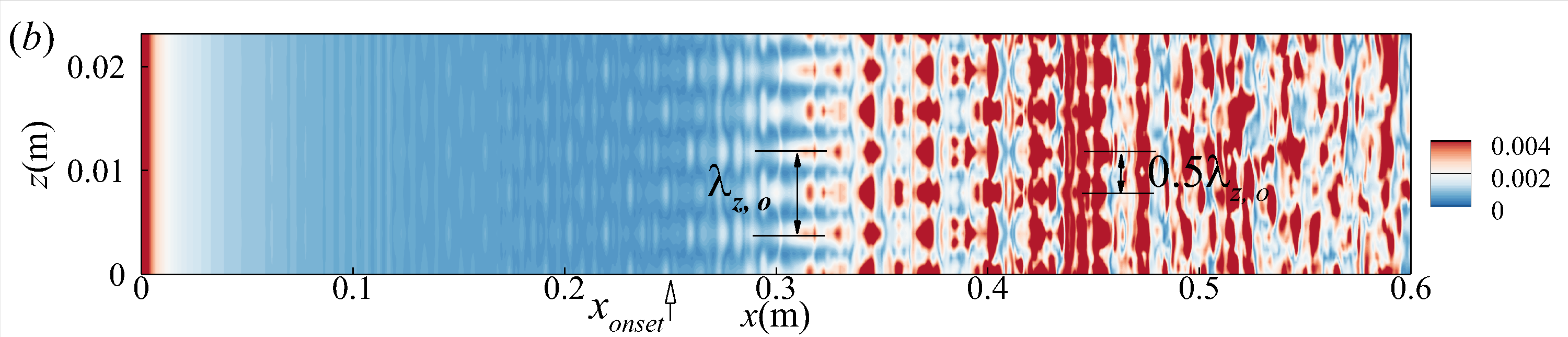}}}
	\centerline{
		\subfigure{\includegraphics[width=13cm]{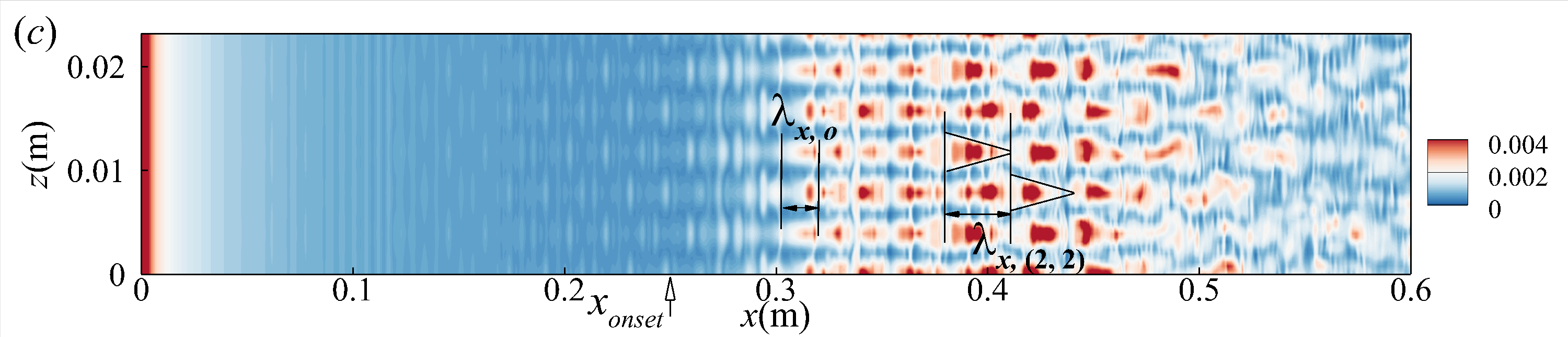}}}
	\caption{ Instantaneous skin friction coefficient ${C_f}$ for ($a$) Case 1, ($b$) Case 2, and ($c$) Case 3, where ${x_\textit{onset}}$ refers to the evaluated starting location of the transition.}
	\label{fig9}
\end{figure}
This subsection serves to facilitate the understanding of the dominant nonlinear interaction mechanism and explore the reason why transition onset is delayed by acoustic metasurface, as well as why wall friction is higher in the late transitional stage with acoustic metasurface (see Case 2 in figure \ref{fig6}). The instantaneous skin friction coefficient ${C_f}$ is plotted and shown in figure \ref{fig9}. The streamwise wavelengths of the planar wave ${\lambda _{x,{\rm{ }}p}}$, the oblique wave ${\lambda _{x,{\rm{ }}o}}$, and large-scale $\Lambda  - $vortices related to the detuned mode ${\lambda _{x,{\rm{ (2,2)}}}}$ can be clearly identified in Case 1, as shown in figure \ref{fig9} ($a$). Notably, the aligned structure of the second mode (at around $x$ = 0.2 m) loses its signature in Case 2 and Case 3 due to the suppression effect of the acoustic metasurface. The detuned mode (2, 2) is generated through
\begin{equation}\label{eq4.22}
\begin{array}{l}
(3,{\rm{ }}1) + (3,{\rm{ }} - 1) \to (6,{\rm{ 0}}),\\[0.8ex]
(10,{\rm{ 0}}) - (3,{\rm{ }} - 1) \to (7,{\rm{ 1}}),\\[0.8ex]
(7,{\rm{ }}1) - (6,{\rm{ 0}}) \to (1,{\rm{ 1}}),\\[0.8ex]
(1,{\rm{ }}1) + (1,{\rm{ }}1) \to (2,{\rm{ 2}}).
\end{array}
\end{equation}
The generation of detuned mode (2, 2) is confirmed by wave vectors  $({\alpha _r},{\rm{ }}\beta )$ of the considered triads (Fezer \& Kloker \citeyear{fezer2000spatial}), and details can be seen in Appendix \ref{Appendix D}.

The detuned mode, manifested as large-scale $\Lambda  - $vortices (Guo \etal \space \citeyear{Guo2023a}), can also be observed in Case 3. This observation indicates that the flow field can be recovered as the wall boundary reverts to the solid wall condition and that the acoustic metasurface only produces a local effect. In contrast, the detuned-mode-related $\Lambda  - $structures disappear in Case 2, where the metasurface persists in the whole domain. In Case 2, the distinct spanwise-aligned structure near $x$ = 0.45 m indicates the importance of planar wave (2, 0), as later demonstrated by figure \ref{fig11}($b$). Furthermore, the spanwise scale approaches $0.5{\lambda _{z,{\rm{ }}o}}$, which is identical to the spanwise wavelength of the detuned mode (2, 2). The enhanced pattern related to detuned modes may be responsible for the increment of skin friction in the late transitional stage with the acoustic metasurface (see Case  2 in figure \ref{fig6}). A further decomposition of skin friction is needed to identify the dominant contributor. In the subsequent investigation, temporal and spanwise bi-Fourier analysis and energy budget analysis will be applied to highlight the key contribution to wall friction.

\begin{figure}
	\vspace{1em}
	\centerline{
		\subfigure{\includegraphics[width=10cm]{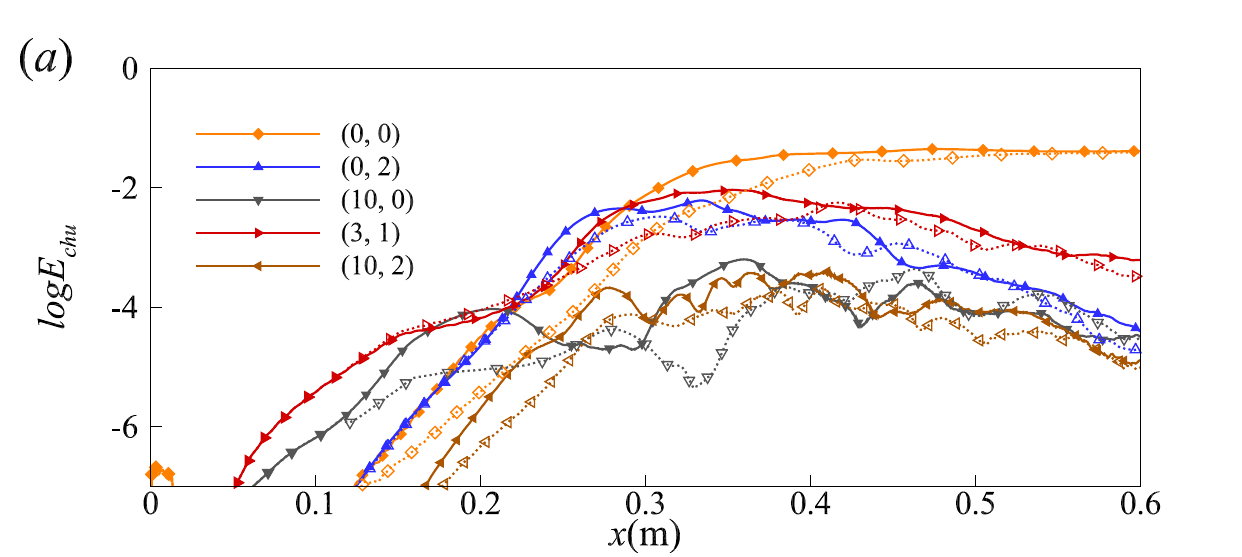}}}
	\centerline{
		\subfigure{\includegraphics[width=10cm]{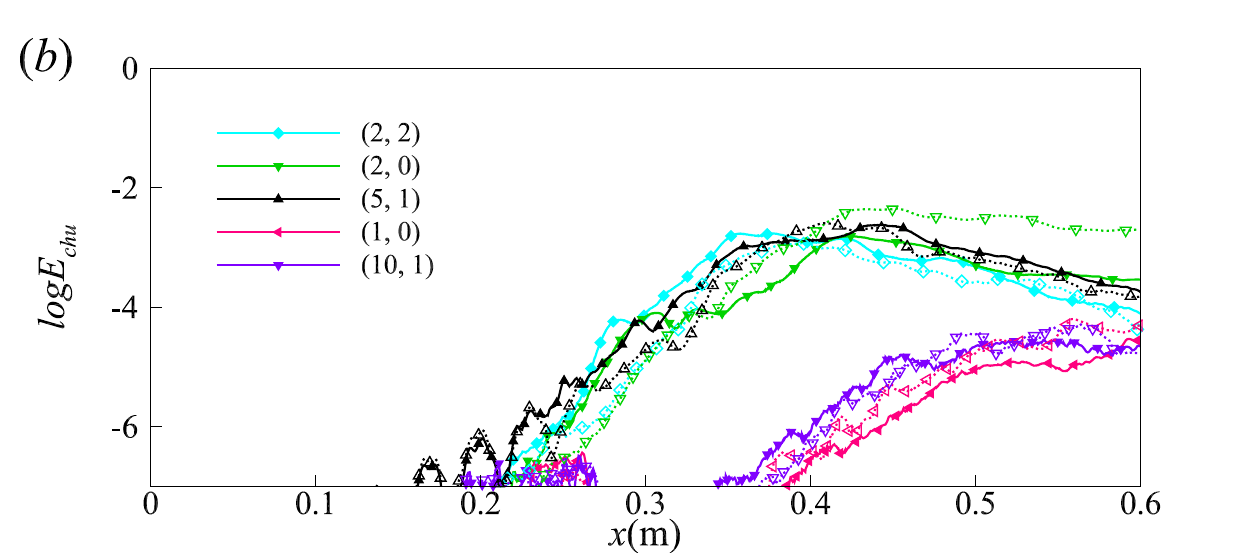}}}
	\centerline{
		\subfigure{\includegraphics[width=10cm]{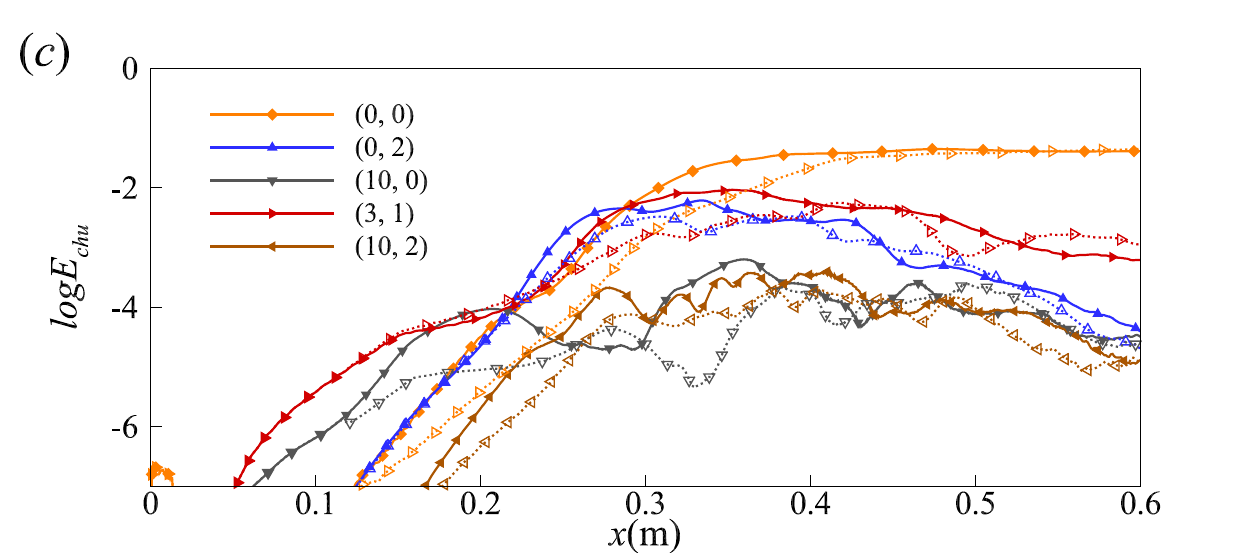}}}
	\centerline{
		\subfigure{\includegraphics[width=10cm]{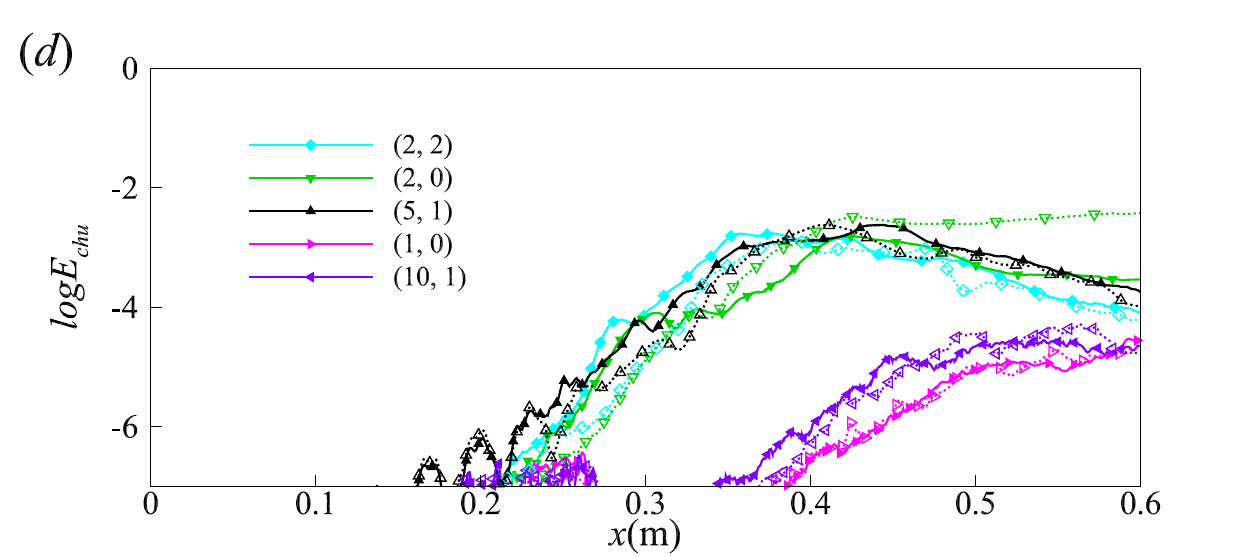}}}
	\caption{Comparison of streamwise development of Chu's energy for ($a$) and ($b$) for Case 1 and Case 2, and ($c$) and ($d$) for Case 1 and Case 3. The modes in Case 1 are represented by solid lines with filled symbols, and modes in Case 2 and Case 3 are represented by dotted lines with open symbols.}
	\label{fig10}
\end{figure}

The Chu's energy integral in the wall-normal direction includes kinetic and internal energy. Its streamwise evolution of multiple Fourier modes can comprehensively reflect the dominant modes in the transitional stage, which already includes the effect of the nonlinear interaction. As depicted in figure \ref{fig10}, the growth of Chu's energy of the second-mode-related mode (10, 0) is suppressed by the acoustic metasurface downstream of $x$ = 0.115 m. Meanwhile, the first mode is slightly promoted near $x$ = 0.2 m, which is consistent with previous theoretical and simulation research (Fedorov \etal \space \citeyear{Fedorov2003b}; Wang \& Zhong \citeyear{Wang2012}). In Case 1, the planar wave (10, 0) saturates at around $x$ = 0.18 m, and its amplitude begins to decrease due to energy transfer via nonlinear interaction with modes (5, 1), (10, 2), and (2, 2) or (2, 0) through the subharmonic resonance, the fundamental resonance and the combination resonance, respectively. The oblique breakdown associated with mode (3, 1), the fundamental resonance associated with mode (10, 0), and the subharmonic resonance associated with mode (10, 0) are attributable to the nonlinear interactions
 \begin{equation}\label{eq4.55}
\begin{array}{l}
(3,{\rm{ }}1) - (3,{\rm{ }} - 1) \to (0,{\rm{ 2}}),\\[0.8ex]
(10,{\rm{ 0}}) + (0,{\rm{ 2}}) \to (10,{\rm{ 2}}),\\[0.8ex]
(10,{\rm{ 0}}) - (5,{\rm{ 1}}) \to (5,{\rm{  - 1}}),
\end{array}
\end{equation}
respectively. The aforementioned modes other than modes $(3, \pm 1)$ and (10, 0), known as secondary instabilities, undergo a rapid growth process and make the flow develop into the late stage of transition and breakdown. The specific nonlinear mode-mode interaction mechanism was systematically studied and reported by Guo \etal \space(\citeyear{Guo2023a}). In this study, the effect of the acoustic metasurface is focused on. In figure \ref{fig10} ($a$), one can see that the second-mode-related mode (10, 0) in Case 2 (dashed line) grows gradually to nearly the same amplitude as that in Case 1 (solid line), but saturate further downstream. Meanwhile, the secondary growth of the primary oblique wave (3, 1) shows a lower growth rate (i.e. slope) in Case 2 at around $x$ = 0.28 m than in Case 1. This result is somewhat consistent with the study of Chen \etal \space(\citeyear{Chen2017}), where the amplitude of the second mode saturates and then the "phase lock" sets in. This phase lock supports a secondary growth stage of the primary low-frequency mode. Meanwhile, the second mode decays due to energy transfer to the rapidly growth secondary modes. Regarding the metasurface effect in this paper, it suppresses the first mode-related (3, 1) when the nonlinearity is considered. This finding differs from the linear case in figure \ref{fig4}, which is probably due to the less energy transferred from the attenuated mode (10, 0) with the metasurface. Another interesting finding is that, the dominant detuned modes (2, 0) are apparently more energetic in Case 2 and Case 3, compared with the solid-wall case. In addition, the low-frequency modes generated in the later transitional stage, such as (1, 0), are also more evident than their counterparts in the solid-wall case. 

 \begin{figure}
	\vspace{1em}
	\centerline{
		\subfigure{\includegraphics[width=13cm]{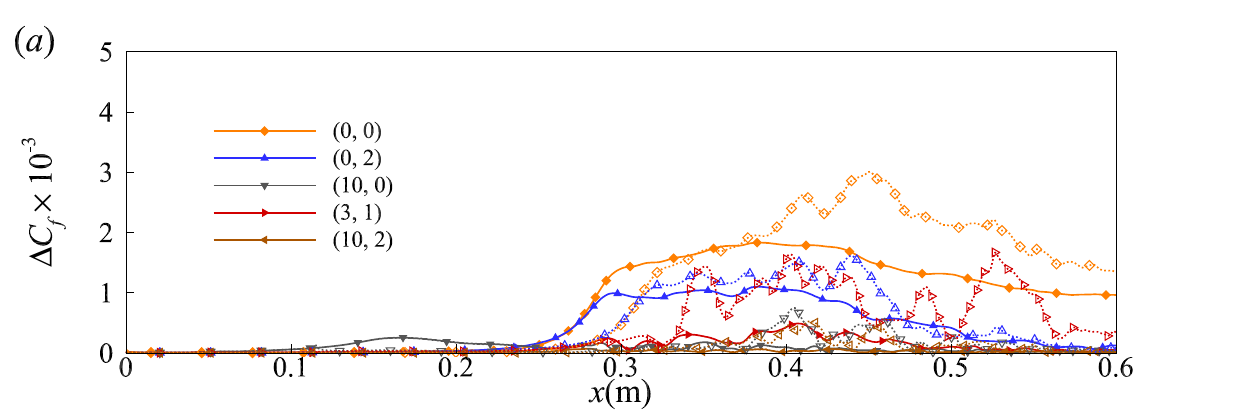}}}
	\centerline{
		\subfigure{\includegraphics[width=13cm]{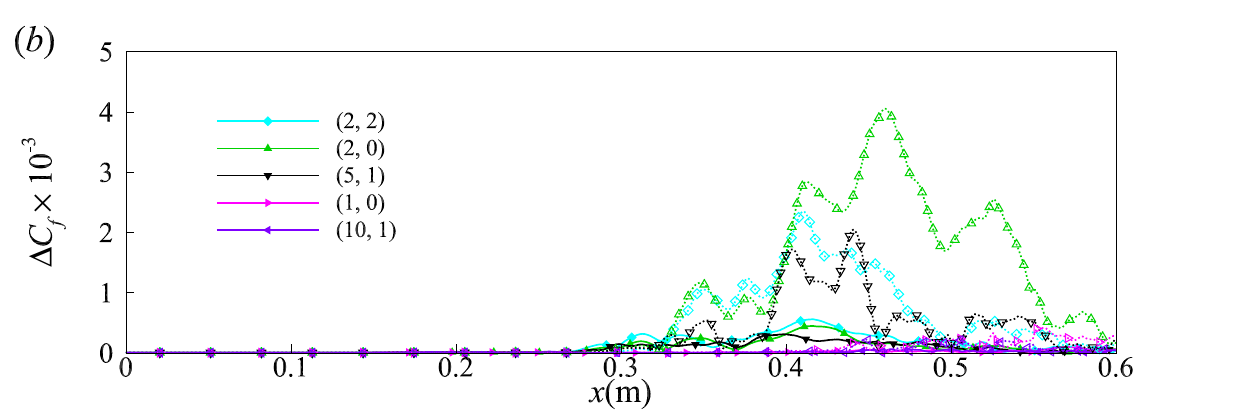}}}
	\centerline{
		\subfigure{\includegraphics[width=13cm]{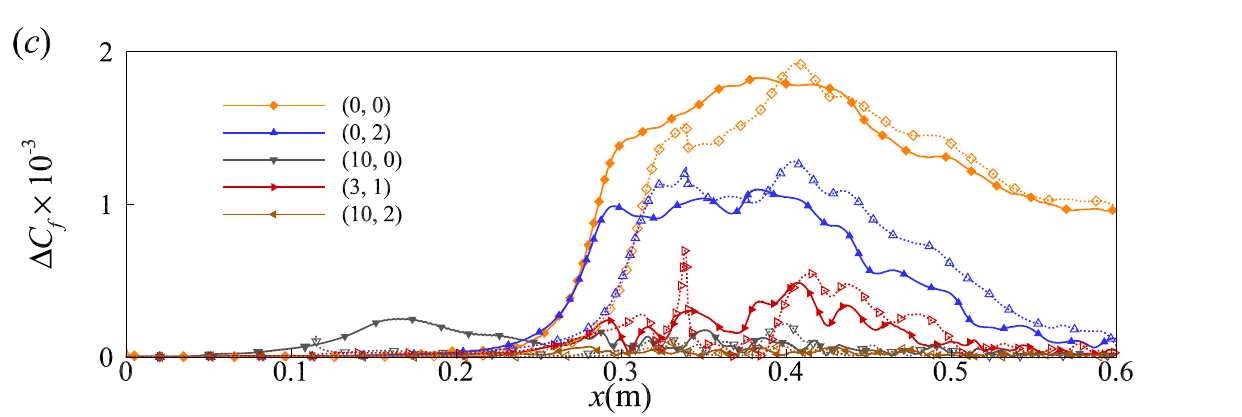}}}
	\centerline{
		\subfigure{\includegraphics[width=13cm]{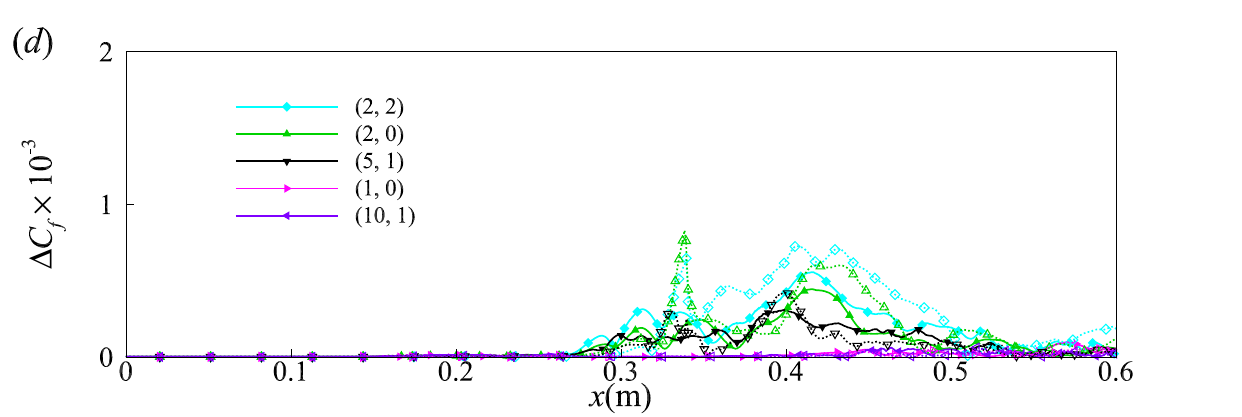}}}
	\caption{Comparison of maximum absolute modal contribution to the instantaneous skin friction coefficient ${C_f}$ for ($a$) and ($b$) for Case 1 and Case 2, and ($c$) and ($d$) for Case 1 and Case 3. The modes in Case 1 are represented by solid lines with filled symbols, and modes in Case 2 and Case 3 are represented by dotted lines with open symbols.}
	\label{fig11}
\end{figure}

\begin{figure}
	\centerline{
		\subfigure{\includegraphics[width=7cm]{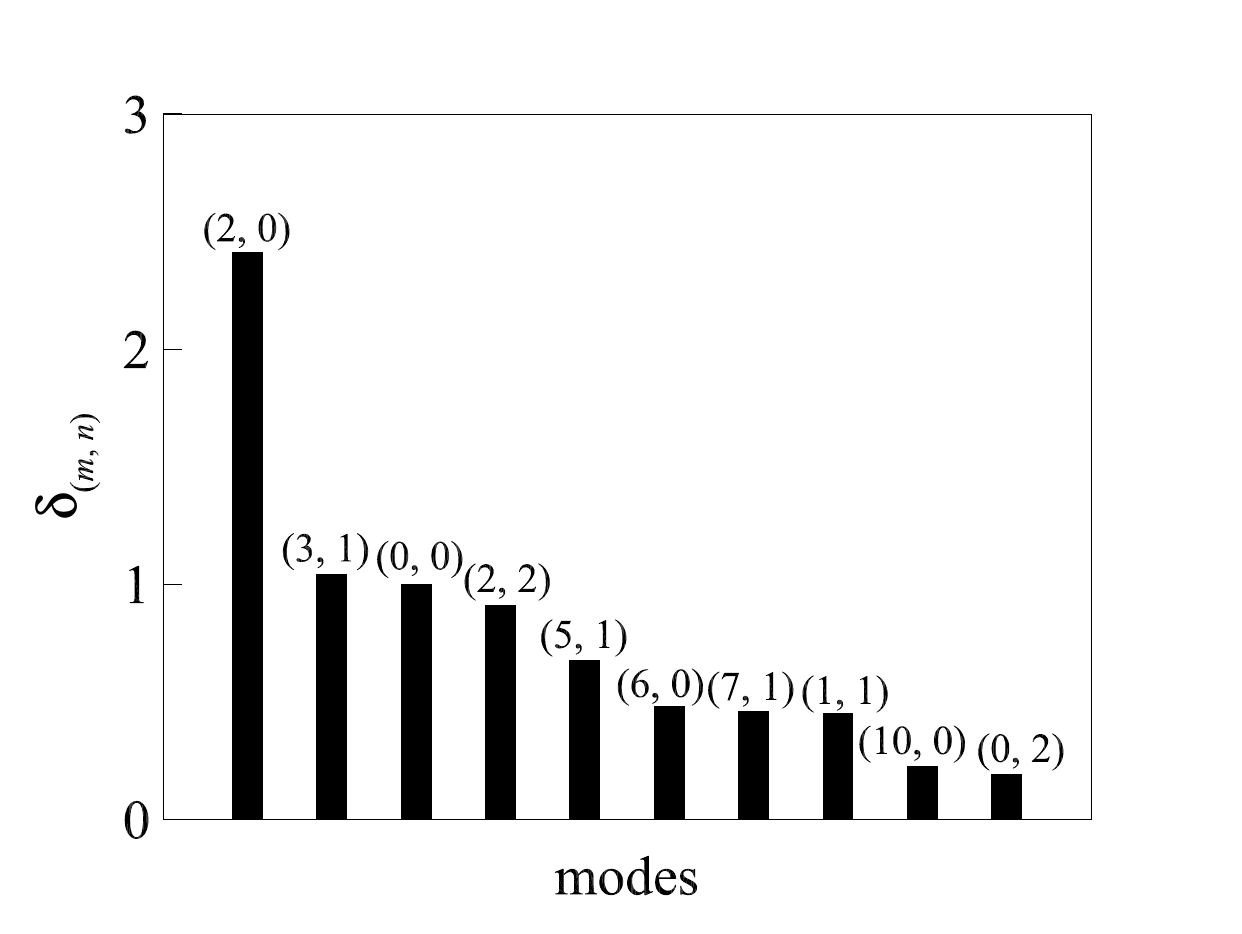}}}
	\caption{Comparision of ${\delta _{(m,\; n)}}$ among  different Fourier modes. The value ${\delta _{(0, \:0)}}$ is utilised for normalisation.}
	\label{figp}
\end{figure}

 We intend to identify the dominant Fourier mode contributing to the increment of higher skin friction in Case 2. To this end, the maximum absolute contribution from mode ($m$, $n$) to the instantaneous skin friction is defined by
 \begin{equation}\label{eq4.2}
\Delta {C_{f,(m,n)}}(x) = \mathop {\max }\limits_{(z,t)} \left| {{C_{f,(m,n),\textit{unsteady}}} - {C_{f,\textit{laminar}}}} \right|,
 \end{equation}
 where ${C_{f,(m,n),\textit{unsteady}}}$ and ${C_{f,\textit{laminar}}}$  are the instantaneous skin friction induced by the laminar flow plus mode ($m$, $ \pm n$) and by the laminar flow alone, respectively. This indicator is defined to highlight the Fourier modes which are the source contributors to the skin friction in figure \ref{fig9}. Figure \ref{fig11} compares the maximum absolute modal contribution to the instantaneous skin friction coefficient ${C_f}$ between cases with (Case 2 and Case 3) and without (Case 1) the acoustic metasurface. Apparently, the second mode (10, 0) is suppressed in both Case 2 and Case 3 with the acoustic metasurface, which agrees with the disappeared aligned structure in the vortex visualization (shown in figure \ref{fig7}) and the instantaneous wall friction (shown in figure \ref{fig9}). As displayed in figure \ref{fig11} ($b$), the detuned modes (2, 2) and (2, 0) demonstrate a considerably higher contribution, indicating a significantly strengthened combination resonance. In addition, the subharmonic-related mode (5, 1) and oblique mode (3, 1) in Case 2 also show a more pronounced contribution than that in Case 1. This result indicates the promotion of these two nonlinear interaction mechanisms (combination and subharmonic resonance) using the acoustic metasurface. Note that there are some differences between the wall friction contributions and the integrated Chu's energy, which resembles the linear case in section \ref{single-frequency disturbance}. For example, the Chu's energies of modes (2, 2) and (2, 0) are decreased and increased by the metasurface in figure \ref{fig10}, respectively. However, the contributions to $C_f$ of both modes are notably augmented by the metasurface. This implies that the wall quantity and the integrated energy across the boundary layer may have different performance. To quantitatively evaluate the effect of the metasurface, the difference of $\Delta C_f$ is calculated between Case 1 and Case 2. The result serves to identify the main contributor to the strengthened skin friction in the late transitional region in Case 2. The overall integral effect is calculated by
  \begin{equation}\label{eq4.2}
{\delta _{(m,\:n)}} = \int_{{x_1}}^{{x_2}} {\left( {\Delta {C_{f,\:(m,n),\:\textit{Case 2}}} - \Delta {C_{f,\:(m,n),\:\textit{Case 1}}}} \right){\rm{ }}dx},
 \end{equation}
 where ${{x_1}}$ = 0.34 m and  ${{x_2}}$ = 0.6 m. The comparision of ${\delta _{(m,\:n)}}$, normalised by ${\delta _{(0,\:0)}}$,  among  different Fourier modes is presented in figure \ref{figp}. The results indicate that the dominant contributor is the detuned planar mode (2, 0), which aligns with the predominant planar structure observed at approximately $x$ = 0.45 m in figure \ref{fig9}($b$). The above analysis answers the question raised by the second research objective in the Introduction.
 
Regarding the question of why transition onset is delayed, the increasing contribution of the mean flow distortion (0, 0) should be discussed. In the range $x \in [0.25, 0.32]$ m where the transition begins in figure \ref{fig6}, the presence of the metasurface leads to the postponed growing contribution of (0, 0) in figure \ref{fig11}($a$). In this range, only the streak mode (0, 2) is notable in terms of the magnitude. As a result, the delayed transition onset can be attributed to the postponed growth of mode (0, 2). This finding is interesting because the manifested control effect by the metasurface does not lie in the second mode directly. Given the nonlinearity, the Fourier mode (0, 2)  is the direct contributor to the delayed transition onset instead. Upstream of $x$ = 0.3 m, the contributions of other secondary modes are not evident in figure \ref{fig11}($a$). As a result, the delayed (0, 2) is mainly caused by the suppressed oblique-wave resonance (see the first interaction equation in equation (\ref{eq4.55})) via the relatively large-amplitude primary instability wave (3, 1). As interpreted in the discussion of figure \ref{fig10}, mode (3, 1) is stabilised by the metasurface in the nonlinear stage, which is probably achieved by the less energy received from the suppressed Mack second mode. Therefore, the second mode plays an indirect role in the transition delay mechanism of the acoustic metasurface.

\subsection{Energy budegt anaysis}
\begin{figure}
	\vspace{1em}
	\centerline{
		\subfigure{\includegraphics[width=6.5cm]{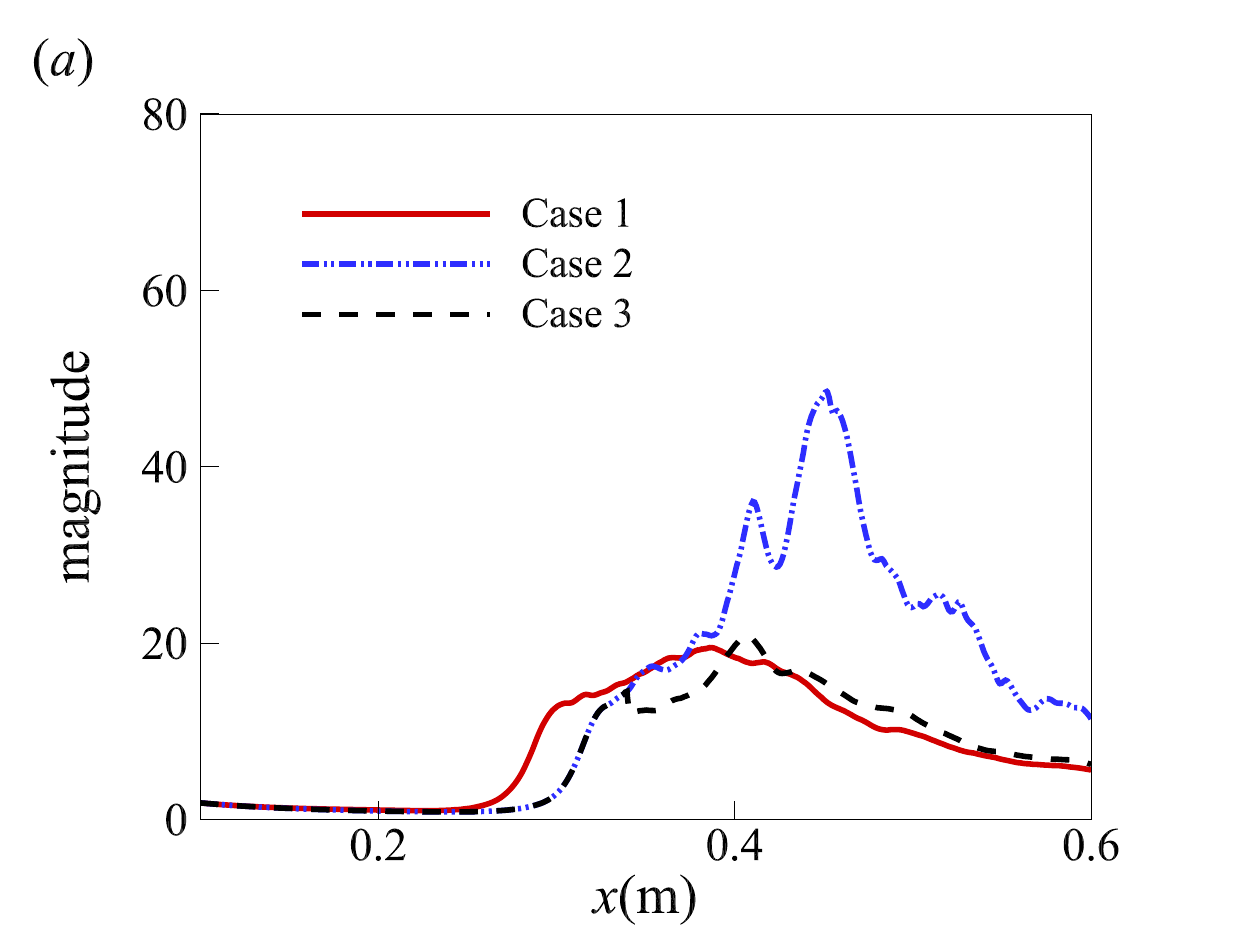}}
		\subfigure{\includegraphics[width=6.5cm]{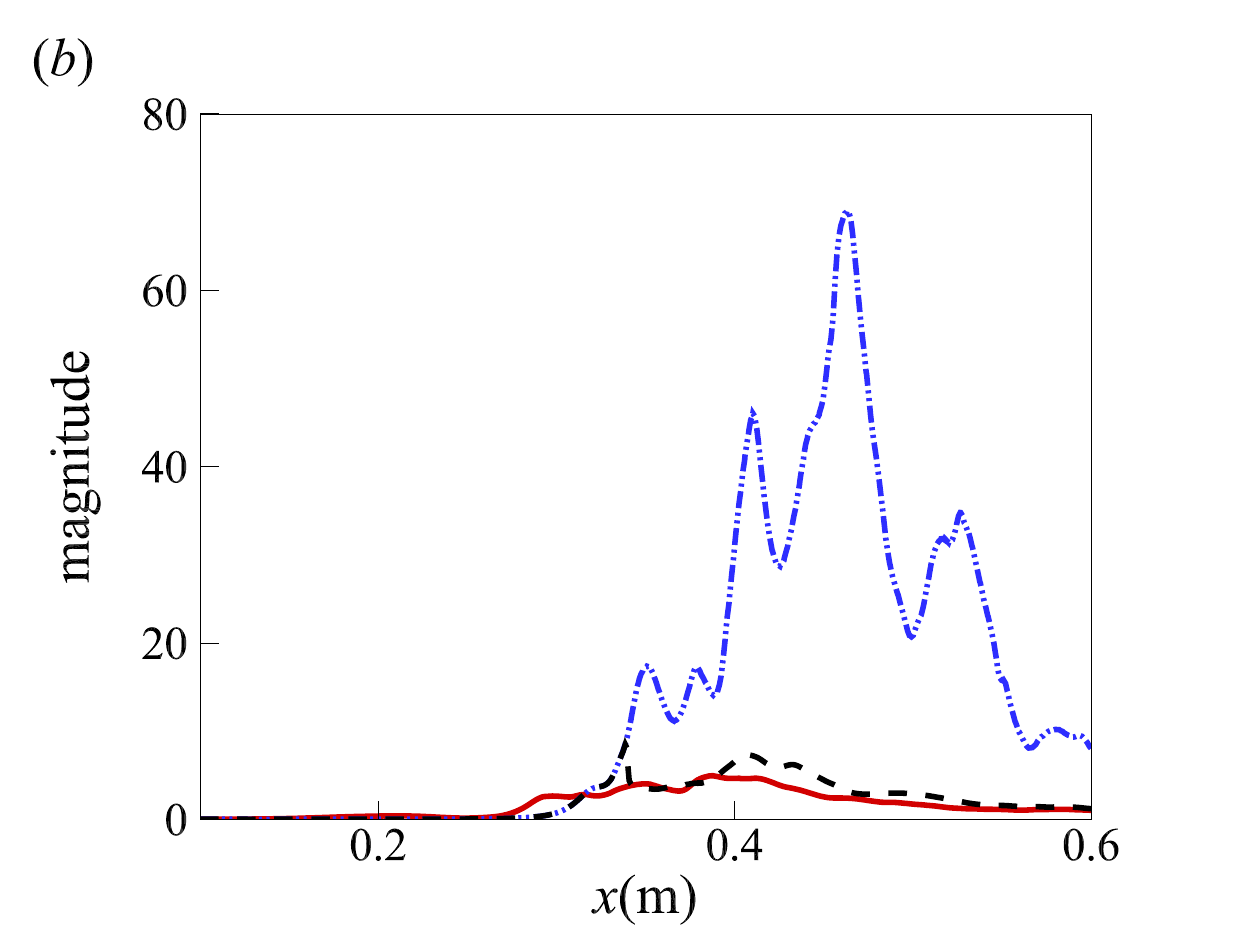}}}
	\centerline{
		\subfigure{\includegraphics[width=6.5cm]{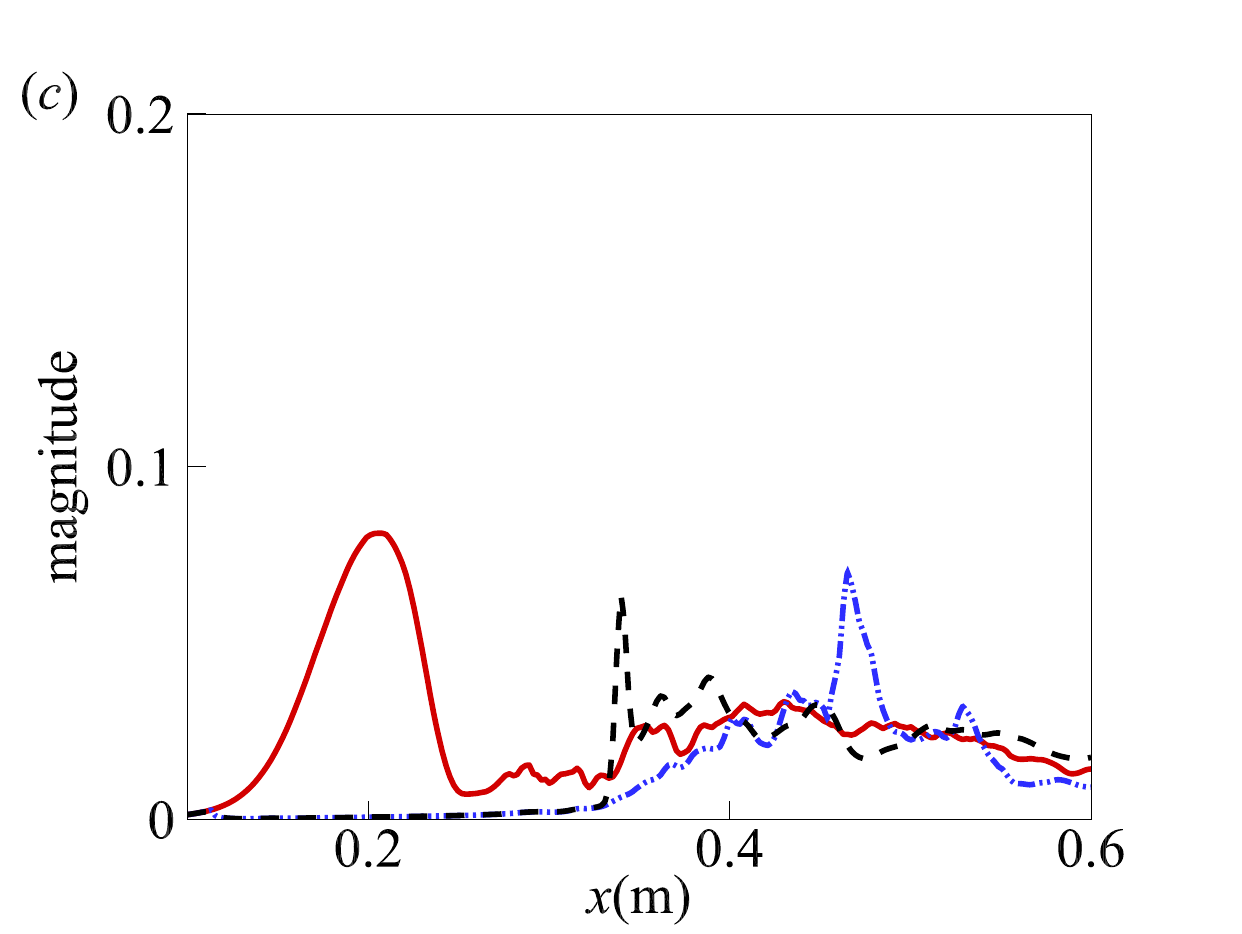}}
		\subfigure{\includegraphics[width=6.5cm]{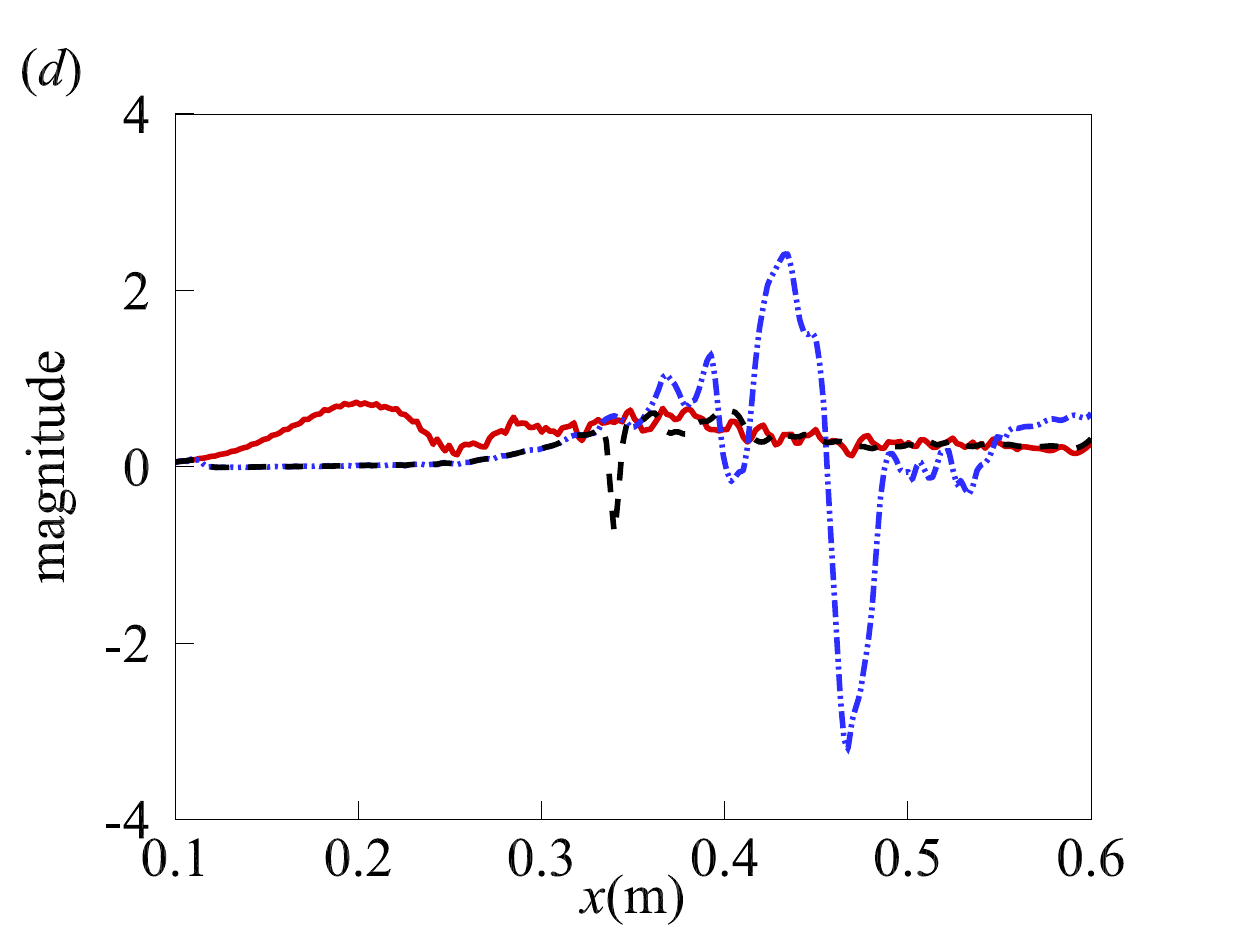}}}

	\caption{Comparison of energy budget terms of ($a$) ${\Phi _{\varpi 0}}$, ($b$) ${\Phi _{\varpi 2}}$, ($c$) ${\Phi _\vartheta }$, and ($d$) ${T_p}$ in the internal energy transport equation on the wall among Case 1, Case 2, and Case 3. Here, ${\Phi _{\varpi 1}}$ and ${\Phi _{\varpi 3}}$ are not shown due to their negligible amplitude in comparison with that of ${\Phi _{\varpi 0}}$ and ${\Phi _{\varpi 2}}$.}
	\label{fig12}
\end{figure}

 To better understand the effect of the acoustic metasurface, the transport of internal energy is of interest (Zhu \etal \space \citeyear{Zhu2016}). The influence of the acoustic metasurface on the energy budget terms is analysed. The viscous dissipation (which is decomposed into the contributions induced by shear ${\Phi _\varpi }$ and by dilatation ${\Phi _\vartheta }$) and pressure dilatation work (${T_p}$) near the wall are analysed herein. The expressions of ${\Phi _\varpi }$ and ${\Phi _\vartheta }$ are
 
  \begin{equation}\label{eq4.3}
{\Phi _\varpi } = \left\langle {\overline {\mu {\varpi _k}{\varpi _k}} } \right\rangle ,\: {\rm{ }}{\Phi _\vartheta } = \left\langle {\overline {\frac{4}{3}\mu {{(\frac{{\partial u}}{{\partial x}} + \frac{{\partial v}}{{\partial y}} + \frac{{\partial w}}{{\partial z}})}^2}} } \right\rangle.
 \end{equation}
 In this subsection, $\overline {( \cdot )} $ and $\overline {\left\langle  \cdot  \right\rangle } $  denote time and spanwise averaging, respectively. For a Cartesian coordinate system, the vorticity component ${{\varpi _k}}$ is expressed as
    \begin{equation}\label{eq4.3}
{\varpi _1} = \frac{{\partial w}}{{\partial y}} - \frac{{\partial v}}{{\partial z}},\:{\rm{ }}{\varpi _2} = \frac{{\partial u}}{{\partial z}} - \frac{{\partial w}}{{\partial x}},\:{\rm{ }}{\varpi _3} = \frac{{\partial v}}{{\partial x}} - \frac{{\partial u}}{{\partial y}}.
  \end{equation}
 Subsequently, the shear-induced dissipation is decomposed into four parts ${\Phi _\varpi } = {\Phi _{\varpi 0}} + {\Phi _{\varpi 1}} + {\Phi _{\varpi 2}} + {\Phi _{\varpi 3}}$, representing the effects of the time-and spanwise-averaged field, the second-order moment of the cross correlation between the fluctuations in dynamic viscosity and vorticity, the second-order moment of the vorticity self-correlation, and the third-order moment, respectively. The specific expressions of these four terms are given by
  \begin{equation}\label{eq4.3}
\begin{array}{l}
{\Phi _{\varpi 0}} = \left\langle {\overline \mu  } \right\rangle \left\langle {\overline {{\varpi _k}} } \right\rangle \left\langle {\overline {{\varpi _k}} } \right\rangle ,\:{\rm{ }}{\Phi _{\varpi 1}} = 2\left\langle {\overline {{\varpi _k}} } \right\rangle \left\langle {\overline {\mu '{{\varpi '}_k}} } \right\rangle ,\\[1.5ex]
{\Phi _{\varpi 2}} = \left\langle {\overline \mu  } \right\rangle \left\langle {\overline {{{\varpi '}_k}{{\varpi '}_k}} } \right\rangle ,\:{\rm{ }}{\Phi _{\varpi 3}} = \left\langle {\overline {\mu '{{\varpi '}_k}{{\varpi '}_k}} } \right\rangle .
\end{array}
  \end{equation}
The dimensionless time- and spanwise-averaged pressure dilatation term (${T_p}$) is written as
   \begin{equation}\label{eq4.4}
{T_p} =  - \left\langle {\bar p} \right\rangle \left\langle {\overline {\frac{{\partial {u_i}}}{{\partial {x_i}}}} } \right\rangle  - \left\langle {\overline {p'\frac{{\partial {{u'}_i}}}{{\partial {x_i}}}} } \right\rangle.
 \end{equation}
The instantaneous quantity is expressed as $\phi  = \left\langle {\overline \phi  } \right\rangle  + \phi '$, where $\left\langle {\overline \phi  } \right\rangle $ represents the base flow superimposed by mean flow distortion (MFD), and $\phi '$ denotes the disturbance excluding MFD. By incorporating MFD into the base flow, the formulation reflects the overall effect of the time- and spanwise-averaged field in the transitional stage.
  
 Figure \ref{fig12} provides the comparison of the streamwise development of the aforementioned terms among Cases 1, 2, and 3. The magnitude of the shear-induced dissipation ${\Phi _\varpi }$ of Case 2 is evidently higher than other cases downstream of $x$ = 0.34 m. The most prominent contribution originates from ${\Phi _{\varpi 2}}$ (shown in figure \ref{fig12}($b$)), which refers to the second-order moment of the vorticity self-correlation $\left\langle {\overline \mu  } \right\rangle \left\langle {\overline {{{\varpi '}_k}{{\varpi '}_k}} } \right\rangle $. Moreover, the reinforcement of time- and spanwise- averaged dissipation ${\Phi _{\varpi 0}} = \left\langle {\overline \mu  } \right\rangle \left\langle {\overline {{\varpi _k}} } \right\rangle \left\langle {\overline {{\varpi _k}} } \right\rangle $  is significant downstream of $x$ = 0.4 m, as shown in figure \ref{fig12}($a$). The other two terms are not displayed due to their negligible contributions. Furthermore, the dissipation induced by dilatation (${\Phi _\vartheta }$) is suppressed by the acoustic metasurface upstream of $x$ = 0.34 m, as illustrated by figure \ref{fig12}($c$). However, the magnitude of ${\Phi _\vartheta }$ is significantly lower than that of the shear-induced dissipation (${\Phi _\varpi }$). These results suggest that the shear-induced dissipation plays a dominant role in the internal energy transport, and that the first-mode oblique breakdown may be crucial in the late-stage transitional flow. As reported in section \ref{interaction}, the oblique breakdown is strengthened by the acoustic metasurface in Case 2. This finding aligns with the transition simulation by Tullio \etal \space (\citeyear{de2010direct}), who directly resolved the flow within the pores of the acoustic metasurface and demonstrated that the first mode was promoted. As for Case 3, the magnitude of energy budget terms gradually agrees with that in Case 1 downstream of $x$ = 0.34 m, as the boundary shifts from the acoustic metasurface to a solid wall condition.
 
 The terms associated with dilatational work, including dilatation-induced dissipation in figure \ref{fig12}($c$) and pressure dilatation work shown in figure \ref{fig12}($d$), are suppressed significantly by the acoustic metasurface at around $x$ = 0.2 m. The underlying physics probably relates to the second mode of acoustic nature (Zhu \etal \space \citeyear{zhu2018}; Chen \etal\space\citeyear{Chen2023b}), whose main energy source near the wall is dilatational work and can be suppressed by the acoustic metasurface.
 
 In summary, dilatation-related budget terms are suppressed while the shear-related terms are strengthened in cases with the acoustic metasurface. The results of the statistical energy budget analysis agree with the energy source analysis using the relative phase analysis (Chen \etal\space\citeyear{Chen2023a}). In that work, it was concluded that the acoustic metasurface stabilises the second mode by suppressing the dilatation near the wall while destabilizing the first mode by strengthening the mean shear.
 
 \begin{figure}
 	\vspace{1em}
 	\centerline{
 		\subfigure{\includegraphics[width=6cm]{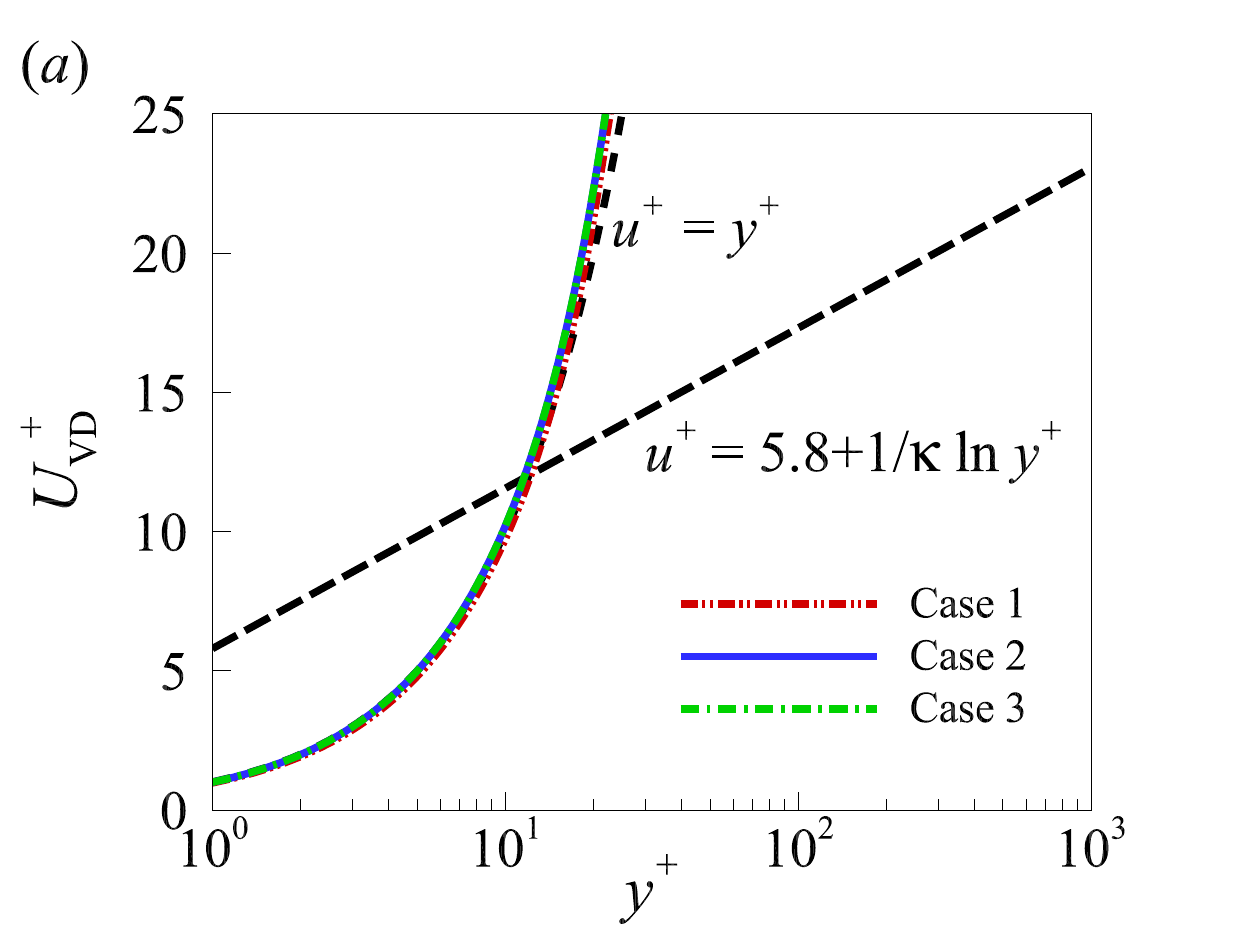}}
 		\subfigure{\includegraphics[width=6cm]{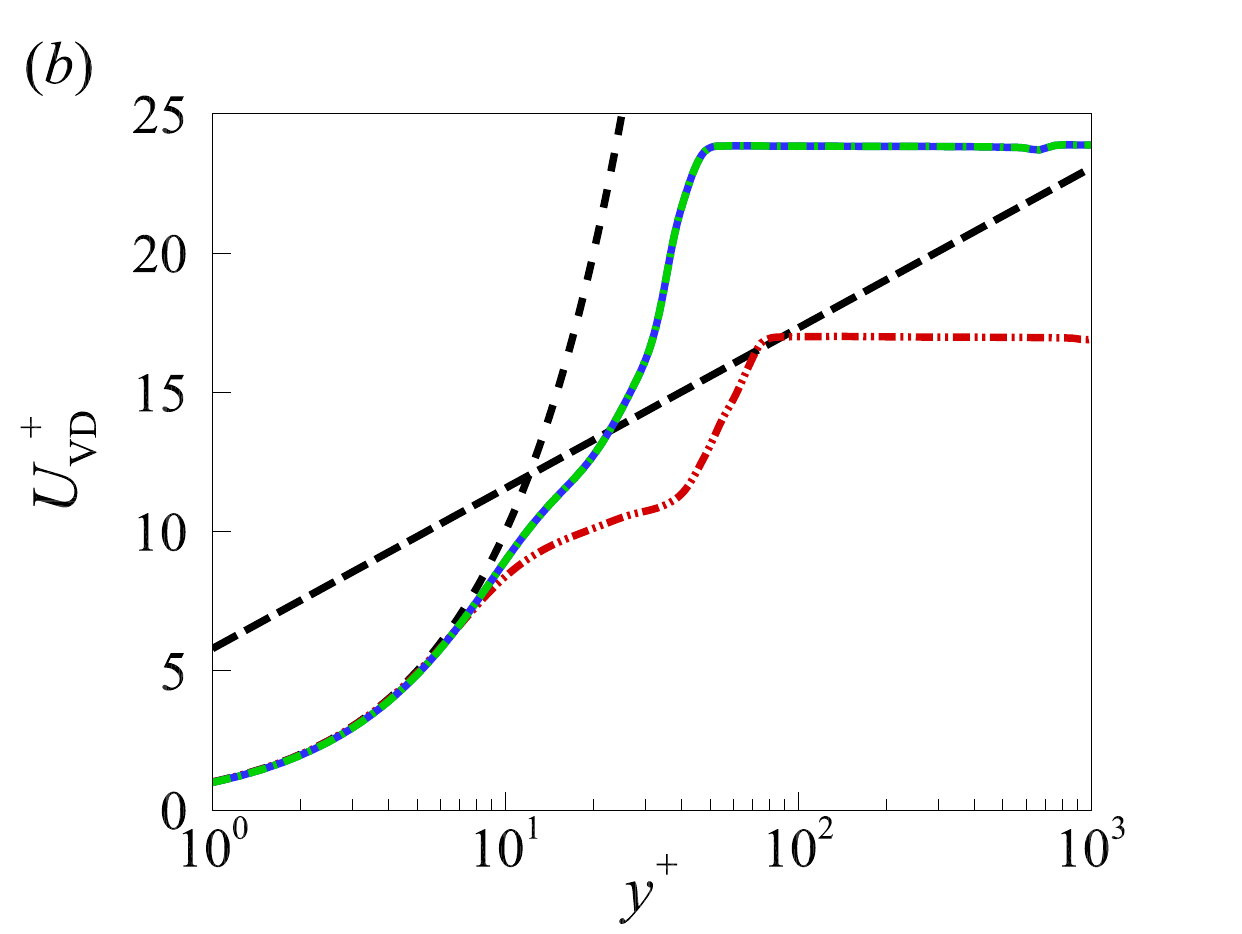}}}
 	\centerline{
 		\subfigure{\includegraphics[width=6cm]{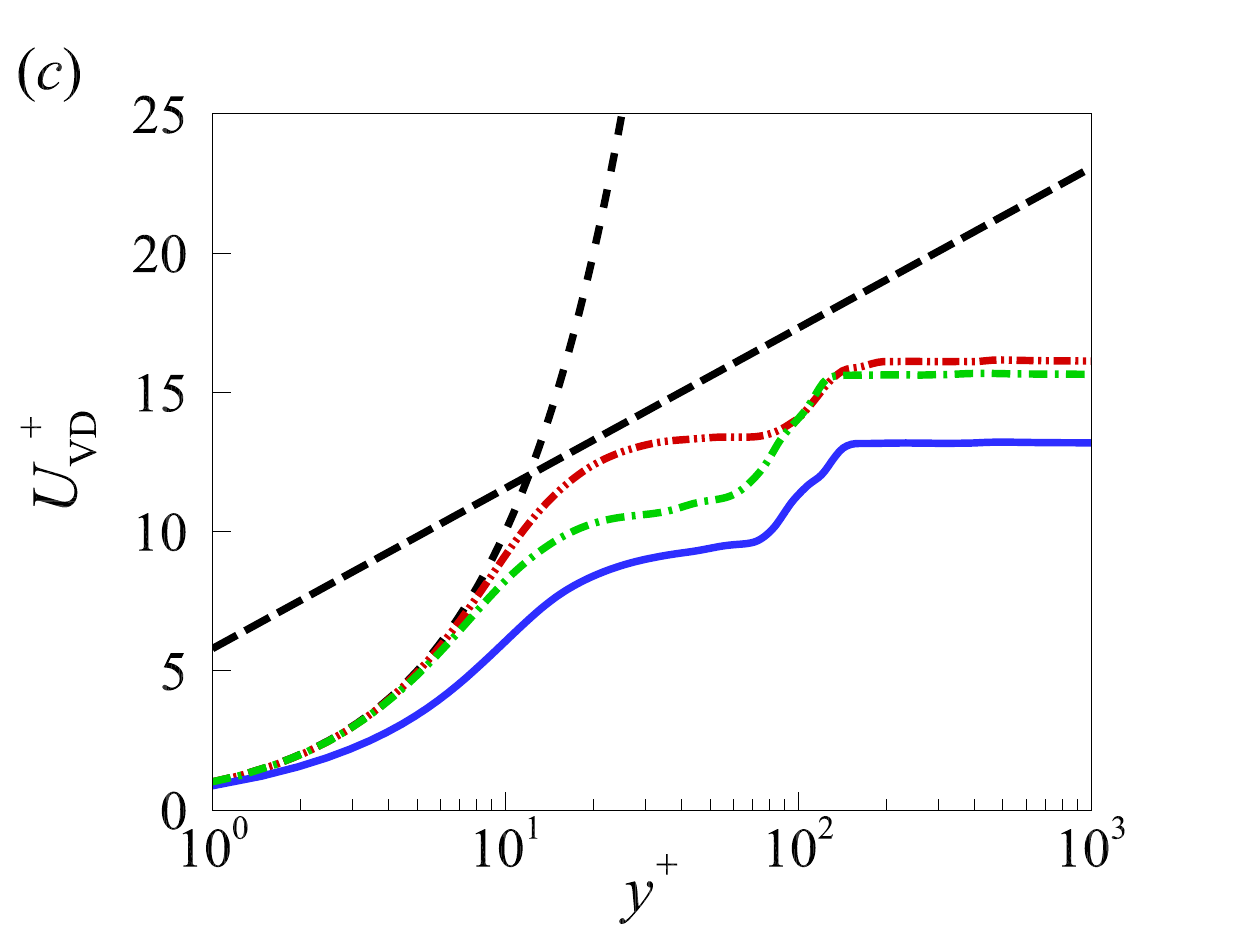}}
 		\subfigure{\includegraphics[width=6cm]{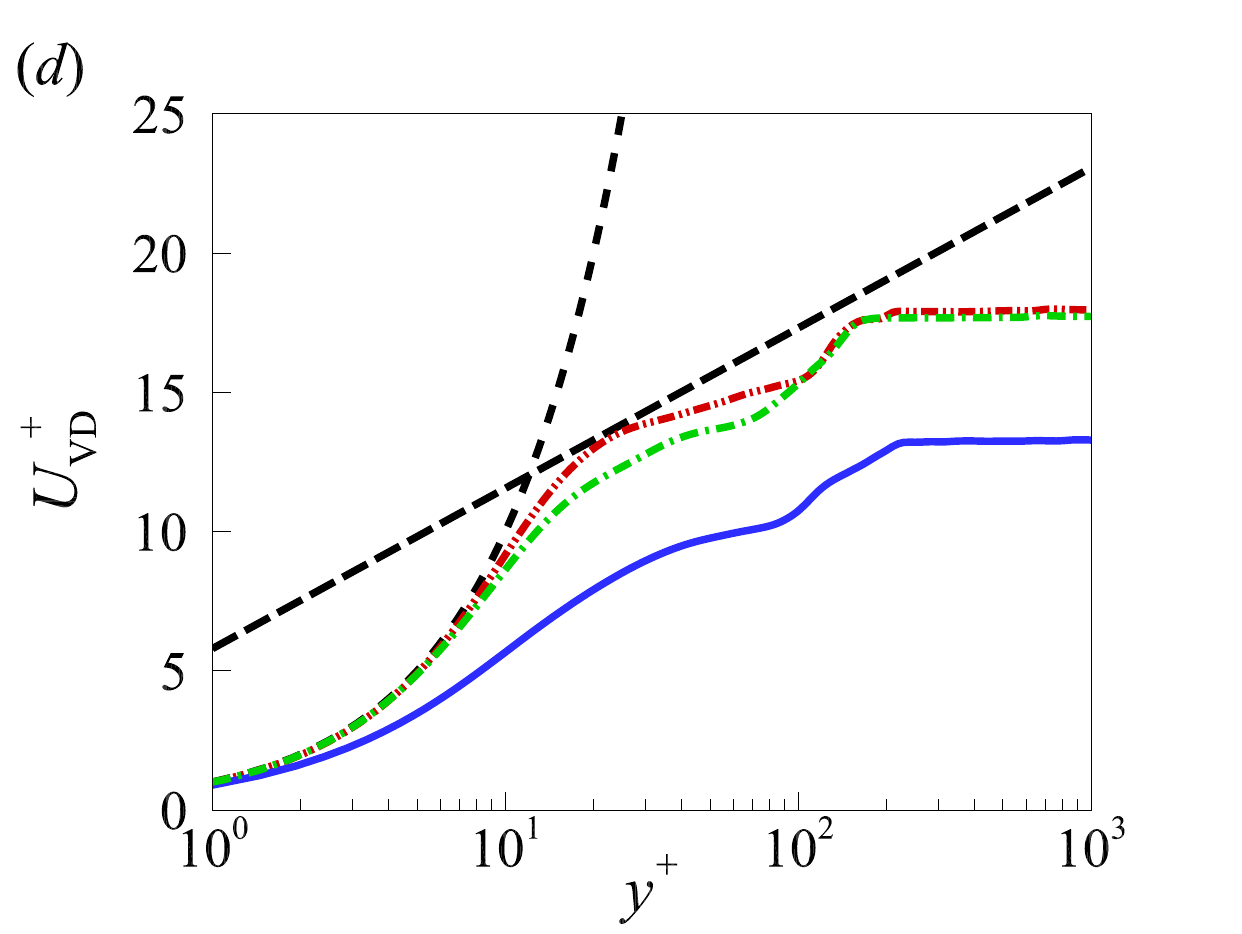}}}
 	\centerline{
 		\subfigure{\includegraphics[width=6cm]{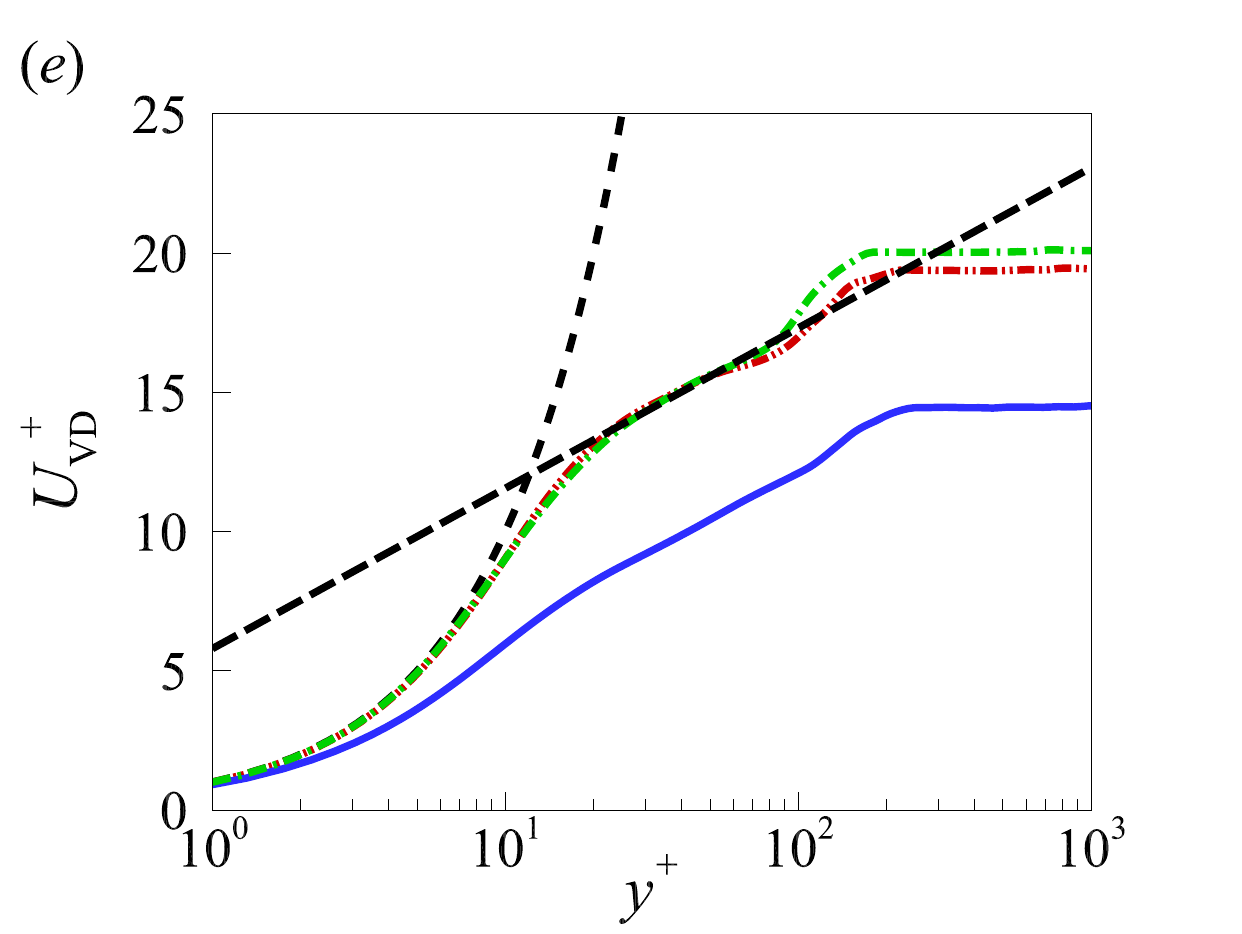}}
 		\subfigure{\includegraphics[width=6cm]{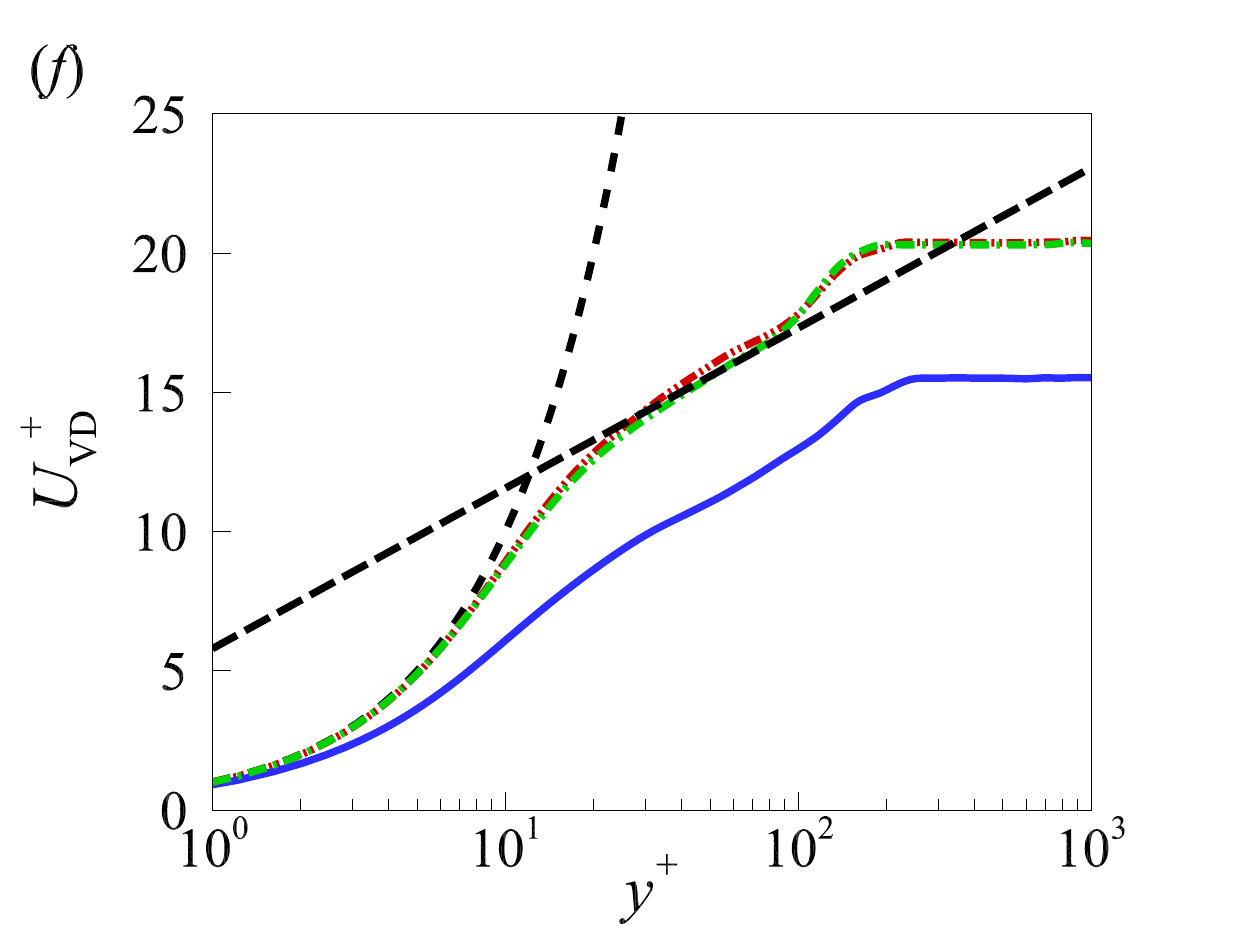}}}
 	
 	\caption{Van Driest velocity profiles for Case 1, Case 2, and Case 3 at ($a$) $x$ = 0.2 m, ($b$) $x$ = 0.3 m, ($c$) $x$ = 0.4 m, ($d$) $x$ = 0.5 m, ($e$) $x$ = 0.55 m, ($f$) $x$ = 0.6 m.}
 	\label{fig8}
 \end{figure}
 
 \subsection{Paths to turbulent flow}
 
 In this section, the spanwise- and time-averaged flow field is focused and analysed. To evaluate the flow state and its progression, the van-Driest-transformed velocity (Van \citeyear{van1951turbulent}) $U_{{\rm{VD}}}^ +$ profiles for three cases at various streamwise stations are plotted in figure \ref{fig8}.  The influence of mean density variations due to the effect of compressibility can be removed by van Driest transformation for the mean velocity
 
 \begin{equation}\label{eq 4.1}
U_{{\rm{VD}}}^ +  = \int_0^{{{\bar u}^ + }} {\sqrt {\frac{{\bar \rho }}{{{{\bar \rho }_w}}}} } d{\bar u^ + }(y),
 \end{equation}
where the friction velocity ${\bar u_\tau }$ is utilised for normalisation
 \begin{equation}\label{eq 4.1}
{\bar u^ + } = \frac{{\bar u}}{{{{\bar u}_\tau }}}, \: {\rm{ }}{\bar u_\tau } = \sqrt {\frac{{{{\bar \tau }_w}}}{{{{\bar \rho }_w}}}}.
\end{equation}

The viscous sublayer law (${u^ + } = {y^ + }$) and the log-layer law (${u^ + } = 5.8 + {1 \mathord{\left/{\vphantom {1 \kappa }} \right.\kern-\nulldelimiterspace} \kappa }\ln {y^ + }$, $\kappa$ = 0.4) are given as references, as black dashed lines. The intercept 5.8 is larger than the incompressible counterpart 5.0, which has been reported by Guo \etal \space (\citeyear{Guo2022b}). At $x$ = 0.2 m, the result of the three cases agrees well with the reference, showing a consistent and well-resolved viscous sublayer. Downstream $x$  = 0.3 m, Case 1 shows its initial establishment of the log layer, while in Case 2 and Case 3, velocity profiles just begin to deviate from sublayer reference at around ${y^ + }$ = 10. Starting from $x$ = 0.55 m, the velocity profile gradually approaches the reference log-layer law,  and the transformed velocity profile is nearly the same for Case 1 and Case 3 at $x$ = 0.6 m. The observations indicate that fully developed turbulence is established, even if the slope shows a little difference due to the high Mach number. A similar difference was also reported in supersonic/hypersonic simulation by Zhou \etal\space(\citeyear{zhou2023interactions}) and Guo \etal \space (\citeyear{Guo2022b}).  As for Case 2, the near wall slope (${{dU_{{\rm{VD}}}^ + } \mathord{\left/{\vphantom {{dU_{{\rm{VD}}}^ + } {d{y^ + }}}} \right.\kern-\nulldelimiterspace} {d{y^ + }}}$) keeps lower due to the acoustic metasurface utilised, compared to solid-wall case downstream $x$ = 0.3 m (see figure \ref{fig8} ($c-f$)). This is qualitatively consistent with the permeable wall over the supersonic boundary layer studied by Chen \& Scalo (\citeyear{Chen2021a}; \citeyear{Chen2021b}).

 \begin{figure}
	\vspace{1em}
	\centerline{
		\subfigure{\includegraphics[width=6cm]{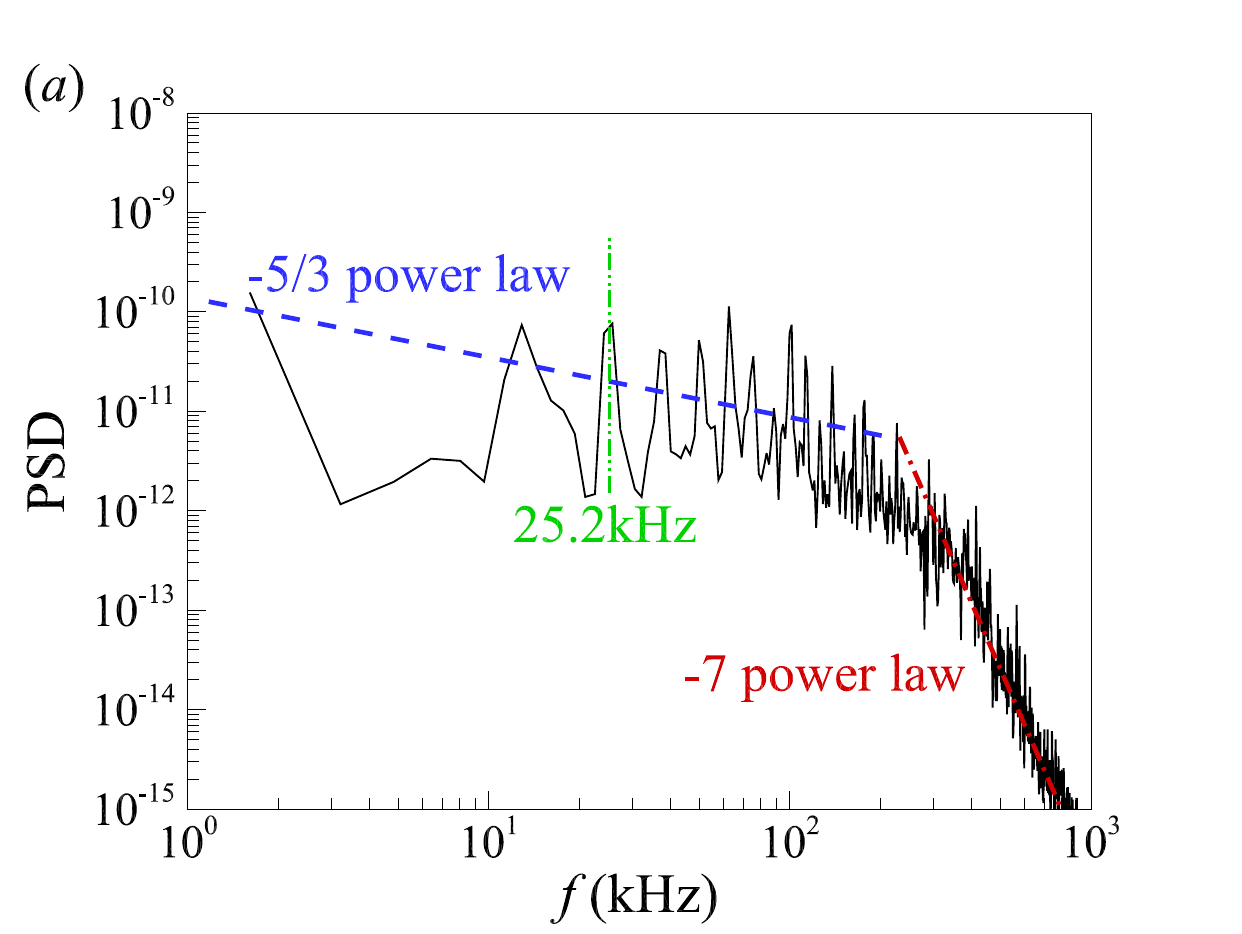}}
		\subfigure{\includegraphics[width=6cm]{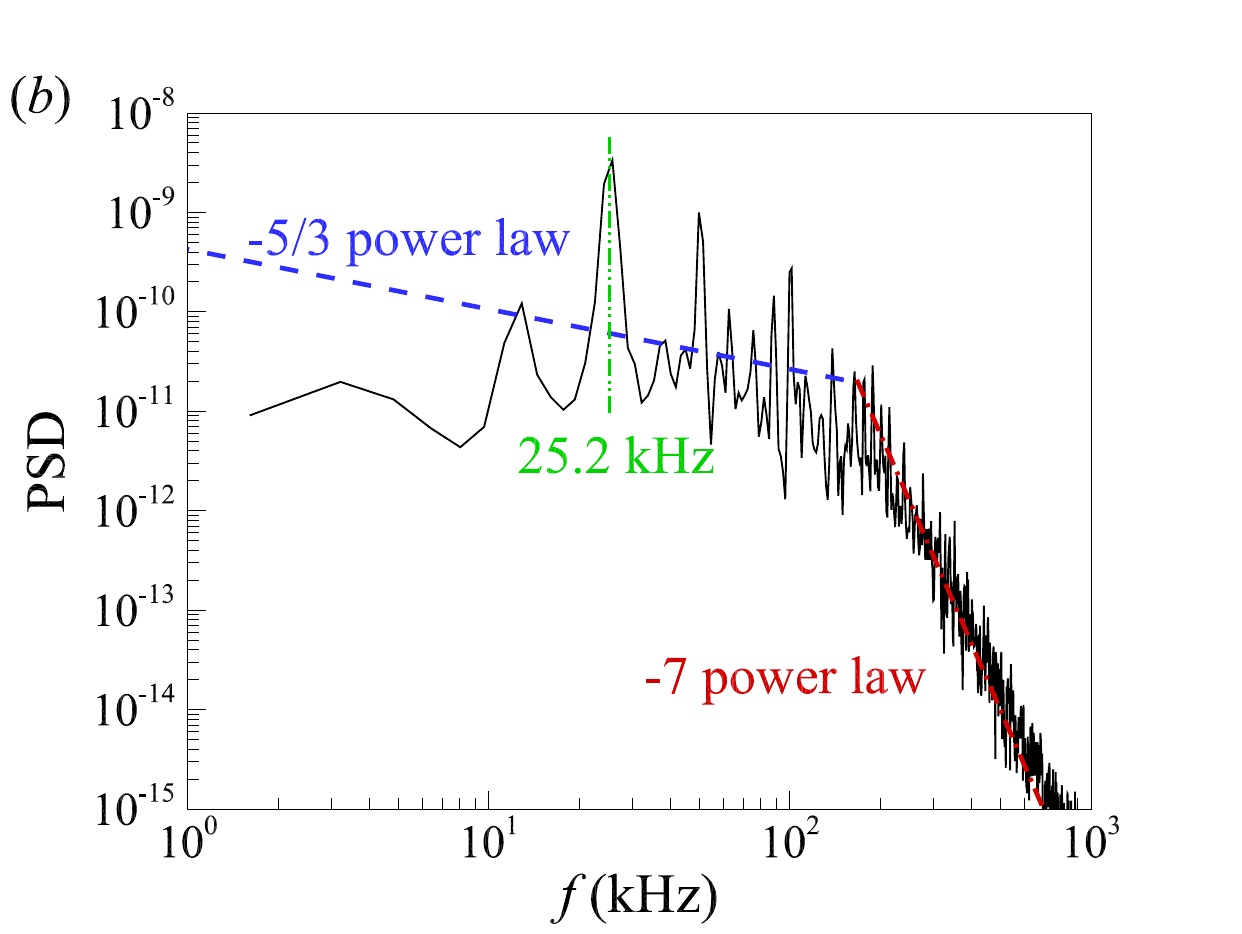}}}
	\centerline{
		\subfigure{\includegraphics[width=6cm]{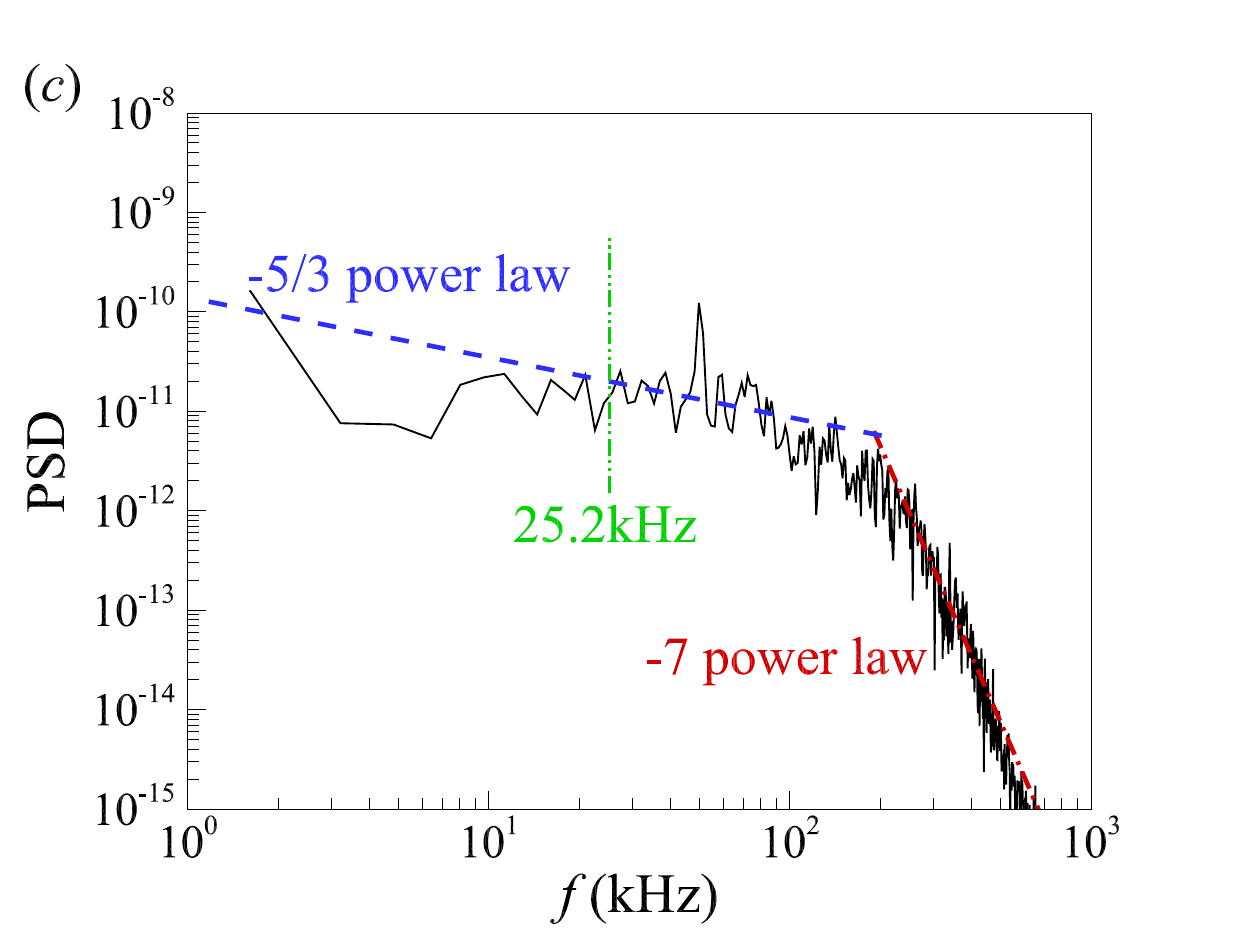}}
		\subfigure{\includegraphics[width=6cm]{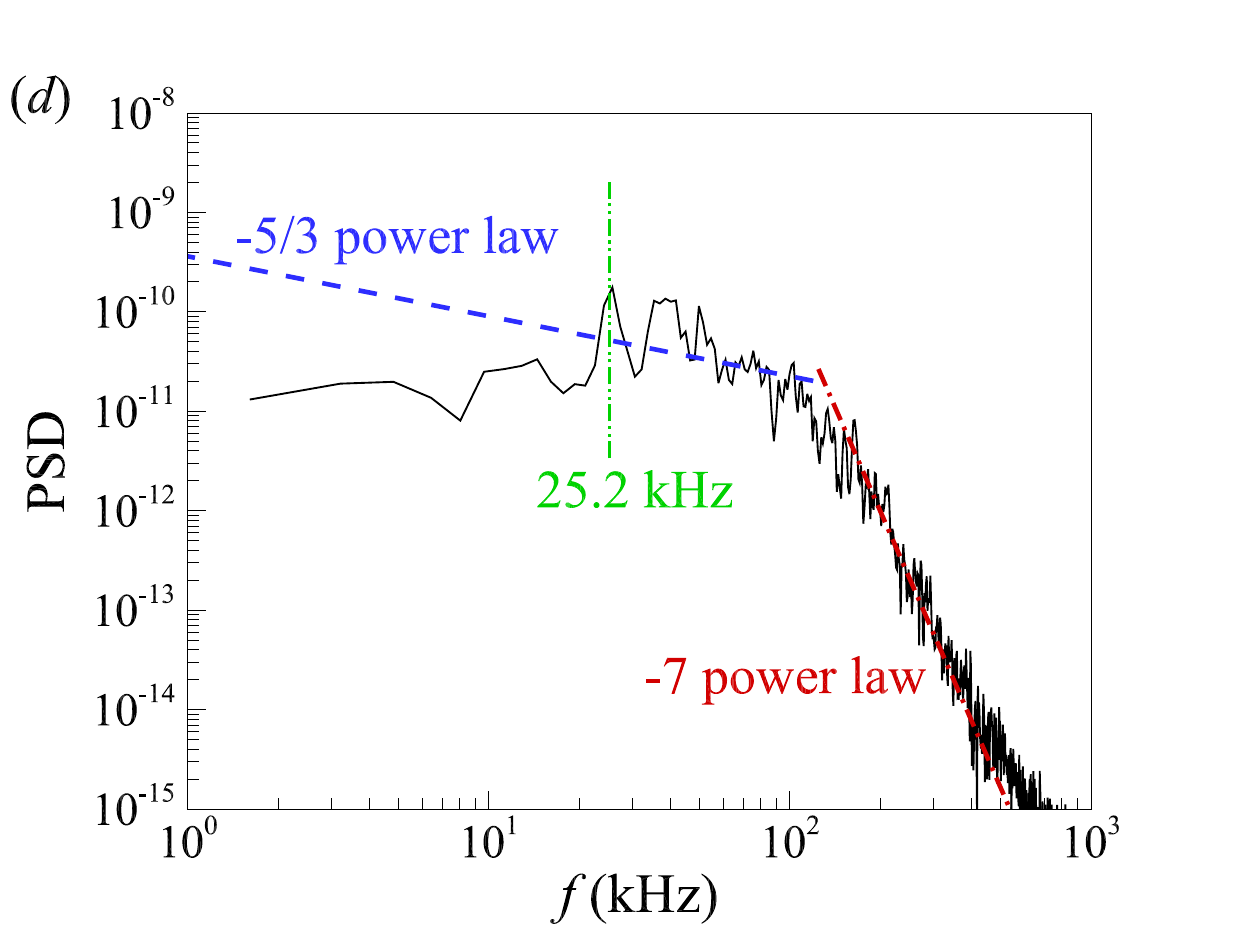}}}
	\caption{Power spectral density (PSD) of wall pressure fluctuation at $x$ = 0.45 m for ($a$) Case 1 and ($b$) Case 2, and at $x$ = 0.6 m for ($c$) Case 1 and ($d$) Case 2, in comparison with the law of  ${\omega ^{ - 5/3}}$  in the inertial subrange and the law of  ${\omega ^{ - 7}}$  in the dissipation scale. The dash dot-dot line marks the frequency corresponding to detuned modes (2, 0) and (2, 2).}
	\label{figPSD}
\end{figure}

 Owing to the augmented near-wall shear-induced dissipation in the late transitional region for Case 2, the van-Driest-transformed velocity profiles deviate from the viscous law near the wall. To evaluate the flow state, the PSD of wall pressure fluctuations at the downstream location is analysed. As illustrated in figure \ref{figPSD}, a good agreement is reached with the reference scale law for Case 1 at  $x$ = 0.6 m, indicating the establishment of a fully developed turbulent state. In contrast, the comparison with the ${\omega ^{ - 5/3}}$ law in the inertial subrange for Case 2 shows less consistency compared to Case 1. Moreover, the peak frequency ($f$ = 25.2 kHz) corresponding to detuned modes (2, 0) and (2, 2) is observed at  $x$ = 0.6 m for Case 2. This indicates that spectral broadening is incomplete at this location, and fully developed turbulence has not yet been established.
 
 In summary, the transition onset is delayed by the acoustic metasurface, and fully developed turbulence, as observed in the solid wall case, does not appear to have been established within the computational domain using the acoustic metasurface in Case 2.
 
\section{Conclusion}\label{Conclusion}
Hypersonic boundary-layer transition control using a promising passive technique, i.e., the acoustic metasurface, is investigated. The purpose of this work is to elucidate the mechanism of transition delay by the acoustic metasurface subject to multiple primary instabilities and to explain the enhancement of skin friction during the late transitional stage. The TDIBC, which models the effect of the acoustic metasurface in the time domain, is successfully incorporated into the Navier-Stokes solver OpenCFD. The resolvent analysis is performed to obtain the optimal disturbances of the boundary layer, and both the low-frequency first mode and the high-frequency second mode are involved in the primary instabilities. 

For the transition induced jointly by the first and second modes, the effectiveness of the metasurface in delaying the onset of transition has been demonstrated through our direct numerical simulations. The vortex visualization indicates that the spanwise-aligned structure associated with the second mode disappears when the acoustic metasurface is employed. The staggered and hairpin structures as well as the streamwise streak are postponed with the presence of the metasurface. The fully developed turbulence observed in the solid-wall case seems not yet established when the acoustic metasurface is utilised. Notably, the acoustic metasurface induces higher wall friction during the late transitional stage, which can be mitigated by replacing the acoustic impedance boundary with a solid wall condition in this region. A bi-Fourier analysis of wall friction is conducted to elucidate the role of the acoustic metasurface in delaying the transition onset and increasing the late-stage skin friction. It turns out that, the saturation of the second mode is delayed in cases involving the acoustic metasurface, which consequently attenuates the secondary growth of the first mode and the growth of the resulting streamwise streak. Eventually, the transition onset is delayed due to the postponed contribution of mean flow distortion. Regarding the reinforced skin friction in the late stage, different secondary modes participate in the physical process. Among them, the detuned modes primarily contribute to the increased wall friction with the metasurface. The mean internal energy transport suggests that the dilatational work is weaker prior to the transition, whereas the dissipation induced by shear is stronger in the cases with the acoustic metasurface. The latter is responsible for the higher wall friction from the perspective of the energy budget. The comparative study implies that the strategic placement of the acoustic metasurface is significant in minimizing the skin friction and probably also the heat flux.
\\	\vspace{1em}

\noindent{\bf Declaration of Interests}\\
The authors report no conflict of interest.
\vspace{1.5em}

\noindent{\bf Aknowlegment}\\
This study was supported by the National Natural Science Foundation of China (NSFC Grants No. 12272049) and the Research Grants Council, Hong Kong, under Contract Nos 15216621, and 15203724. We would like to express our sincere appreciation to Prof. Li Xinliang for his generosity in providing the direct numerical simulation codes, and we sincerely thank Dr. Hao Jiaao for his kind provision of the resolvent analysis codes.

\appendix
\section{Poles and residues for pole base functions}\label{Appendix A}
Table \ref{label2} lists the 20 pairs of dimensionless poles and residues employed to calculate softness using equation (\ref{eq2.15}) with ${C_0}$ = -0.05. The reference length is 1 m here. The resulting softness, dependent on the frequency, is illustrated in figure \ref{fig1}.
\begin{table}
	\centering
	\caption{Conjugate pairs of dimensionless poles and residues.}
	\begin{tabular}{ccc}
		\text{Pair number ($k$)} & \text{Poles (${p_k}$)} & \text{Residues (${\mu _k}$)} \\ 
		\midrule
        1 & ( -1504294.553, 4810.197) & (677409.898, -158275.251) \\
2 & (-466.785, 52.763) & (1161.854, 4177.756) \\
3 & (-419.365, 1007.668) & (596.328, 365.519) \\
4 & (-517.687, 1882.378) & (1153.962, -317.971) \\
5 & (-513.669, 2834.064) & (-119.973, -572.596) \\
6 & (-1032.603, 2120.59) & (-2407.724, 3249.423) \\
7 & (-445.944, 3554.34) & (64.41, -212.655) \\
8 & (-75, 14.263) & (-489.092, -992.871) \\
9 & (-1443.028, 5837.405) & (-78.771, 5.575) \\
10 & (-75.01, 34.641) & (387.44, 191.596) \\
11 & (-173.145, 4272.261) & (-19.664, 22.645) \\
12 & (-107.785, 4217.54) & (9.436, 28.795) \\
13 & (-102.821, 4091.752) & (-90.437, -39.358) \\
14 & (-75, 4102.434) & (-69.528, -19.905) \\
15 & (-78.478, 4165.123) & (19.763, 22.55) \\
16 & (-138.227, 4111.577) & (3.371, -12.674) \\
17 & (-75, 4122.214) & (26.314, 2.057) \\
18 & (-75.867, 4118.516) & (36.017, 4.191) \\
19 & (-85.535, 4099.064) & (49.002, 2.039) \\
20 & (-81.259, 4095.32) & (61.568, 1.498) \\
	\end{tabular}
\label{label2}
\end{table}

\section{Verification of TDIBC code}\label{Appendix B}
\begin{figure}
	\vspace{1em}
	\centerline{
		\subfigure{\includegraphics[width=6.8cm]{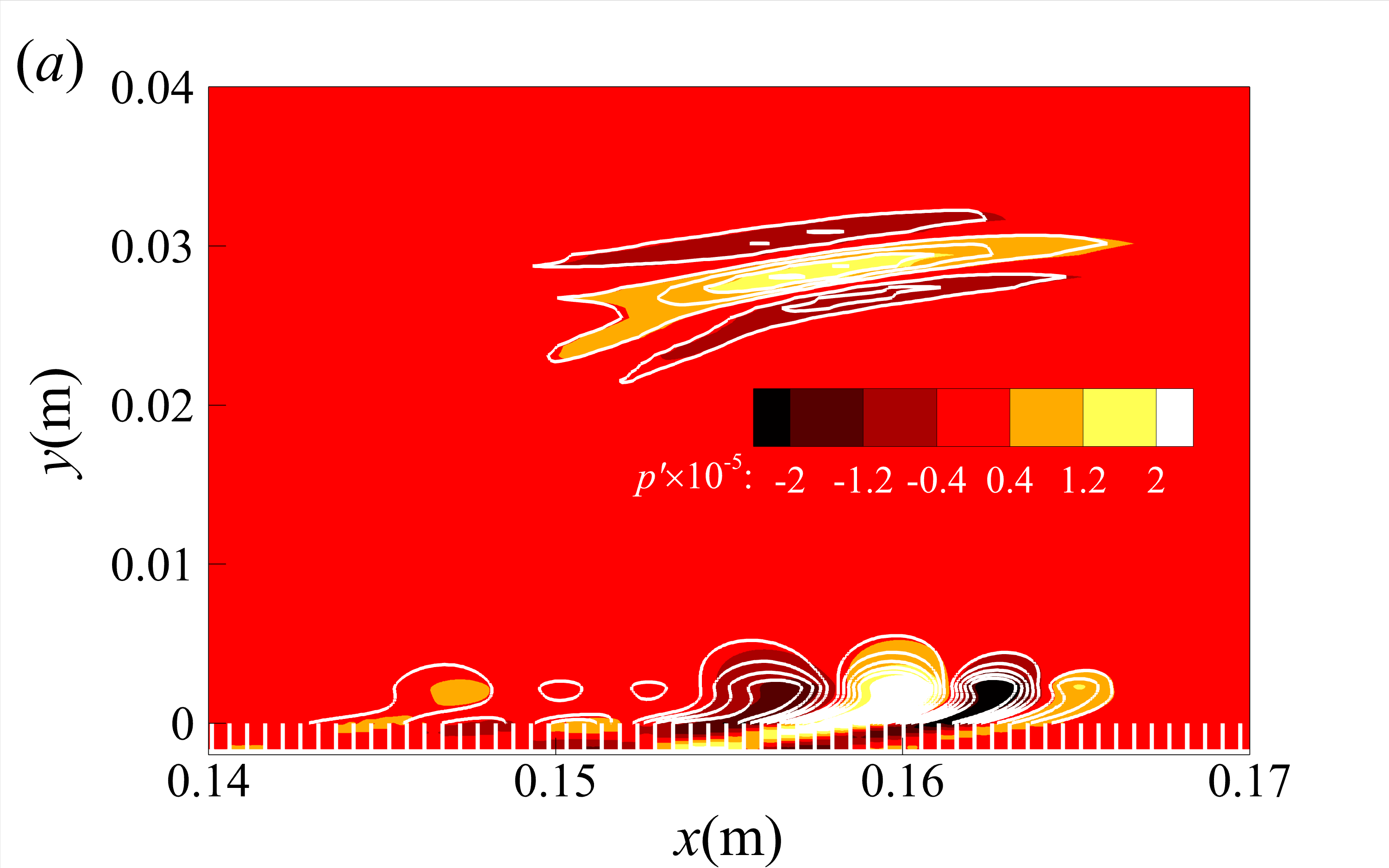}}
		\subfigure{\includegraphics[width=6.8cm]{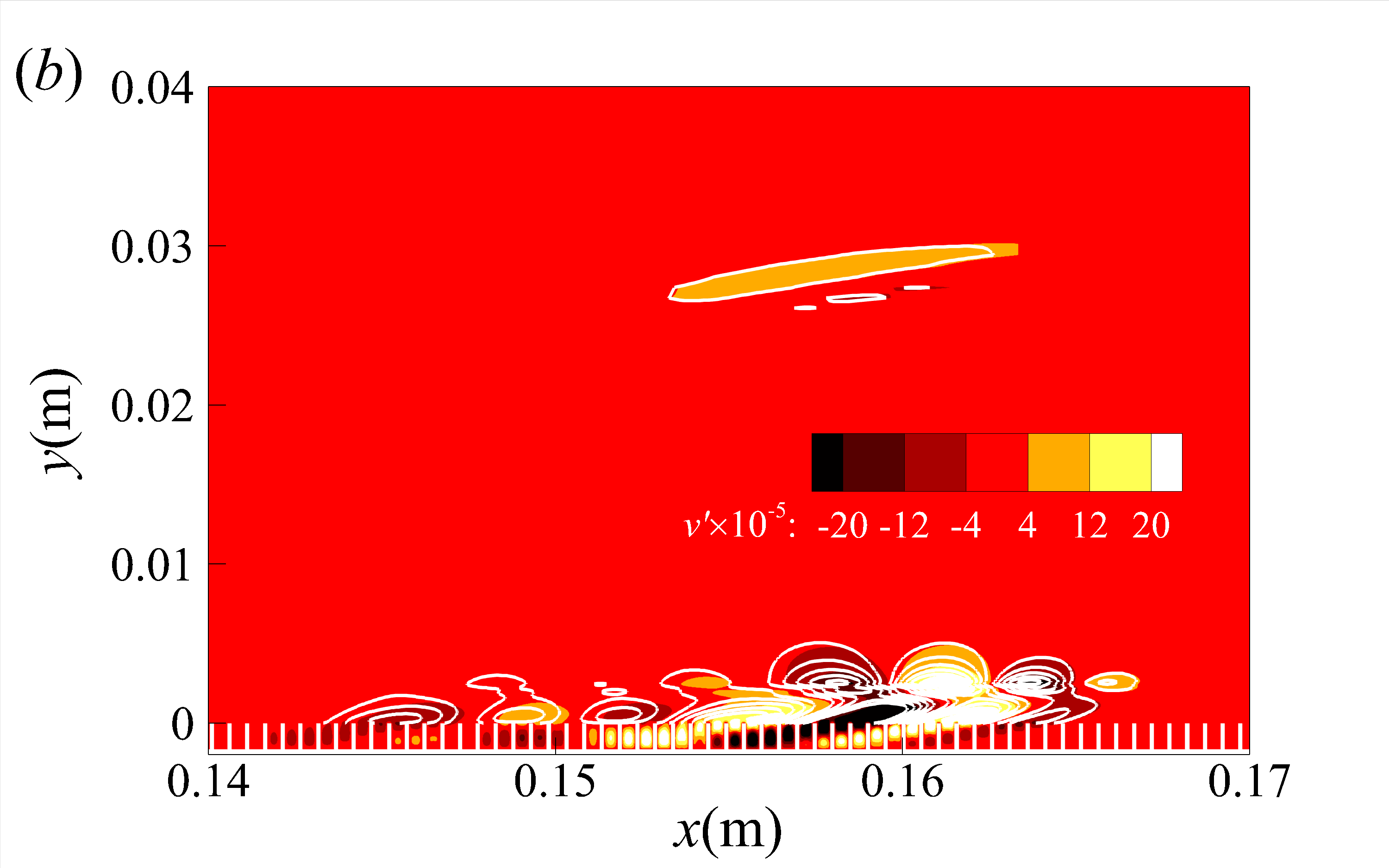}}}
	\caption{Comparison of ($a$) dimensionless pressure fluctuations and ($b$) wall-normal velocity fluctuations between the results using the meshed cavity (contour) and modelled TDIBC (solid line) subject to broadband disturbances.}
	\label{fig14}
\end{figure}

\begin{figure}
	\vspace{1em}
	\centerline{
		\subfigure{\includegraphics[width=6.5cm]{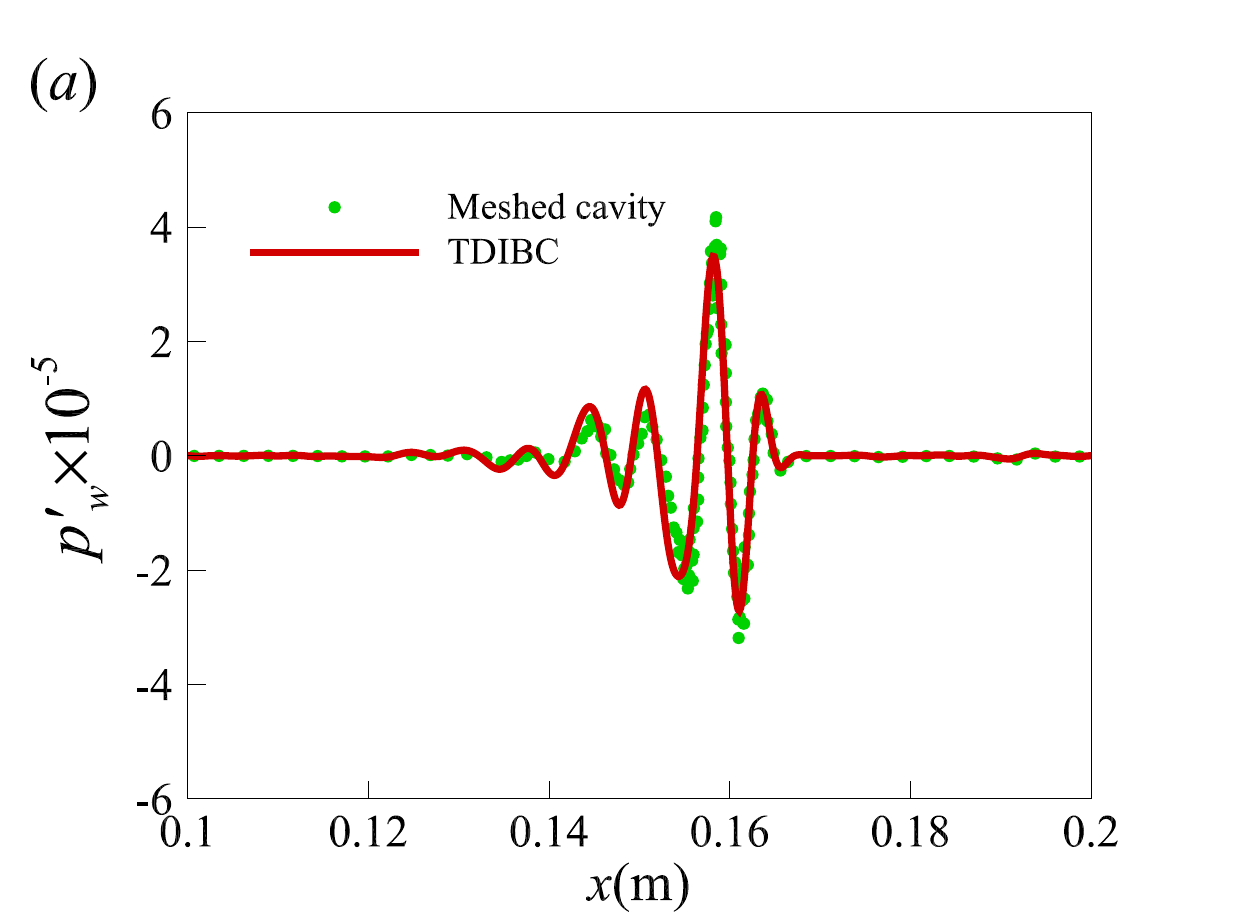}}
		\subfigure{\includegraphics[width=6.5cm]{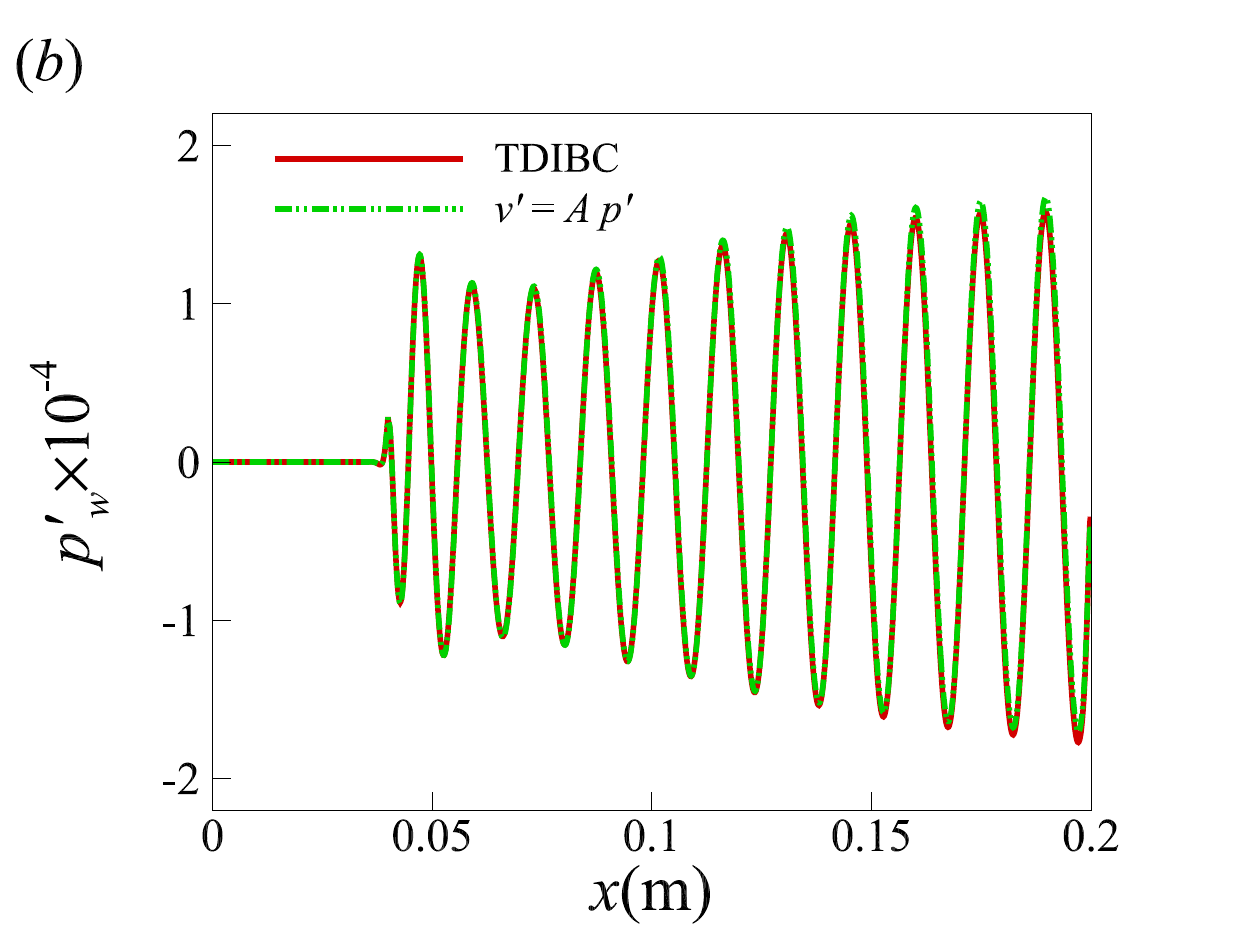}}}
	\caption{Comparison of dimensionless wall fluctuation pressure between results ($a$) using meshed cavity and modelled TDIBC subject to broadband disturbances, and ($b$) between results of using $v = Ap'$ and TDIBC for Fourier mode (3.8, 1.25).}
	\label{fig15}
\end{figure}

The contours of pressure and wall-normal velocity fluctuations are compared between the results of the meshed metasurface simulation and the TDIBC for broadband disturbances in figure \ref{fig14}. The parametric setup for broadband disturbances is the same as that in Guo \etal \space(\citeyear{Guo2023b}). It is indicated that the TDIBC-embedded solver can reproduce nearly the same flow structure and amplitude (see figure \ref{fig15}($a$)). Therefore, TDIBC can serve as an efficient tool in studying the effect of metasurface on broadband disturbances. This is because TDIBC avoids directly meshing the cavity, thus reducing the computational cost.
For three-dimensional simulations with oblique waves, the result between a constant admittance model ($v' = Ap'$) and TDIBC for oblique wave (3.8, 1.25) is plotted in figure \ref{fig15}($b$), where the admittance $A$ = (-17.041809, -2.57$ \times {10^{ - 4}}$) at this frequency. Mode (3.8, 1.25) is representative as it approaches the peak of the oblique wave in the gain contour of figure \ref{fig2}($a$). Note that the constant admittance model has been verified in previous research and utilised for simulations with single-frequency disturbances (Egorov \etal \space\citeyear{egorov2007direct}; Zhao \etal \space\citeyear{Zhao2019}). The good agreement indicates the accuracy of TDIBC in modelling the effect of the acoustic metasurface on oblique waves in three-dimensional simulations.

\section{Verification of initialisation of OpenCFD}\label{Appendix C}
\begin{figure}
	\vspace{1em}
	\centerline{
		\subfigure{\includegraphics[width=6.8cm]{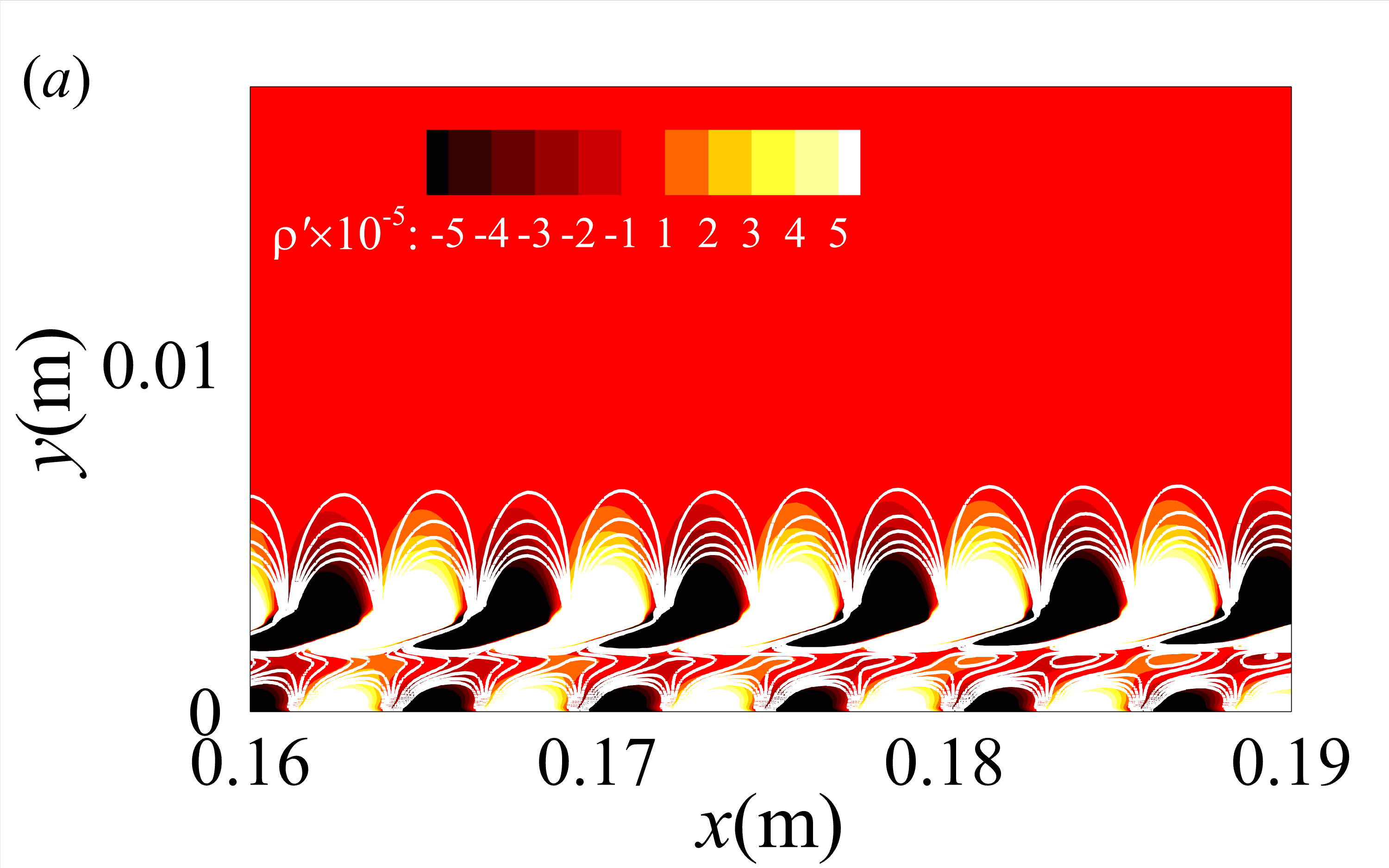}}
		\subfigure{\includegraphics[width=6.8cm]{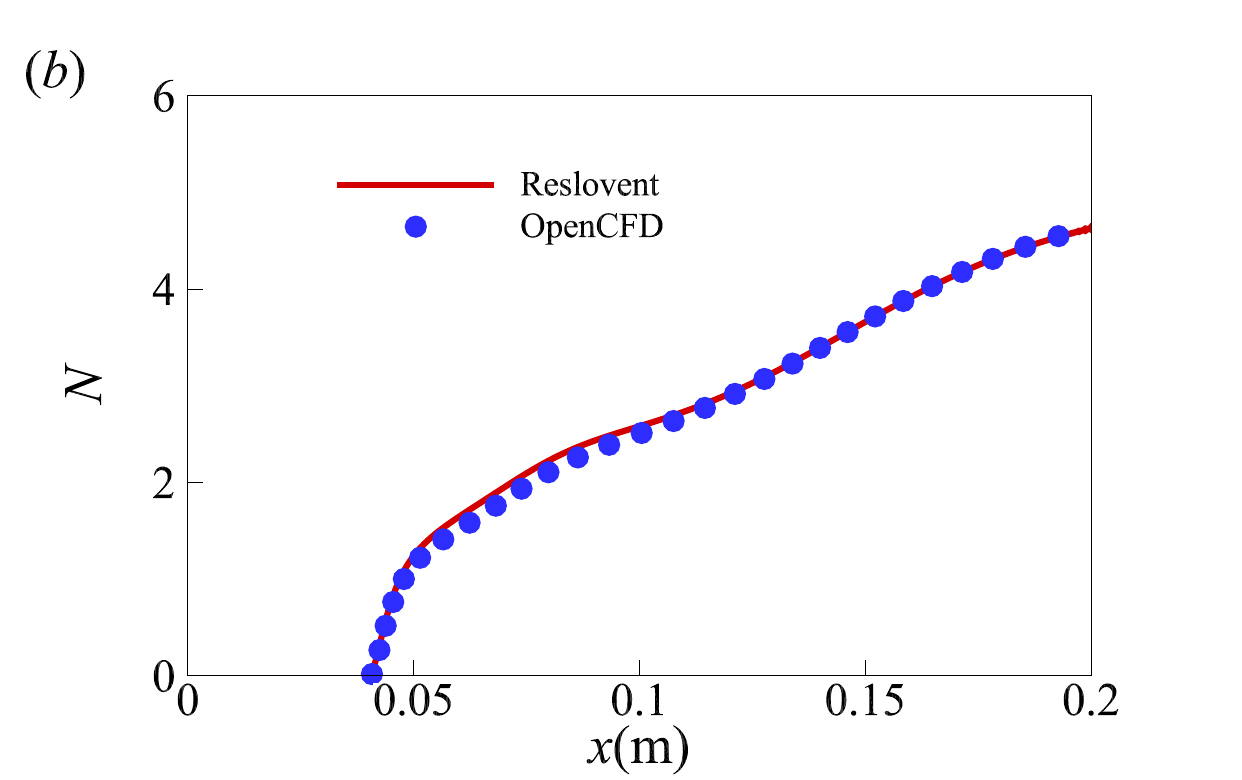}}}
	\caption{Comparison of ($a$) dimensionless density fluctuation and ($b$) $N$-factor between the results of resolvent analysis (contour) and OpenCFD (solid line) for optimal mode (10, 0). No acoustic metasurface is applied.}
	\label{fig13}
\end{figure}

The evolutions of optimal responses between the result of resolvent analysis and the N--S solver agree well with each other in the density fluctuation and the integrated $N$-factor of Chu energy density, as shown in figure \ref{fig13}.

\section{The resonate state}\label{Appendix D}
\begin{figure}
	\vspace{1em}
	\centerline{
		\subfigure{\includegraphics[width=5cm]{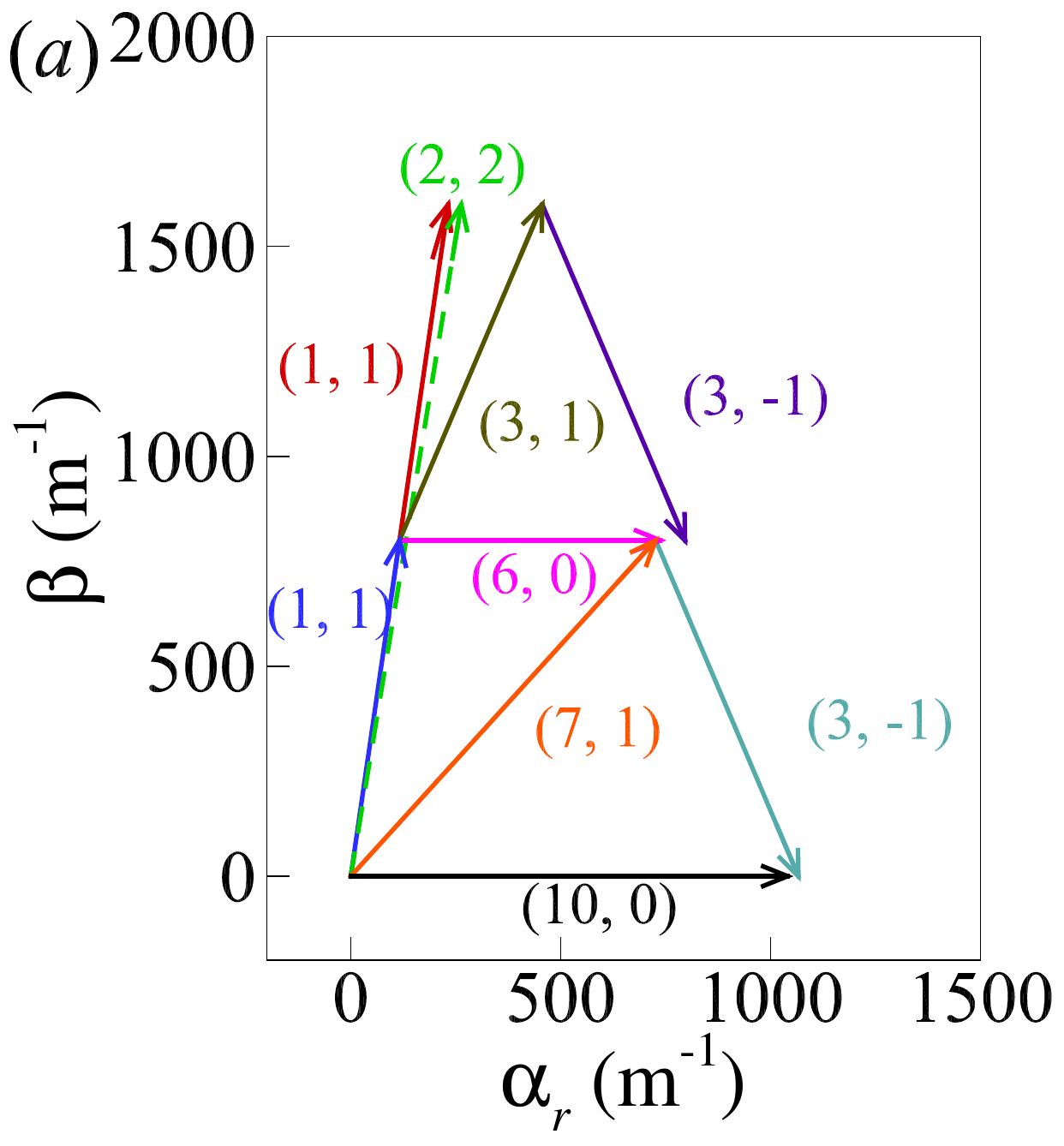}}
		\subfigure{\includegraphics[width=5cm]{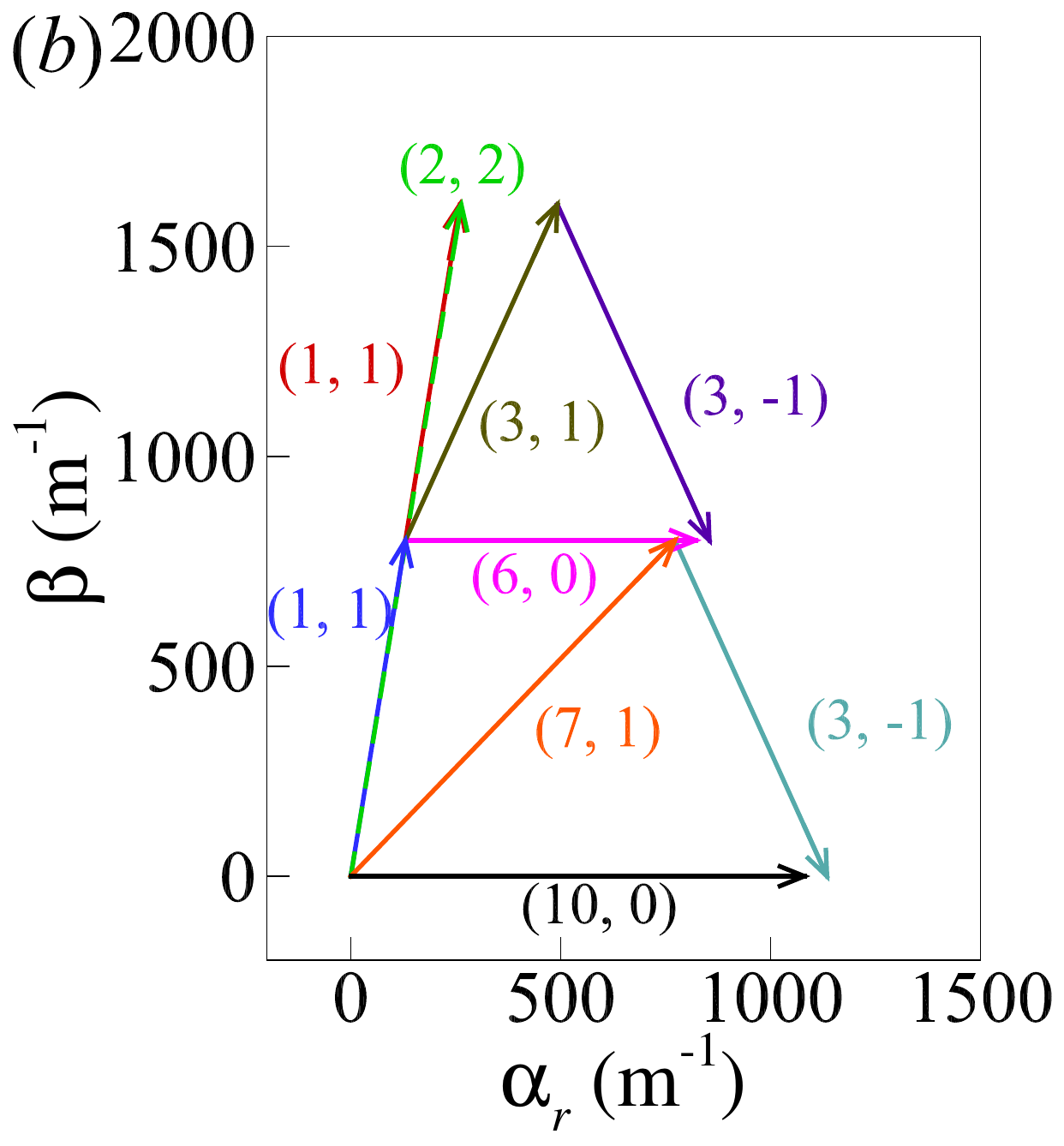}}}
	\caption{Wave vectors $({\alpha _r},{\rm{ }}\beta )$ of the modes in \ref{eq4.22} at $x$ = 0.27 m for ($a$) Case 1 and ($b$) Case 2.}
	\label{figv}
\end{figure}

The resonant state can be examined by plots of wave vectors of the Fourier modes. The resonance condition is satisfied when three wave vectors $({\alpha _r},{\rm{ }}\beta )$ of the considered triad constitute a closed form (Craik \citeyear{craik1971non}):

\begin{equation}\label{eq4.29}
{\chi _1} + {\chi _2} = {\chi _3},{\rm{ }}\chi  = {\alpha _r},{\rm{ }}\beta ,{\rm{ }}\omega.
\end{equation}
The streamwise wavenumber is obtained via ${\alpha _{r,(m,n)}} = {{\partial {\theta _{(m,n)}}} \mathord{\left/{\vphantom {{\partial {\theta _{(m,n)}}} {\partial x}}} \right.\kern-\nulldelimiterspace} {\partial x}}$, where ${{\theta _{(m,n)}}}$ is the phase angle of Fourier transform of the wall pressure for mode ($m$, $n$) using DNS data. Figure \ref{figv} depicts the wave vectors of the four interactions in \ref{eq4.22} at $x$ = 0.27 m for Case 1 and Case 2. The resonance condition for the generation of detuned mode (2, 2) is generally fulfilled at $x$ = 0.27 m for both Case 1 and Case 2.

\vspace{1.5em}

\nocite{*}
\bibliographystyle{jfm}

\bibliography{manuscript_jfm_2025_TDIBC}


\end{document}